\documentclass[reprint,aps,prl,twocolumn,letterpaper]{revtex4-2}
\usepackage{amssymb,amsfonts,amsmath}
\usepackage{times}

\usepackage{graphicx}

\usepackage{color}
\usepackage{float}
\newlength{\figurewidth}
\newlength{\pagewidth}
\begin{document}

\title{Structural Motifs and Bonding in Two Families of Boron Structures Predicted  at Megabar Pressures}
\author{Katerina P. Hilleke}
\affiliation{Department of Chemistry, State University of New York at Buffalo, Buffalo, NY 14260-3000, USA}
\author{Tadashi Ogitsu}
\affiliation{Lawrence Livermore National Laboratory, Livermore, CA, 94550, USA}
\author{Shuai Zhang}
\affiliation{Laboratory for Laser Energetics, University of Rochester, Rochester, NY, 14623, USA}
\author{Eva Zurek}
\email{ezurek@buffalo.edu}
\affiliation{Department of Chemistry, State University of New York at Buffalo, Buffalo, NY 14260-3000, USA}

\begin{abstract}
The complex crystal chemistry of elemental boron has led to numerous proposed structures with distinctive motifs as well as contradictory findings. Herein, evolutionary structure searches performed at 100~GPa have uncovered a series of potential new metastable phases of boron, and bonding analyses were carried out to elucidate their electronic structure.  These polymorphs, dynamically stable at 100~GPa, were grouped into two families. The first was derived from the thermodynamic minimum at these conditions, $\alpha$-Ga, whereas channels comprised the second.  Two additional intergrowth structures were uncovered, and it was shown they could be constructed by stacking layers of $\alpha$-Ga-like and channel-like allotropes on top of each other. A detailed bonding analysis revealed networks of four-center $\sigma$-bonding functions linked by two-center B-B bonds in the $\alpha$-Ga based structures, and networks that were largely composed of three-center $\sigma$-bonding functions in the channel-based structures. Seven of these high pressure phases were found to be metastable at atmospheric conditions, and their Vickers hardnesses were estimated to be $\sim$36~GPa.
\end{abstract}

\maketitle
\section{Introduction}
The potential for structural diversity in boron clusters \cite{Zhai:2003a,Alexandrova:2006a}, sheets \cite{Li:2019a,Sun:2017a} and crystalline lattices \cite{Albert:2009,Ogitsu:2013,Shirai:2017a}, arises from boron's propensity to adopt novel multicenter bonding schemes to accomodate its three valence electrons. Various stable and metastable polymorphs of elemental boron have been proposed as superhard \cite{Veprek:2011,Simak:2009,Zhou:2010a}, superconducting \cite{Zhang:2020,Eremets:2001,Ma:2004,Li:2014a}, and even topological materials \cite{Gao:2018}. Furthermore, because of its higher density and tensile strength as compared to plastics, this low-$Z$ material offers an option as an ablator in inertial confinement fusion and high energy density experiments, and its phase behavior as a function of temperature and pressure is of extreme interest \cite{PhysRevE.98.023205,ZhangShuai:2020}.  The structural chemistry of boron, including metastable phases that could be created in experiment, must be understood in order to accurately model its behavior under such conditions. 

\begin{figure}
\begin{center}
\includegraphics[width=\figurewidth]{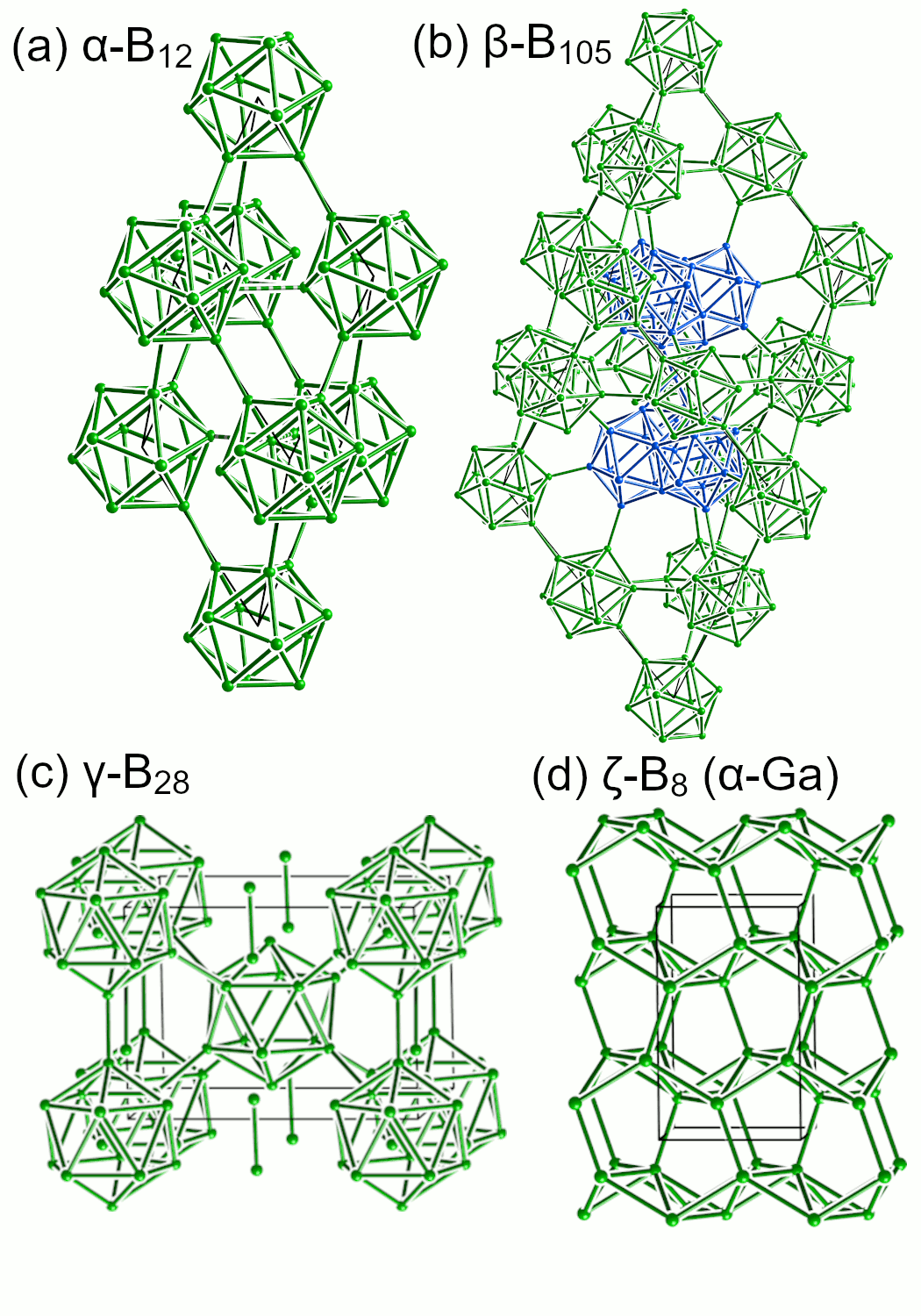}
\end{center}
\caption{Select experimentally observed allotropes of boron. At ambient pressures the (a) $\alpha$- and (b) $\beta$-rhombohedral forms predominate, with transitions to the (c) $\gamma$-orthorhombic, and (d) $\alpha$-Ga phases at 19 and 89 GPa, respectively. The characteristic icosahedra of boron-based structures (green: free standing; blue: face- and vertex-sharing) are evident in the phases that are preferred at lower pressures.
\label{fig:Boron_allotropes.png}}
\end{figure}

Some of the synthesized forms of elemental boron are illustrated in Figure \ref{fig:Boron_allotropes.png}. 
At ambient pressures two elemental phases, both based on packings of B$_{12}$ icosahedra, are known. The unit cell of the $\alpha$-rhombohedral polymorph, which can be synthesized at $T<1300$~K, contains a single icosahedron linked to neighbors by classic two-center two-electron (2c-2e) bonds between polar atoms, and three-center two-electron (3c-2e) bonds between equatorial atoms. The idealized 105-atom unit cell of the $\beta$-rhombohedral polymorph, which can be accessed at higher temperatures, contains an outer layer of 8 icosahedra at the corners and 12 icosahedra along the edges, with two fused icosahedra in the middle connected by an interstitial boron atom. Multiple computational studies have explored the relative energetics of these two polymorphs~\cite{Masago:2006a, Masago:2004, Masago:2005, Prasad:2005, Shang:2007, Siberchicot:2007, deGroot:2007, Mikhalkovic:2008a, Mikhalkovic:2008b, Ogitsu:2009, Ogitsu:2010, Ogitsu:2012, Ogitsu:2015}. Much emphasis has been placed on unravelling the role of defects on the stability of the $\beta$ phase \cite{Ogitsu:2013,Ogitsu:2009, Ogitsu:2010}. With these defects taken into account, first-principles calculations identify it as being the most stable boron allotrope at ambient pressure~\cite{deGroot:2007,Mikhalkovic:2008b,Ogitsu:2009}, although the $\alpha$ structure is close enough to make experimental differentiation of their energetics intractable~\cite{Ogitsu:2015}. Tetragonal boron allotropes have also been synthesized. It has been questioned whether one of these, T-50 \cite{Laubengeyer:1943,Hoard:1951, Hoard:1958, Hoard:1960}, can be stable in its pure form, whereas it is generally accepted that the other, T-192 \cite{Talley:1960,Vlasse:1979a,Vlasse:1979b},  is the only allotrope other than $\alpha$- or $\beta$- rhombohedral boron that is synthesizable at atmospheric pressures, although others have been reported \cite{Parakhonskiy:2013,Karmodak:2017,An:2016a}.

Computations have predicted the following sets of structural phase transitions for boron: $\alpha\text{-B}_{12}\rightarrow \gamma\text{-B}_{28}\rightarrow\alpha\text{-Ga}$ at 19 and 89~GPa \cite{Oganov:2009}. The orthorhombic unit cell of $\gamma\text{-B}_{28}$ features B$_{12}$ icosahedra and B$_2$ dumbells arranged in a rock-salt pattern. The structure of this phase, which was first synthesized in 1965 by Wentorf \cite{Wentorf:1965}, was elucidated via experiment and theory more than 40 years after it was made \cite{Zarechnaya:2008,Oganov:2009}. The disputed question of electron transfer between the dumbbells and icosahedra, and the possibility of ionic behavior in an elemental phase have led to numerous studies of the electronic structure and bonding in this allotrope~\cite{Oganov:2009, Simak:2009, Haussermann:2010,Mondal:2011}. 

von Schnering and Nesper were the first to propose that boron could adopt a $Cmca$ symmetry $\alpha$-Ga structure that does not contain any icosahedral motifs \cite{vonSchnering:1991}. First-principles calculations have verified the stability of this phase under pressure \cite{Haussermann:2003,Segall:2003}, and predicted its superconducting behavior \cite{Ma:2004}. It is suspected that the synthesized superconducting boron phase whose critical temperature, $T_c$, was measured to be as high as 11~K at 250~GPa \cite{Eremets:2001} corresponds to the $\alpha$-Ga structure. Recently, the synthesis of this polymorph has been reported at 115~GPa and 2100~K, and the phase has been dubbed $\zeta$-B \cite{Chuvashova:2017}. We note that static compression of $\beta$-B$_{105}$ leads to amorphization at $\sim$100~GPa, not to the lower enthalpy $\alpha$-Ga phase. This may be a result of the slow kinetics of boron phase transitions \cite{Sanz:2002}. Such kinetic effects are believed to be important in $\alpha$-B$_{12}$ where static compression of both doped and undoped samples showed a superconducting phase transition without a structural transformation~\cite{Shirai:2009,Shirai:2011,Nagatochi:2011}. Theoretical calculations predict a hitherto unobserved transformation to a non-icosahedral $P6_3/mcm$ symmetry structure above 375~GPa, whose $T_c$ was estimated to be 44~K at 400~GPa \cite{Li:2014a}. 

The Thomas-Fermi-Dirac model predicts that compression increasingly leads to bonding delocalization and the formation of simpler crystal structures with large coordination numbers. However, recent theoretical and experimental discoveries have shown that the behavior of elements under pressure may be much more complex~\cite{Zurek:2019k}, as in the case of lithium \cite{Nelmes:2011a,Tsuppayakorn-aek:2018}, sodium \cite{Gregoryanz:2008a,Ma:2009a} and aluminum \cite{Pickard:2010a}. Moreover, experimental variables such as the temperature and compression mechanism (static vs.\ dynamic) can affect the phases that are formed. For boron, in particular, discrepancies between the measured and calculated shock Hugoniot, as well as the computed melting temperatures suggest that new post-$\alpha$-Ga phases can be formed at megabar pressures \cite{ZhangShuai:2020}.

With this in mind, herein we employ evolutionary algorithms coupled with first-principles calculations to predict metastable phases of boron at 100~GPa. Two families of allotropes with distinct structural and bonding motifs, one based on derivatives of the $\alpha$-Ga structure and the other on phases containing channels, whose enthalpies are within 100~meV/atom of the thermodynamic minimum are found. The results suggest that many more related allotropes of boron can be constructed based on simple building blocks arranged and rearranged with different periodicity and alignments, as well as intergrowths of these two structure types. In addition to $\alpha$-Ga, six of the discovered polymorphs are metastable at atmospheric pressures, and their estimated Vickers hardnesses are comparable to rhombohedral $\alpha$-B and orthorhombic $\gamma$-B$_{28}$\cite{Veprek:2011,Solozhenko:2008,Zhang:2018}.

\section{Computational Details}

Structure searches were carried using the \textsc{XtalOpt} evolutionary algorithm (EA) version r11 \cite{Lonie:2011,Avery:2018} on unit cells containing 8, 12 and 20 boron atoms at 100~GPa. The initial set of random symmetric structures was produced via the \textsc{RandSpg} algorithm \cite{Avery:2017}, and  duplicates were identified and removed from the gene pool using the \textsc{XtalComp} algorithm \cite{Lonie:2012}. Structures within 150~meV/atom of the most stable phase found, $\alpha$-Ga, were fully optimized, and those within 100~meV/atom (a total of 10 phases) were chosen for further analysis. The structural coordinates, relative enthalpies, equations of states (EOS), and other computed parameters are provided in the Supporting Information (SI). 

Geometry optimizations and electronic structure calculations (densities of states, band structures, electron localization functions \cite{ELF}) were performed via density functional theory (DFT) as implemented in the  Vienna \emph{Ab Initio} Simulation Package (VASP) \cite{kresse1996efficient,kresse1999ultrasoft} using the Perdew-Burke-Ernzerhof (PBE) gradient-corrected exchange and correlation functional~\cite{Perdew:1996}. The projector augmented wave (PAW) method \cite{blochl1994projector} described the core states, and the boron 2s$^{2}$2p$^{1}$ electrons were treated explicitly. In the structure searches the geometries were optimized with the accurate precision setting in a two-step process. First, the atomic positions were allowed to relax within a set unit cell, followed by a full relaxation. A plane-wave basis set with an energy cutoff of 600~eV, and a $\Gamma$-centered Monkhorst-Pack scheme \cite{monkhorst1976special} where the number of divisions along each reciprocal lattice vector was chosen such that its product with the real lattice constant was 30~\AA{} was employed. These values were increased to 800~eV and 50~\AA{} for precise reoptimizations and electronic structure calculations.
Phonon densities of states and band structures were obtained using the supercell approach as implemented in the PHONOPY code~\cite{Togo:2015} in concert with VASP. 

Helmholtz free energy data under finite temperature conditions for the $\alpha$-Ga and Channel-I phases, including the electron thermal and vibrational contributions, were calculated under the quasiharmonic approximation as implemented in the phonopy-qha script. The data at various volume were fitted to a volume integrated third-order Birch-Murnaghan EOS \cite{Birch:1952}, which then gave the pressure and the Gibbs free energies were obtained thereafter.

The LOBSTER package~\cite{Deringer:2011, Maintz:2016} was used to calculate the crystal orbital Hamilton population (COHP)~\cite{Dronskowski:1993}, and the negative of the COHP integrated to the Fermi level (-iCOHP) between select atom pairs. A bonding analysis was performed using the Solid State Adaptive Natural Density Partitioning (SSAdNDP) method~\cite{Galeev:2013}, an extension of the AdNDP method \cite{Zubarev:2008} for molecules to solids. A periodic implementation of the Natural Bond Orbital (NBO) technique \cite{NBO,Dunnington:2012} interfaced with SSAdNDP was used to project the plane-wave density onto the 6-31G basis set. Calculations performed with the cc-PVTZ basis \cite{Dunning:1989} on a subset of the structures yielded similar results (see the SI). The visualization module of the SSAdNDP code was employed to generate Gaussian cube files for each bond, which were visualized using the VESTA software package \cite{Momma:2011}. Powder X-ray diffraction patterns for each structure were generated using the Diamond software package \cite{diamond}.

\section{Two Families of Metastable Boron Phases}

The evolutionary searches carried out at 100~GPa identified $\alpha$-Ga as the most stable phase, in agreement with previous theoretical studies \cite{Haussermann:2003,Segall:2003,Fan:2014,Oganov:2009}, and the recent synthesis of this phase in a diamond anvil cell at 115~GPa and 2100~K \cite{Chuvashova:2017}. Provided the energy or enthalpy of a metastable phase is not too high, it could potentially be made by a judicious choice of the temperature, heating method, pressure, and starting material, or by using synthesis techniques to access phases far away from equilibrium \cite{Parija:2018a}. A data mining study of the Materials Project high throughput database, which contains DFT energies computed at $T=0$~K of structures found in the Inorganic Crystal Structure Database (ICSD), was recently carried out to quantify the thermodynamic scale of metastability \cite{materialsproject}. The 90$^{th}$ percentile of the DFT-calculated metastability of all of the known inorganic crystalline materials was found to be $\sim$70~meV/atom. However, examples of known polymorphs whose energy was $\sim$150~meV/atom above the ground state illustrated that  a low energy of metastability did not necessarily correspond with synthesizability. With this in mind, we chose to analyze the phases identified by the EA search whose enthalpies, computed within the static lattice approximation, were within 100~meV/atom of the $\alpha$-Ga phase. Several recent noteable examples of metastable materials that have been synthesized under pressure include superconducting PH$_n$ \cite{drozdov2015superconductivity,Zurek:2015j,Flores-Livas:2016,Zurek:2017c}  and CSH$_x$ \cite{Snider:2020a,Zurek:2020b,Sun:2020a} phases.

Table I lists the space-groups and enthalpies (with and
without zero-point-energy, ZPE, corrections) with respect to the $\alpha$-Ga phase of the structures chosen for further analysis. Phonon calculations confirmed they were dynamically stable at 100 GPa. Generally speaking, phases that were related to $\alpha$-Ga had a lower ZPE
than the channel structures, with the $\alpha$-Ga having the second lowest ZPE and the Channel-I the highest. The quasiharmonic approximation was used to calculate the relative Gibbs free energies of the $\alpha$-Ga and Channel-I phases (see Fig S17-S20). These calculations revealed that finite temperature conditions stabilize the Channel-I structure more than the $\alpha$-Ga phase, although the $\alpha$-Ga phase remains thermodynamically preferred at 100 GPa up to 1000 K. However, the Channel-I phase (to be described below) is
within $k_{B}T$, where $k_{B}$ is the Boltzmann constant, of $\alpha$-Ga at room temperature. While distinct from one another, a number of morphological connections could be drawn between these phases, resulting in two main families -- those based on $\alpha$-Ga, and those featuring channels. Additionally, two phases arising from the intergrowth of the $\alpha$-Ga and channel structures were identified.

\begin{table}
     \centering
     \def\arraystretch{1}
     \caption{Enthalpies (meV/atom)  of the ten most stable boron allotropes predicted at 100~GPa with and without zero-point-energy (ZPE) corrections relative to the enthalpy of the $\alpha$-Ga phase. }
      \setlength{\tabcolsep}{3mm}{
        \begin{tabular}{c c c c}
\hline
Structure & Space group  & $\Delta H$ & $\Delta H$+ZPE  \\
\hline
\hline
$\alpha$-Ga & $Cmca$  & 0.0 & 0.0 \\
Channel-I & $C2/m$  & 12.5 & 15.1 \\
Channel-II & $C2/m$  & 45.8 & 47.5 \\
Intergrowth-I & $C2/m$  & 51.1 & 52.7 \\
Channel-III & $C2/m$  & 61.9 & 63.2 \\
$\alpha$-Ga'-I & $P\bar{1}$  & 63.9 & 65.7 \\
$\alpha$-Ga'-II & $P\bar{1}$  & 70.8 & 71.9 \\
$\alpha$-Ga'-III & $P4_{2}2_{1}$2  & 77.1 & 76.6 \\
Intergrowth-II & $P\bar{1}$  & 85.3 & 85.9 \\
Channel-IV & $Ima2$  & 89.2 & 90.2 \\

\hline
\end{tabular}
\label{tab:enthalpies}}
\end{table}

\begin{figure}
\begin{center}
\includegraphics[width=\figurewidth]{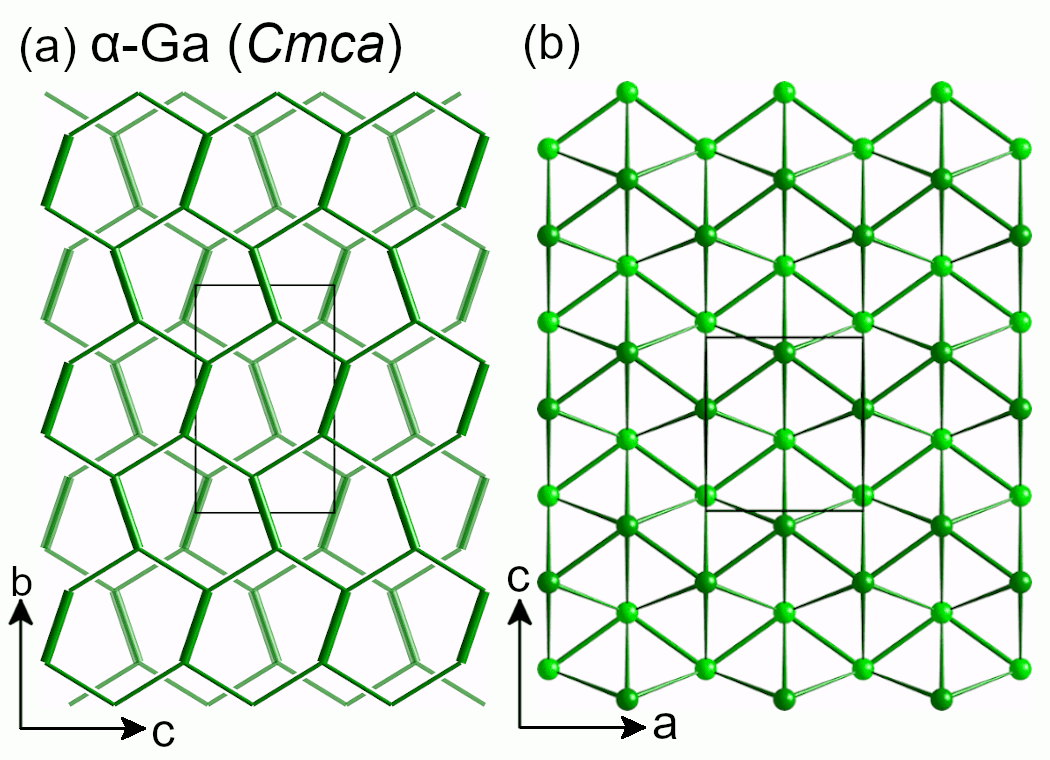}
\end{center}
\caption{Illustration of the $\alpha$-Ga structure of elemental boron ($\zeta$-B) \cite{Chuvashova:2017}. It can be viewed as (a) a layered structure consisting of sheared honeycomb nets, with thicker contacts along short B-B distances, and (b) as layers of buckled trigonal nets connected by the short B-B contacts highlighted in (a). The light green and dark green boron atoms do not lie in the same plane.
\label{fig:alpha_Ga_structural_figure_v2.png}}
\end{figure}

The first group of structures is derived from the  $\alpha$-Ga phase, which is the thermodynamic minimum. As shown in Figure \ref{fig:alpha_Ga_structural_figure_v2.png}(a) this $Cmca$ symmetry structure is composed of layers of sheared honeycomb nets stacked in an ABAB... pattern along the $a$ axis.  Figure \ref{fig:alpha_Ga_structural_figure_v2.png}(b) highlights another viewpoint: layers of buckled triangular sheets of boron atoms that are stacked along the $c$ axis and connected by short B-B contacts measuring 1.63~\AA{} comprise this phase. Within the buckled sheets, the B-B distances are slightly longer, ranging from 1.67-1.75 \r{A}.

\begin{figure}
\begin{center}
\includegraphics[width=\figurewidth]{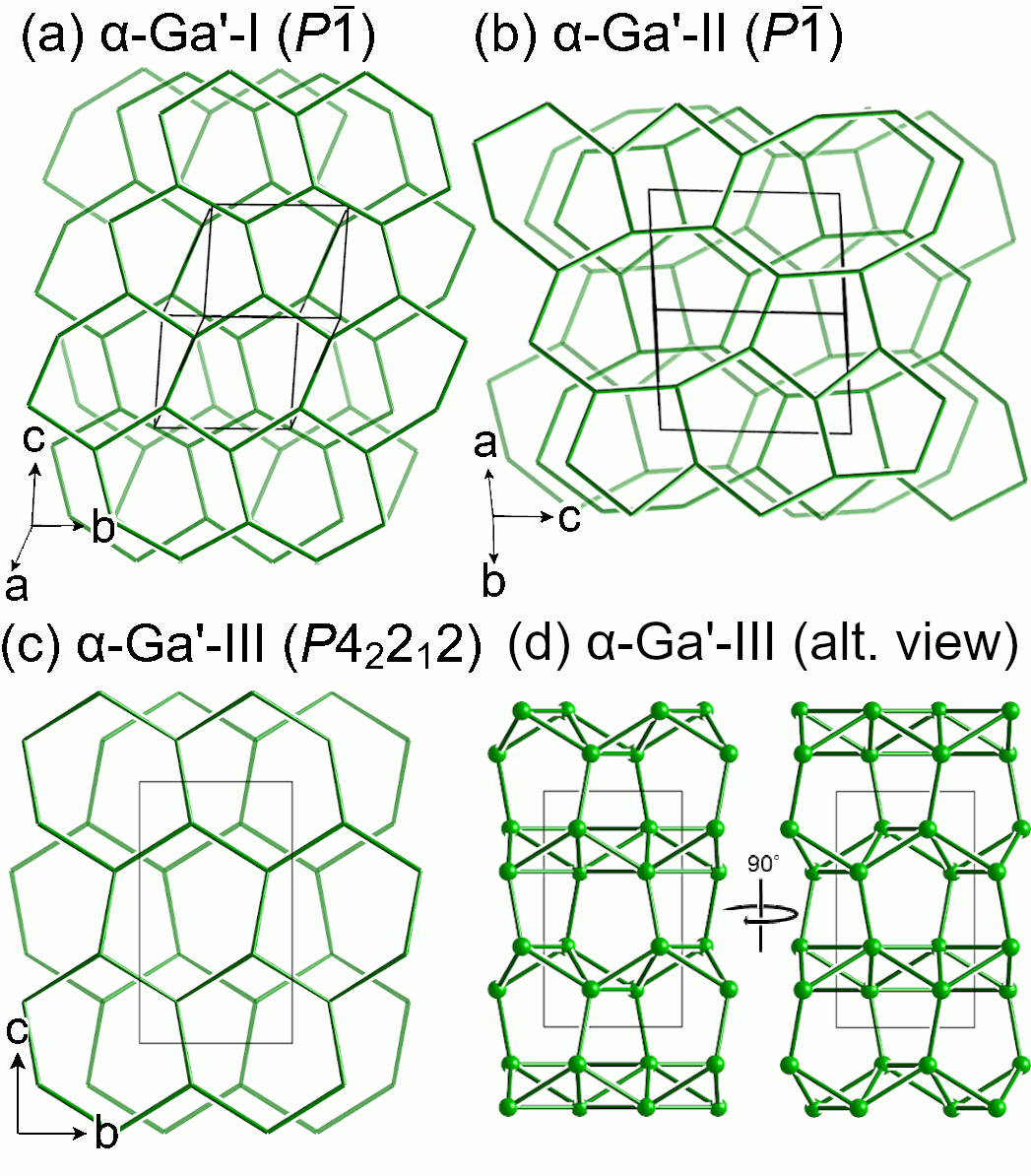}
\end{center}
\caption{Boron allotropes related to $\alpha$-Ga. The (a) $\alpha$-Ga'-I,  and (b) $\alpha$-Ga'-II  structures are based on stacks of sheared honeycomb nets with different stacking orders and shearing patterns. The (c,d) $\alpha$-Ga'-III phase can be directly derived from $\alpha$-Ga by rotating every buckled layer by 90$^\circ$ from the layer above it.
\label{fig:alpha_Ga_derivatives_struct.png}}
\end{figure}

The structural features of three of the metastable phases found, denoted as $\alpha$-Ga'-I, II, and III, place them within the same family as $\alpha$-Ga. As shown in Figure \ref{fig:alpha_Ga_derivatives_struct.png}, the first of these polymorphs, with P$\bar{1}$ symmetry, is constructed via an ABCABC... stacking of sheared honeycomb nets. The shearing pattern differs from that of the parent $\alpha$-Ga structure, resulting in a changed repeat period and a much lower symmetry. However, rotating the crystal by 90$^\circ$ still reveals stacked layers of buckled triangular nets of boron,  a consistent feature in these $\alpha$-Ga-based structural derivatives. The $\alpha$-Ga'-II phase, also of P$\bar{1}$ symmetry, consists of yet another ABCABC... stacking pattern, this time not only of hexagons but also of pentagons and octagons. Lastly, the $\alpha$-Ga'-III phase, with P4$_{2}$2$_{1}$2 symmetry, is an ABAB... stacking of sheared hexagonal layers very similar to that of the parent $\alpha$-Ga structure. However, viewing it as layers of buckled triangular boron nets stacked along the $c$ axis reveals that every other net is rotated by 90$^\circ$ relative to the ones above and below, resulting in a 4$_{2}$ screw axis. 

One could imagine multiple other derivatives of the $\alpha$-Ga structure based on different shearing or stacking patterns of hexagonal nets; indeed, several more such polymorphs were produced by \textsc{XtalOpt}, although out of the range of enthalpies we considered. Moreover, the $Imma$ symmetry $o$-B$_{16}$ phase proposed by Fan et al. \cite{Fan:2014} is built from sheared honeycomb nets. At 100~GPa we calculate its enthalpy as being 33~meV/atom higher than that of $\alpha$-Ga, in-line with the structures presented here, although it was not found in our EA searches.

\begin{figure*}
\begin{center}
\includegraphics[width=\pagewidth]{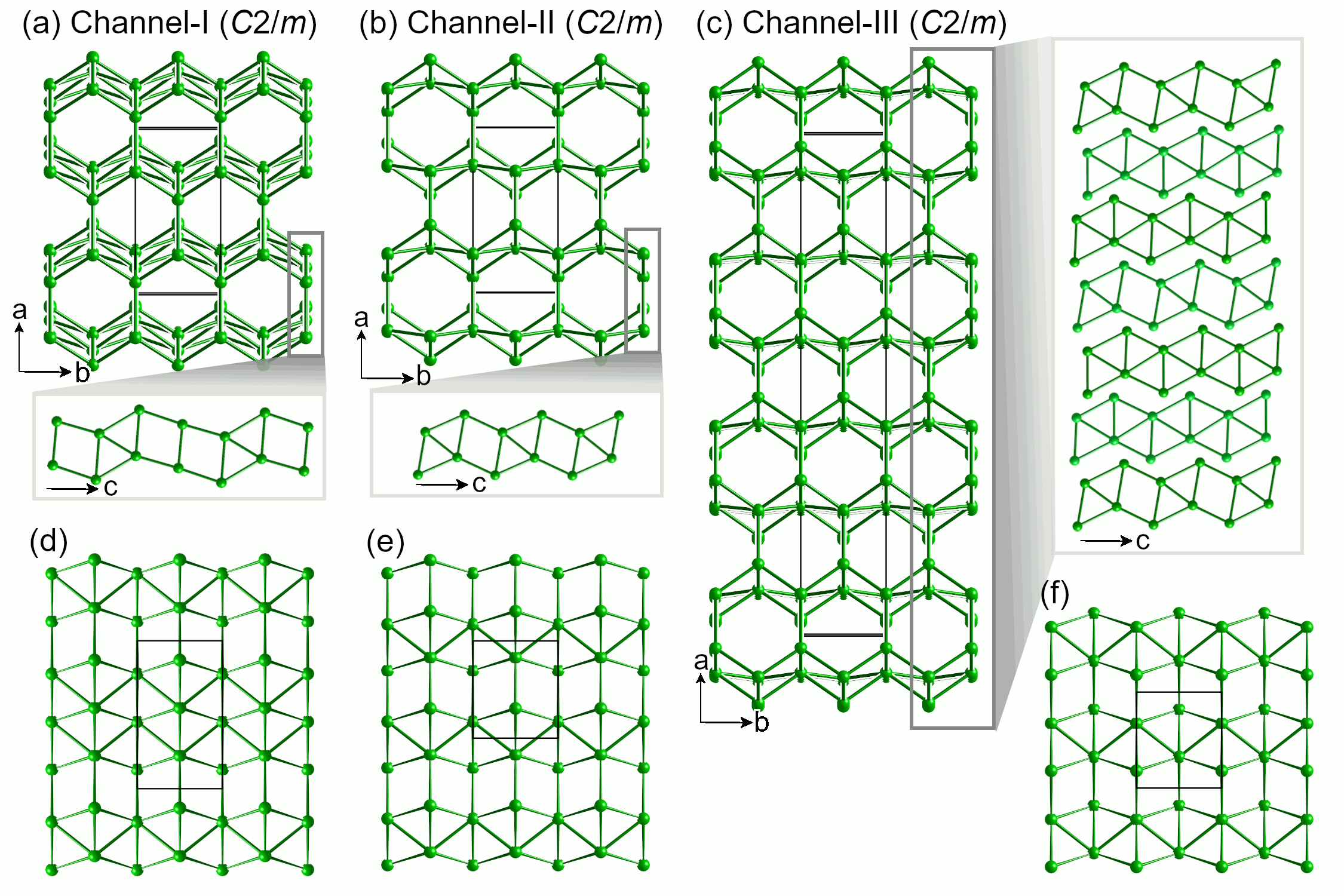}
\end{center}
\caption{Channel-based family of boron structures that contain diamond-like B$_{4}$ motifs made from two edge-sharing triangles, along with more open parallelograms (strips connecting buckled layers are shown in the insets), with different arrangements producing different phases: (a) Channel-I, (b) Channel-II, and (c) Channel-III. The buckled layers comprising these phases are made from: two adjacent rows of diamond-like motifs separated by a single row of parallelogram units in (d) Channel-I, or alternating strips of diamond-like B$_4$ units and parallelograms as in the (e) Channel-II, and (f) Channel-III structures. 
\label{fig:Channel_struct.png}}
\end{figure*}

The second family of phases found in the EA searches features large open channels. The three most stable members of this group, dubbed Channel-I, II, and III are shown in Figure \ref{fig:Channel_struct.png}. As in the case of $\alpha$-Ga and its derivatives, they can be viewed as stacked layers of buckled boron nets. However, unlike the triangular nets in the $\alpha$-Ga derivatives, parallelograms and pairs of triangles join to form diamond-like motifs that comprise the nets. Instead of being connected by B-B bonds, as in the $\alpha$-Ga family, the buckled layers in the channel structures are joined by strips that also contain these parallelogram and diamond motifs, as shown in the insets. In the Channel-I phase, two open parallelogram units are adjacent in these connecting strips (Figure \ref{fig:Channel_struct.png}(a), inset). The Channel-II phase can be viewed as a distorted version of the Al network comprising the $I4/mmm$ BaAl$_4$ structure type \cite{Shatruk:2019a}.  In fact, a recent theoretical study suggested that a boron allotrope with this symmetry could be made by removing the Na atoms from an $I4/mmm$ NaB$_4$ phase that was predicted to be stable under pressure, and quenching to 1~atm \cite{Zhang:2020}. Placing this open-channel boron framework under pressure might yield the channel-based structures highlighted here. Another predicted compound that features similar, albeit larger channels based on networks of B$_4$ diamond and square motifs is a $Pnnm$-B$_{16}$ phase whose electronic structure reveals exotic topological states \cite{Dong:2018}. 

\begin{figure}
\begin{center}
\includegraphics[width=\figurewidth]{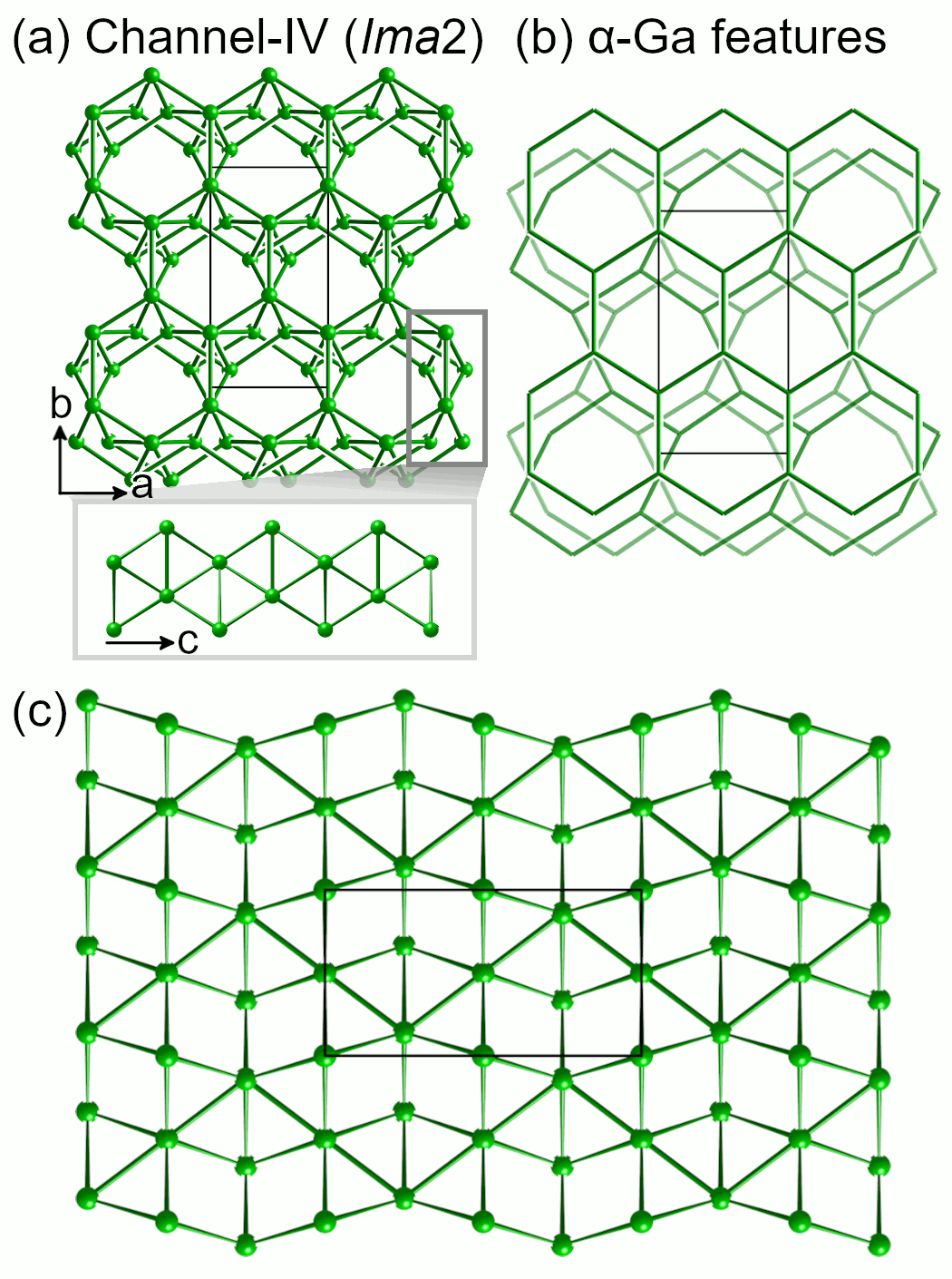}
\end{center}
\caption{(a) The Channel-IV structure is built up of corrugated strips of B triangles (inset) connecting buckled layers of B atoms, leading to open channels in the phase. (b) It can be alternately viewed as another derivative of the $\alpha$-Ga structure built up of sheared hexagonal nets. (c) The buckled layers of this structure reveal a complex pattern of diamond-like B$_{4}$ and open parallelogram motifs. 
\label{fig:Fig5_ChannelIV_alphaGa.png}}
\end{figure}

Changing patterns and repeat periods of strips of diamonds and squares in the buckled nets give rise to the Channel-I, II, and III structures, which all possess $C2/m$ symmetry. In comparison to these, the Channel-IV phase shown in Figure \ref{fig:Fig5_ChannelIV_alphaGa.png}, which assumes the $Ima2$ spacegroup, is more complex. Whereas the buckled layers in the three $C2/m$ symmetry phases are connected by planar strips composed of diamonds and squares, the connections in the Channel-IV structure are crimped,  resulting in a visually less open channel. As illustrated in Figure \ref{fig:Fig5_ChannelIV_alphaGa.png}(b), the Channel-IV phase can also be viewed as a derivative of $\alpha$-Ga that is constructed from layers of hexagonal nets in such a way that leads to the formation of channels.

\begin{figure}
\begin{center}
\includegraphics[width=\figurewidth]{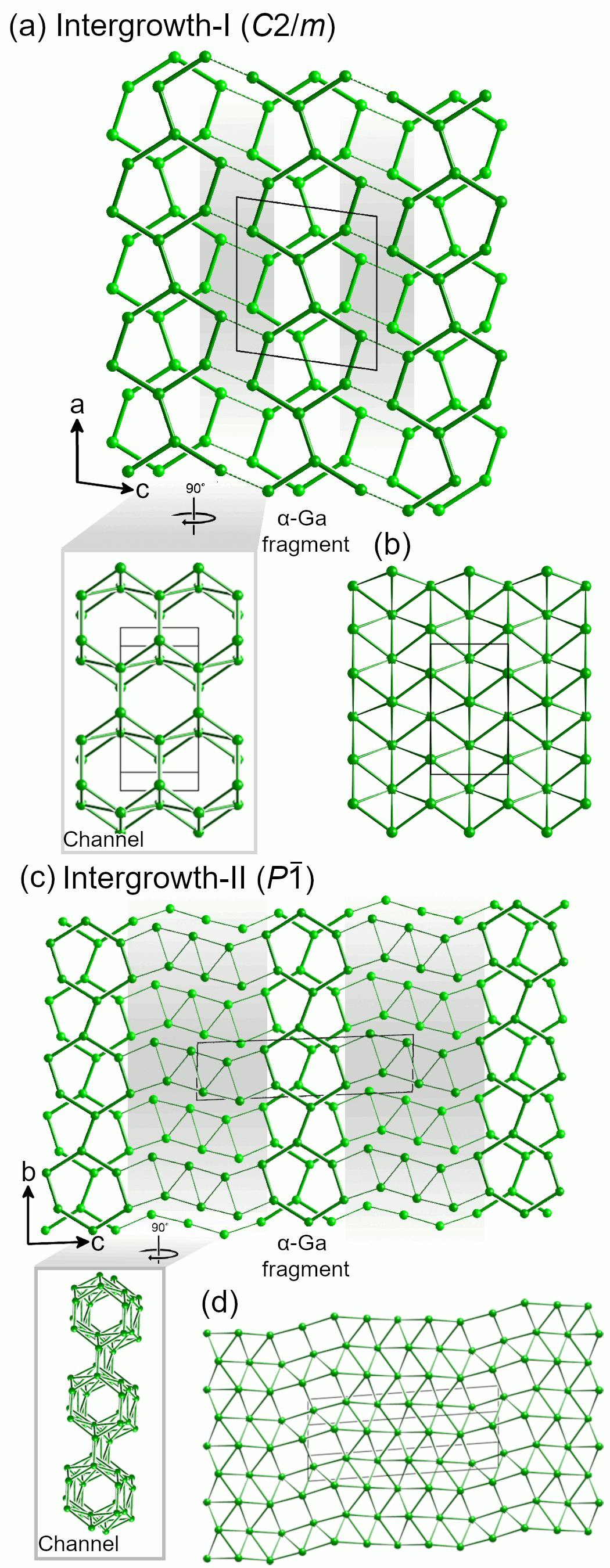}
\end{center}
\caption{(a) The Intergrowth-I structure, and one of its (b) buckled layers. (c) The Intergrowth-II structure, and (d) one of its buckled layers.  Both phases are  based on the $\alpha$-Ga and channel structures, with thicker bonds highlighting $\alpha$-Ga-based regions. Grey backgrounds span channel-based regions (shown in the insets), which are larger in the Intergrowth-II structure.  
\label{fig:Fig6_Intergrowths.png}}
\end{figure}

Finally, the two structures shown in Figure \ref{fig:Fig6_Intergrowths.png}, dubbed Intergrowth-I and Intergrowth-II, which contained distinct $\alpha$-Ga and channel-like regions were identified. In the first of these, with $C2/m$ symmetry, the two regions of near equal size are intergrown along the $c$ axis. In Figure \ref{fig:Fig6_Intergrowths.png}(a) regions of the $\alpha$-Ga structure are drawn with thicker connections between atoms, highlighting the sheared honeycomb motifs, whereas thinner lines connect boron atoms within the channel-based regions (highlighted in grey; see inset). In Intergrowth-I, a view of the buckled layers reveals that each boron atom is linked in a purely triangular network. The open parallelograms seen in purely channel-based structures are thus not present, due to the small spatial extent of the channel-based regions in the intergrown phase. In Intergrowth-II, with P$\bar{1}$ symmetry, the two regions are again intergrown along the $c$ axis, although with the channel structure-based regions (highlighted in grey) being much larger than in Intergrowth-I. Consequently, the distortions of the buckled layers present in Intergrowth-II are similar to those seen in the polymorphs that only contain channels. Additionally, as with the multiple possible adaptations that could be imagined for the $\alpha$-Ga derivatives, one can anticipate a plethora of possible intergrowths between the $\alpha$-Ga and channel structures, with differing repeat periods and relative amounts of each phase being incorporated. Indeed, one might imagine blurring the line between an intergrowth structure as shown here and two crystals with $\alpha$-Ga and channel morphology interfaced with one another.

\section{Factors Affecting Stability}

In the static lattice approximation at 0~K the thermodynamic variable governing relative phase stability is the enthalpy, which is the sum of the internal energy and the pressure volume term, given as $H=U+PV$. At high pressures the magnitude of the $PV$ term, which favors denser structures, can be greater than that of the $U$ term, which is a reflection of the bonding present in the system. We therefore wondered if the geometric pecularities between the two families of phases could be traced to different factors affecting their stability? To explore the driving forces for stabilization we calculated the aforementioned quantities for the nine metastable phases predicted here, and compared them with those obtained for the $\alpha$-B$_{12}$, $\beta$-B$_{105}$, and $\gamma$-B$_{28}$ phases up to 300~GPa.

\begin{figure*}
\begin{center}
\includegraphics[width=\pagewidth]{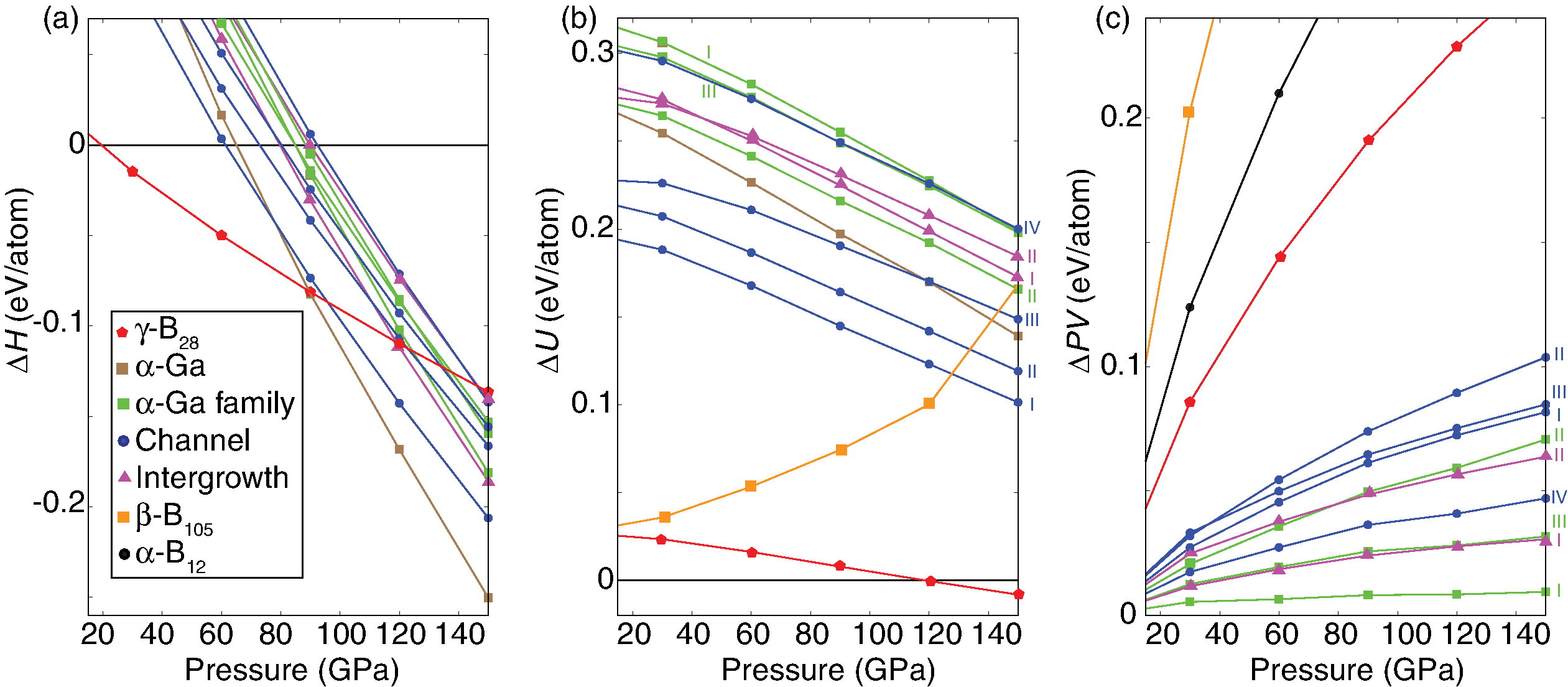}
\end{center}
\caption{Enthalpies of select phases of boron, including those listed in Table 1, with respect to that of $\alpha$-B$_{12}$ as a function of pressure. (b) The internal energy of these boron phases relative to $\alpha$-B$_{12}$. (c) The $PV$ contribution to the enthalpy of these boron phases relative to that of the $\alpha$-Ga polymorph. Channel-based phases tend to have a lower internal energy, while those based on $\alpha$-Ga have smaller $PV$ contributions to the enthalpy.
\label{fig:Fig7_enthalpy-terms_new.png}}
\end{figure*}

For ease of visualization, the enthalpies are plotted relative to that of the $\alpha$-B$_{12}$ phase in Figure \ref{fig:Fig7_enthalpy-terms_new.png}(a). At 0 GPa $\alpha$-B$_{12}$ and a model of the  $\beta$-B$_{105}$ structure wherein all of the atomic positions are fully occupied are within 26~meV/atom of each other. $\alpha$-B$_{12}$ remains the thermodynamic minimum until $\sim$19~GPa, at which point $\gamma$-B$_{28}$ becomes the thermodynamic minimum. The $\gamma$-B$_{28}$ polymorph persists until 89~GPa, when the $\alpha$-Ga phase becomes lower in enthalpy \cite{Oganov:2009}. Remarkably, until $\sim$70~GPa, the Channel-I phase is actually lower in enthalpy than $\alpha$-Ga, although neither is more stable than $\gamma$-B$_{28}$.  At higher pressures the $\gamma$-B$_{28}$ phase becomes progressively more destabilized as compared to the polymorphs found within the EA search, a reflection of the poor space-filling abilities of icosahedra. 

Examination of Figure \ref{fig:Fig7_enthalpy-terms_new.png}(b) illustrates that the stability of the Channel-I structure can be linked to its low internal energy. With increasing pressure the internal energies of  $\gamma$-B$_{28}$, $\alpha$-Ga, as well as the phases generated in the EA search approach and even begin to overtake  $\alpha$-B$_{12}$. Generally speaking, the channel polymorphs tend to have lower internal energies than the $\alpha$-Ga derivatives, with those of the intergrowths being intermediate between the two. Meanwhile, as shown in Figure \ref{fig:Fig7_enthalpy-terms_new.png}(c), the $\alpha$-Ga structure has the lowest $PV$ contribution to the enthalpy, with the $\alpha$-B$_{12}$, $\gamma$-B$_{28}$, and $\beta$-B$_{105}$ having the highest values, respectively, reflecting the low packing density achieved by icosahedra. Unsurprisingly, the $\alpha$-Ga derivatives tend to be denser than the channel phases, with the intergrowth phases being intermediate.

Thus, the low enthalpies of the channel structures can be traced back to their large negative internal energies, which are offset by the large positive $PV$ terms. In comparison, $\alpha$-Ga and its derivatives have smaller volumes, but the magnitude of their internal energies are also smaller. The Channel-IV allotrope, whose $PV$ term is less than that of both the Intergrowth-II and $\alpha$-Ga'-II phases, is an exception. As previously discussed, although this polymorph resembles the channel-based structures, it can also be considered as an $\alpha$-Ga derivative. Its relatively low $PV$ term and high internal energy are in-line with the latter interpretation, however a closer look (Figure ~\ref{fig:Fig5_ChannelIV_alphaGa.png}(a)) shows that,  rather than the planar chains of square-and-diamond boron motifs connecting buckled layers of boron, the chains running along the $c$ axis are themselves corrugated, leading to a smaller relative volume.

\section{Bonding Analysis}

\begin{figure}
\begin{center}
\includegraphics[width=\figurewidth]{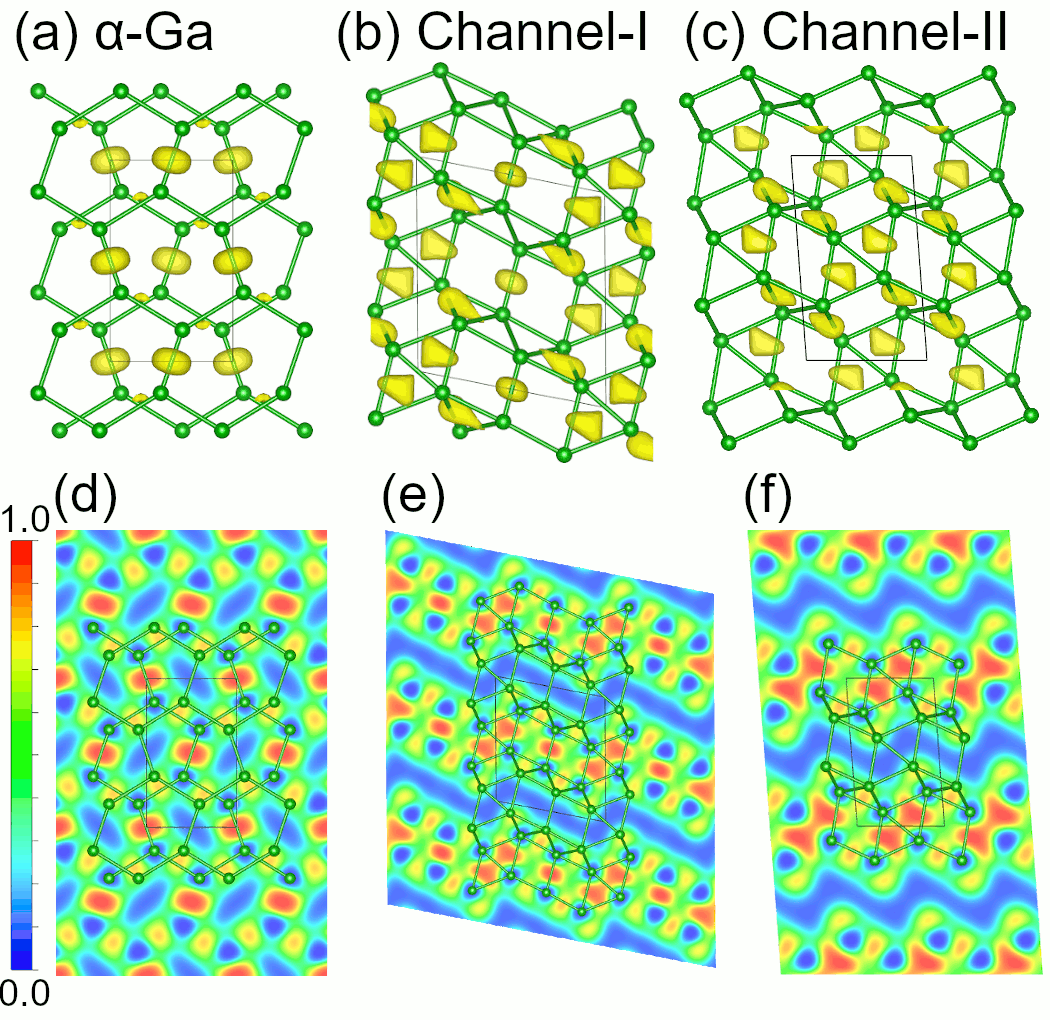}
\end{center}
\caption{Isosurface (isovalue = 0.8) plots of the ELF for the (a) $\alpha$-Ga, (b) Channel-I, and (c) Channel-II structures at 100 GPa. Plots of the ELF in the (d) (100) plane of $\alpha$-Ga, (e) (010) plane of Channel-I, and (f) (010) plane of Channel-II are also provided.
\label{fig:Fig8_ELF.png}}
\end{figure}

The prevalence of peculiar bonding schemes in boron-based materials \cite{Longuet-Higgings:1955,vonSchnering:1991,Prasad:2005,Jemmis:2001,Fujimori:1999,Wade:1976,Mingos:1984} including the $\gamma$-B$_{28}$ phase \cite{Haussermann:2010,Mondal:2011,Oganov:2009}, which frequently involve multicenter bonds, led us to wonder what sorts of curious features were present in the discovered polymorphs, and if their sorting into structural families heralded similar bonding arrangements? To investigate this we began by calculating the Electron Localization Function (ELF) for each polymorph; a summary of the results obtained for the $\alpha$-Ga, Channel-I, and Channel-II phases at 100~GPa is presented in Figure \ref{fig:Fig8_ELF.png}. For $\alpha$-Ga, ELF maxima are observed along the short 1.63~\AA{} B-B contacts connecting the buckled nets that lie in the $ac$ plane, suggestive of classic two-center two-electron (2c-2e) bonding regions. Such maxima are also present in the ELF plot for the Channel-I phase along B-B contacts 1.62~\AA{} apart, on the edges shared by two parallelograms within the chain of parallelograms and diamonds that connect the buckled nets in the $bc$ plane. ELF maxima are also present within the triangles comprising the B$_4$ diamond-like motifs, and along the buckled nets, although these are spread over a larger region. Finally, the ELF maxima within the Channel-II structure are located within the B$_{4}$ diamonds, similar to the Channel-I phase, although no signatures of 2c-2e bonding interactions are detected. As shown in the plots provided in the SI (Fig. S8-S14) these general features are present in the remaining structures, with the $\alpha$-Ga derivatives displaying ELF maxima indicative of 2c-2e bonds formed along the short contacts connecting buckled nets, and the maxima in the channel-based compounds arising within the diamond motifs. 

To gain a deeper understanding of the bonding features present in these systems, we used the SSAdNDP bonding analysis \cite{Galeev:2013}, an extension of the NBO technique. Within this method, the plane-wave DFT electronic structure is projected onto a Natural Atomic Orbital basis and the resulting density matrix is used to locate putative bonding functions with occupancies above a user-defined cutoff. In this way, a chemically meaningful, localized bonding scheme can be derived containing lone pairs and classical 2c-2e bonds, as well as the more complex multicenter bonding motifs that may be required for an accurate description of the bonding in extended systems. Indeed, a hypothetical planar boron allotrope, the $\alpha$-sheet, was used as a proof of concept for the applicability of the SSAdNDP analysis, with bonding functions ranging from 3c-2e to 7c-2e being required to account for all valence electrons of the system \cite{Galeev:2013}. In addition, a novel phase consisting of stacked wiggle $\alpha$-sheets, 3D-$\alpha$' boron, was proposed to be the first three-dimensional topological boron structure, with a spindle nodal chain arising from the intersection of nodal lines and rings in momentum space \cite{Gao:2018}. Its curious electronic structure was linked to the same sorts of $\pi$-bonds present in the 2D $\alpha$-sheet analogue. 

\begin{figure}
\begin{center}
\includegraphics[width=\figurewidth]{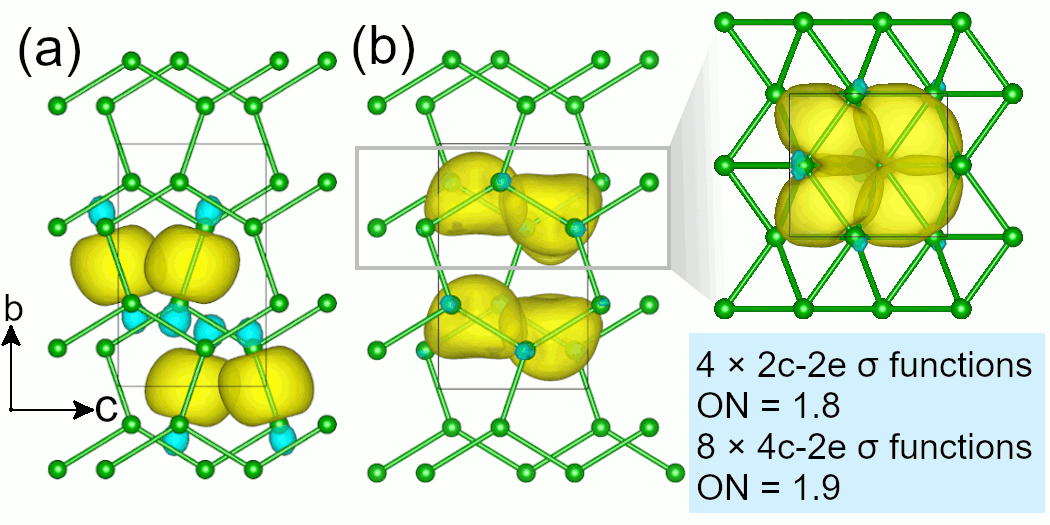}
\end{center}
\caption{Isosurfaces (isovalue=0.08~a.u.) of the bonding functions obtained with the SSAdNDP method for $\alpha$-Ga at 100~GPa. (a) Four 2c-2e $\sigma$-bonds between buckled layers of boron atoms, and (b) eight 4c-2e $\sigma$-bonds within buckled layers of boron atoms account for all 24 valence electrons in the structure. 
\label{fig:Fig9_aGa_SSAdNDP.png}}
\end{figure}

Let us begin with the simplest boron allotrope considered herein, the $Cmca$ $\alpha$-Ga phase whose unit cell contains eight boron atoms, resulting in a total of 24 valence electrons to be accounted for. The SSAdNDP analysis assigns eight of these to four 2c-2e $\sigma$-bonding functions with occupation numbers (ONs) of 1.77 $\vert$e$\vert$ that connect the buckled nets, as shown in Figure \ref{fig:Fig9_aGa_SSAdNDP.png}(a). These functions span the same region of space as the ELF maxima, reflecting the high degree of localization in these classical bonds. The remainder of the electrons are apportioned to eight 4c-2e $\sigma$-bonds with ON = 1.92 $\vert$e$\vert$, see Figure \ref{fig:Fig9_aGa_SSAdNDP.png}(b), that span the buckled nets present in the $\alpha$-Ga phase. Thus, each boron atom takes part in one 2c-2e $\sigma$-bond with a boron atom on a neighboring buckled net, and four 4c-2e $\sigma$-bonds that are shared among the surrounding eight boron atoms in the buckled net. Each boron atom can then be assigned one electron from the 2c-2e function and half an electron from each of the 4c-2e functions, yielding three electrons in total, in-line with what is expected for a phase of boron.

\begin{figure}
\begin{center}
\includegraphics[width=\figurewidth]{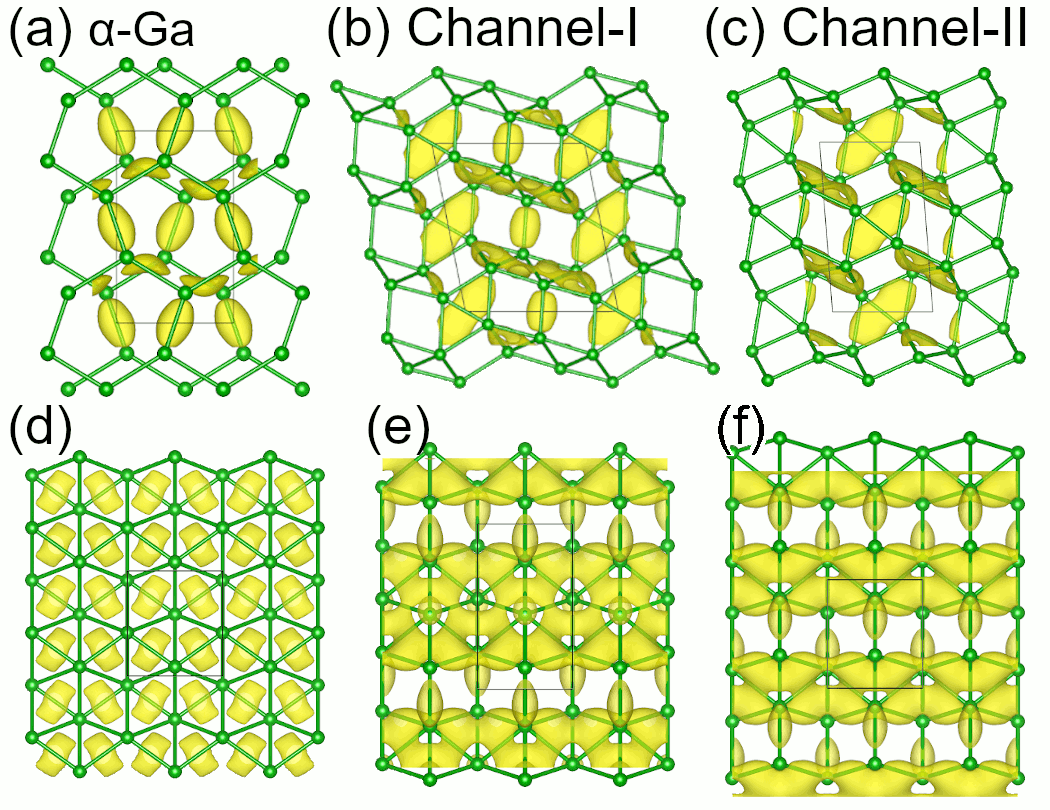}
\end{center}
\caption{Electron density difference isosurface plots (positive isovalue of 0.015 a.u. is shown) for the (a) $\alpha$-Ga, (b) Channel-I, and (c) Channel-II structures. Bottom: buckled layers; top: chains connecting them.
\label{fig:Fig10_diffden.png}}
\end{figure}

Figure \ref{fig:Fig10_diffden.png} plots the difference between the converged electron density from a self-consistent calculation and that of the noninteracting atoms for the $\alpha$-Ga, Channel-I and Channel-II structures. The resulting map is positive in regions where electron density accumulates in proceeding from the non-interacting picture to the density in the compound, that is, where bonds might be expected. The results for $\alpha$-Ga are in-line with the ELF, and also with the SSAdNDP analysis: significant electron accumulation is observed along the B-B contacts that correspond to 2c-2e bonds, whereas more delocalized regions are found along the B$_4$ diamond like motifs, reflective of the 4c-4e bonds found to span the buckled net.

\begin{figure}
\begin{center}
\includegraphics[width=\figurewidth]{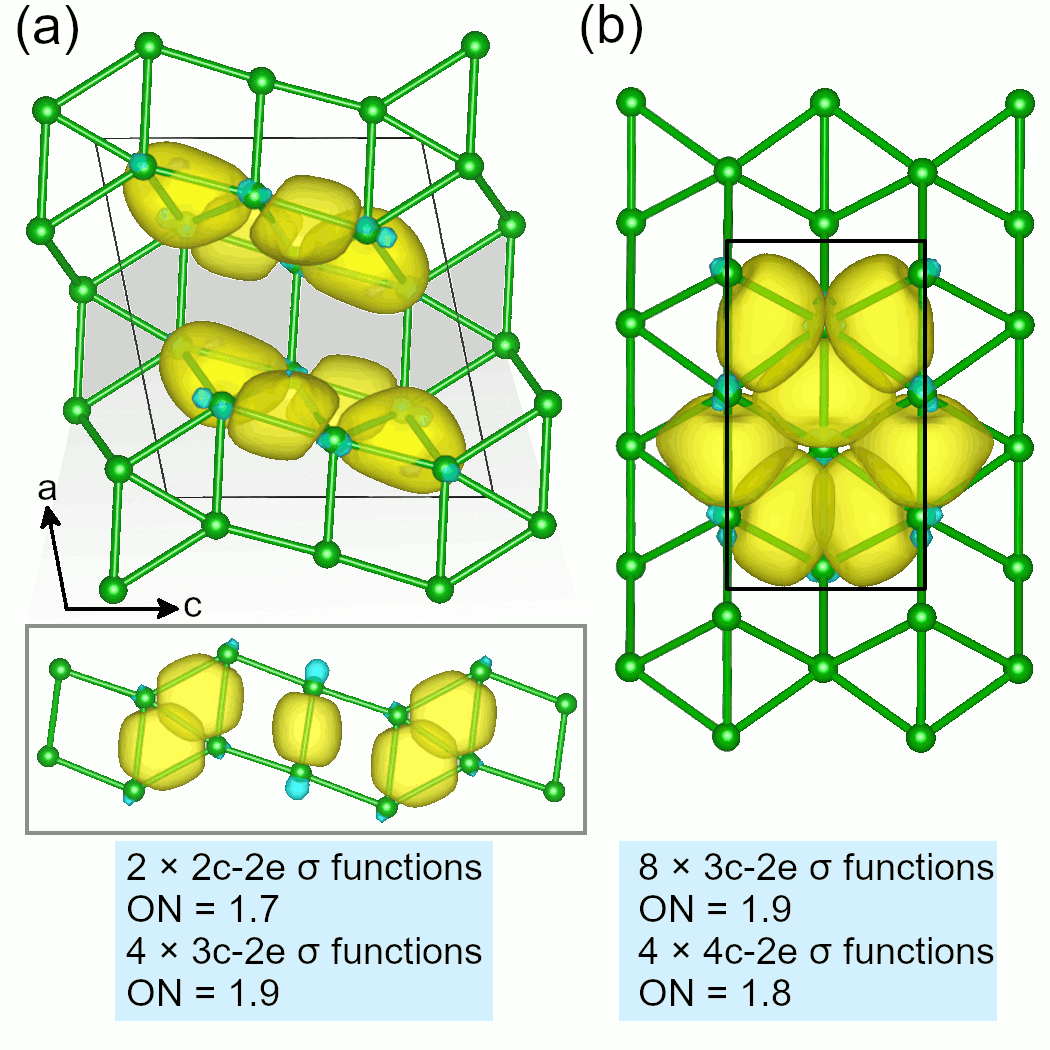}
\end{center}
\caption{Isosurfaces (isovalue=0.08~a.u.) of the bonding functions obtained for the Channel-I structure of boron at 100 GPa. (a) Two 2c-2e and four 3c-2e $\sigma$-bonds between buckled layers of boron atoms, and (b) eight 3c-2e and four 4c-2e $\sigma$-bonds within buckled layers of boron atoms account for all 36 valence electrons in the phase.
\label{fig:Fig11_Channel1_SSAdNDP.png}}
\end{figure}

Next we turn our attention to the bonding framework of the Channel-I and Channel-II polymorphs. In the former, the ELF and density difference analysis both indicate the presence of 2c-2e bonds in the chains running perpendicular to and connecting the buckled layers. The SSAdNDP analysis confirms this characterization, with two such bonds of ON = 1.7~$\vert$e$\vert$ per unit cell, as seen in Figure \ref{fig:Fig11_Channel1_SSAdNDP.png}. The diamond-like fragments comprising the remainder of these chains are spanned by four 3c-2e bonds with ON = 1.9 $\vert$e$\vert$, with two of these found in each B$_4$ diamond motif, so that twelve electrons in total are allocated to the chains. In the buckled layer, strips of triangular nets alternate with strips of corrugated open squares. The bonding can be decomposed into eight 3c-2e bonds per unit cell with ON = 1.9~$\vert$e$\vert$ that lie along the edges of the triangular net strip, and four 4c-2e bonds per unit cell with ON = 1.8 $\vert$e$\vert$ that lie in the middle, totalling 24 electrons. 

The Channel-I structure is peculiar among the channel phases in the placement of two open parallelogram units adjacent to one another in the chains connecting the buckled layers, which is then spanned by a two-center bond. Consequently, the local coordination of the boron atoms in the two-center bond is very similar to that of the boron atoms in the $\alpha$-Ga structure, which each participate in one two-center bond connecting buckled layers and four four-center bonds in the buckled layers. The atoms involved in the two-center bond in the Channel-I phase participate in four- and three-center bonds in the buckled layers. Additional repetition of adjacent open parallelogram units could extend this network of boron atoms involved in two-center bonds between layers and four-center bonds within layers, leading to the bonding scheme exhibited by the $\alpha$-Ga based phases, rather than the three-center bonds found in the channel-based structures.

\begin{figure}
\begin{center}
\includegraphics[width=\figurewidth]{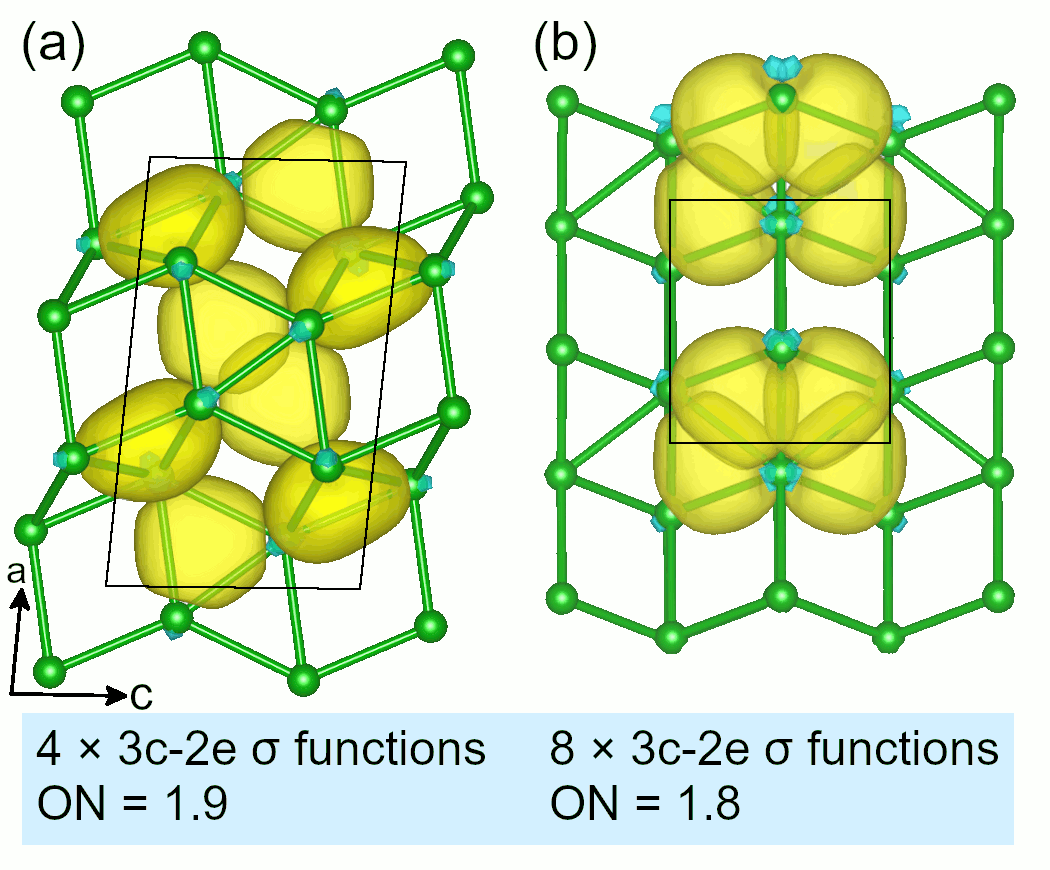}
\end{center}
\caption{Isosurfaces (isovalue=0.08~a.u.) of the bonding functions obtained for the Channel-II structure of boron at 100 GPa. (a) Four 3c-2e $\sigma$-bonds between buckled layers of boron atoms, and (b) eight 3c-2e $\sigma$-bonds within buckled layers of boron atoms account for all 24 valence electrons in the phase.
\label{fig:Fig12_Channel2_SSAdNDP.png}}
\end{figure}

In the less stable Channel-II phase the SSAdNDP analysis, whose results are plotted in Figure \ref{fig:Fig12_Channel2_SSAdNDP.png}, reveals that the chains linking buckled layers composed of diamond-like fragments can be described by 3c-2e bonds with ON = 1.9~$\vert$e$\vert$, and do not contain any 2c-2e bonds. Within the buckled layer, the strips of triangular nets are narrower than in the Channel-I structure and are spanned entirely by 3c-2e functions with ON = 1.8~$\vert$e$\vert$. With eight 3c-2e functions in the buckled layers and four in the connecting chains, all 24 electrons in the unit cell are accounted for. In essence, this bonding scheme matches the one derived for Channel-I but with the 2c-2e and 4c-2e functions deleted.
 
The prevalence of 3c-2e bonds in both channel phases can be traced to features in their electron difference density maps (bottom of Figure \ref{fig:Fig10_diffden.png}). The map for the $\alpha$-Ga phase reveals regions of electron accumulation in the buckled layers that are spread across four B atoms in diamond motifs. However, the difference density in the triangular nets of the channel structures, where three-center bonding functions were detected by the SSAdNDP analysis, displays greater concentration along the edges of the triangular regions rather than being contained within B$_{4}$ units as in the $\alpha$-Ga phase. This may reflect the greater amount of electron localization within a region constrained by three atoms rather than by four. 

To further support the assignment of the bonding to 2c-2e, 3c-3e and 4c-4e functions in the $\alpha$-Ga and two channel phases, we calculated the negative of the crystal orbital Hamilton populations integrated to the Fermi level, -iCOHP, which reflects the strength of a bond between select atom pairs. The results, provided in Table \ref{tab:icohp}, highlight the relative strengths of interaction between pairs of boron atoms involved in 2c, 3c, or 4c bonds as identified by the SSAdNDP analysis. B-B contacts of more than 2.0~\AA{} were neglected, as their -iCOHP values were an order of magnitude smaller than the remaining pairs of atoms.

By far the largest -iCOHP value, at 7.59~eV/bond, corresponds to the 2c-2e bond in the $\alpha$-Ga structure, with the second-largest value of 6.81~eV/bond arising from the two-center contact identified in the Channel-I phase. Between pairs of boron atoms involved in 3-center bonding functions, the -iCOHP values range from 6.58~eV/bond (in the functions connecting buckled layers in the Channel-I phase) to 4.68~eV/bond (within the buckled layers in Channel-I), with most of them between 5 to 6~eV/bond. While robust, these interactions are not as strong as those between the pairs of boron atoms involved in two-center bonds. Finally, regions containing four-center bonds tended to have the weakest bond strengths, with the highest value of 5.38~eV/bond being found within the buckled layer of the $\alpha$-Ga structure and most lying between 4 and 5~eV/bond. In this way, the -iCOHP analysis provides additional delineation between the bonding motifs identified by SSAdNDP and the difference density analyses.

\begin{table}
    \centering
    \def\arraystretch{1}
    \caption{The crystal orbital Hamilton populations integrated to the Fermi level (-iCOHP) for select B-B contacts in the $\alpha$-Ga, Channel-I, and Channel-II phases at 100 GPa. The B-B contacts are labeled according to the type of bonding function they comprise according to the SSAdNDP analysis (2c, 3c, 4c, or between 3c and 4c regions).}
     \setlength{\tabcolsep}{3mm}{        
       \begin{tabular}{c c c}
\hline
\hline
Location & Distance ({\AA}) & -iCOHP (eV/bond) \\
\hline
\hline
$\alpha$-Ga & & \\
\hline
Interlayer (2c) & 1.63 & 7.59 \\
Buckled (4c) & 1.67 & 5.38 \\
Buckled (4c) & 1.72 & 4.77 \\
Buckled (4c) & 1.75 & 4.09 \\
\hline
Channel-I & & \\
\hline
Interlayer (2c) & 1.62 & 6.81 \\
Interlayer (3c) & 1.58 & 6.58 \\
Interlayer (3c) & 1.70 & 5.18 \\
Interlayer (3c) & 1.72 & 4.68 \\
Buckled (3c) & 1.66 & 5.84 \\
Buckled (3c) & 1.63 & 5.32 \\
Buckled (3c/4c) & 1.68 & 4.88 \\
Buckled (4c) & 1.67 & 4.88 \\
Buckled (4c) & 1.71 & 4.48 \\
\hline
Channel-II & & \\
\hline
Interlayer (3c) & 1.59 & 6.14 \\
Interlayer (3c) & 1.68 & 5.01 \\
Interlayer (3c) & 1.70 & 5.11 \\
Interlayer (3c) & 1.64 & 5.11 \\
Buckled (3c) & 1.66 & 5.66 \\
Buckled (3c) & 1.67 & 4.77 \\
\hline
\end{tabular}
\label{tab:icohp}}
\end{table}

The SSAdNDP analysis was carried out for the remaining structures, and the results are summarized in Table \ref{tab:ssadndp} with further details provided in the SI (Fig. S8-S14). The bonding in the $\alpha$-Ga'-I, II, and III phases can be described by 4c-2e functions spanning the buckled triangular nets of boron, with 2c-2e bonds bridging the buckled nets. The diamond-like motifs that comprise the Channel-III structure are filled with 3c-2e bonds.The Channel-IV polymorph, also interpretable as an $\alpha$-Ga derivative, contains two-, three-, and four-center bonds in arrangements reflecting its relationships to both $\alpha$-Ga and channel polymorphs. Finally, the bonding within the two intergrowths of alternating $\alpha$-Ga and channel-like layers is characterized by two- and four-center networks in the $\alpha$-Ga based regions, while the channel-based regions of the Intergrowth-I and -II polymorphs contain four- and three-center bonds, respectively. The three-center bonds in the channel-like regions of the Intergrowth-II phase are analogous to the 3c bonding networks in the parent channel-based structures, while the four-center bonds found in the Intergrowth-I structure reflect the narrowness of the channel-like regions in that structure  in comparison to those based on $\alpha$-Ga, which contain four-center bonding functions in the buckled nets.

\begin{table}
    \centering
    \def\arraystretch{1}
    \caption{Summary of SSAdNDP analyses of boron phases at 100~GPa.}
     \setlength{\tabcolsep}{3mm}{        
       \begin{tabular}{c c c c c}
\hline
\hline
Structure & Formula units & 2c-2e & 3c-2e & 4c-2e \\
\hline
\hline
$\alpha$-Ga & 8 & 4 & 0 & 8 \\
Channel-I & 12 & 2 & 12 & 4 \\
Channel-II & 8 & 0 & 12 & 0 \\
Intergrowth I & 12 & 6 & 0 & 12 \\
Channel-III & 24 & 0 & 36 & 0 \\
$\alpha$-Ga'-I & 12 & 6 & 0 & 12 \\
$\alpha$-Ga'-II & 8 & 4 & 0 & 8 \\
$\alpha$-Ga'-III & 8 & 4 & 0 & 8 \\
Intergrowth-II & 12 & 4 & 4 & 10 \\
Channel-IV & 16 & 4 & 16 & 4 \\

\hline
\end{tabular}
\label{tab:ssadndp}}
\end{table}

\section{Properties at 1 Atmosphere}
Phonon calculations were carried out to determine which boron allotropes could conceivably be quenched to atmospheric pressures (effectively 0~GPa in our calculations). In addition to the $\alpha$-Ga structure, the Channel-I and -II, $\alpha$-Ga'-I and -II, and Intergrowth-I and -II phases were found to be dynamically stable. Their enthalpies range from 199~meV/atom (Channel-I) to 287~meV/atom (Intergrowth-I) relative to the $\alpha$-B$_{12}$ phase. Band structure calculations within the PBE functional suggested that Channel-I and Intergrowth-I were both small gap semiconductors, whereas the remainder were metallic. However, the PBE functional is known to underestimate band gaps, and inclusion of Hartree-Fock exchange would likely open up the band gap in some of these. 

Covalent compounds that contain boron atoms are known to behave as hard materials. Therefore, the Vickers Hardness, $H_\text{v}$, of these phases at ambient pressure were obtained using the Teter equation \cite{Teter:1998a,Chen:2011a} combined with shear moduli obtained via a machine-learning model \cite{Isayev:2017a} trained on the materials within the AFLOW repository \cite{Curtarolo:2012b,Zurek:2017o}. This method was shown to provide hardness estimates that correlated well with experiment for a broad range of compounds \cite{Zurek:2019b}. The estimated $H_\text{v}$s were 39.6~GPa for $\alpha$-Ga, 32.8~GPa for the $\alpha$-Ga'-I structure, and between 35.5 and 36.6~GPa for the remainder, all of which are slightly higher than the $H_\text{v}$s of 30.5, 31.5 and 28.4~GPa obtained for the known $\alpha$-B$_{12}$, $\gamma$-B$_{28}$ and $\beta$-B$_{105}$ phases. The $m$-B$_{16}$ and $o$-B$_{16}$ allotropes predicted in Ref.\ \cite{Fan:2014}, were estimated to be somewhat harder, whereas the $I4/mmm$-B$_4$ and $Pm$-B$_{17}$ phases from Ref.\ \cite{Zhang:2020} were not quite as hard. However, it is well known that both the computational and experimental methods employed to guage hardness are imperfect, with the former often yielding values that can differ by more than 10\% \cite{Gao:2010a}. Therefore, we suggest that the $H_\text{v}$s of these various allotropes of boron at 1~atm do not differ significantly from each other.

\section{Conclusions} 

Evolutionary crystal structure prediction searches were employed to identify a series of metastable polymorphs of elemental boron at 100~GPa. Potentially synthesizable phases, which were within 100~meV/atom of the thermodynamic minimum at this pressure, the $\alpha$-Ga structure, were further analyzed. These could be grouped into families based upon prevalant structural motifs: three were clearly derived from the $\alpha$-Ga phase, three were based on structures that contained channels of boron atoms, and two were layered intergrowths of the $\alpha$-Ga structure with the channel-based phases. One additional structure bears a clear resemblance with the channel-based compounds, but can also be interpreted as another derivative of $\alpha$-Ga based on layers of sheared honeycomb nets. 

An examination of the factors stabilizing these allotropes revealed that the $PV$ contribution to the enthalpy favored the denser $\alpha$-Ga-derived structures, while the internal energies of the channel-based phases were lower. We further found that finite temperature contributions stabilize the Channel-I phase more than the $\alpha$-Ga structure. This effect is magnified at lower pressures owing to the changes in the relative $PV$ and Helmholtz free energy terms, suggesting that a viable route to synthetically accessing the Channel-I structure could be a high-pressure high-temperature synthesis followed by quenching to lower-pressure conditions.

To further understand the bonding in these phases we employed the Solid State Adaptive Natural Density Partitioning (SSAdNDP) method, coupled with calculations of the electron localization functions (ELF), crystal orbital Hamilton populations (COHP), and density difference analysis. In the $\alpha$-Ga structure, four-center two-electron (4c-2e) bonds were found to span the buckled layers with classic two-center two-electron (2c-2e) bonds connecting these layers. All of the $\alpha$-Ga derived phases were characterized by these same types of 4c-2e and 2c-2e bonds. The channel-based polymorphs, meanwhile, were largely characterized by three-center two-electron (3c-2e) bonds distributed across the diamond-like B$_4$ motifs present in the phases. The intergrowth structures contained features from both $\alpha$-Ga and channel-based parents. The $\alpha$-Ga structure, along with six of the predicted structures remained dynamically stable at atmospheric pressures, with estimated Vickers hardnesses of $\sim$36~GPa.

The plethora of known and hypothesized allotropes of boron reflect the complexity of its crystal chemistry, suggesting a wealth of materials for further study under appropriate experimental conditions. Furthermore, the structural relations revealed in this study indicate a possibility of numerous additional metastable variants of elemental boron based on minor modifications or intergrowths of a few basic parent structures. We look forward to further experimental and theoretical explorations of the complex crystal chemistry of elemental boron, in particular, the twin roles of pressure and temperature in the stability of various allotropes, including the effects of static versus dynamic compression.

\section{Acknowledgments}

K.H. is thankful to the U.S.\ Department of Energy, National Nuclear Security Administration, through the Capital-DOE Alliance Center under Cooperative Agreement DE-NA0003975 for financial support. Calculations were performed at the Center for Computational Research at SUNY Buffalo \cite{ccr}. Part of this work was performed under the auspices of the U.S.\ Department of Energy by Lawrence Livermore National
Laboratory under Contract DE-AC52-07NA27344. T.O.notes LLNL-JRNL-817818.

\bibliography{Boron}

\begin{thebibliography}{108}%
\makeatletter
\providecommand \@ifxundefined [1]{%
 \@ifx{#1\undefined}
}%
\providecommand \@ifnum [1]{%
 \ifnum #1\expandafter \@firstoftwo
 \else \expandafter \@secondoftwo
 \fi
}%
\providecommand \@ifx [1]{%
 \ifx #1\expandafter \@firstoftwo
 \else \expandafter \@secondoftwo
 \fi
}%
\providecommand \natexlab [1]{#1}%
\providecommand \enquote  [1]{``#1''}%
\providecommand \bibnamefont  [1]{#1}%
\providecommand \bibfnamefont [1]{#1}%
\providecommand \citenamefont [1]{#1}%
\providecommand \href@noop [0]{\@secondoftwo}%
\providecommand \href [0]{\begingroup \@sanitize@url \@href}%
\providecommand \@href[1]{\@@startlink{#1}\@@href}%
\providecommand \@@href[1]{\endgroup#1\@@endlink}%
\providecommand \@sanitize@url [0]{\catcode `\\12\catcode `\$12\catcode
  `\&12\catcode `\#12\catcode `\^12\catcode `\_12\catcode `\%12\relax}%
\providecommand \@@startlink[1]{}%
\providecommand \@@endlink[0]{}%
\providecommand \url  [0]{\begingroup\@sanitize@url \@url }%
\providecommand \@url [1]{\endgroup\@href {#1}{\urlprefix }}%
\providecommand \urlprefix  [0]{URL }%
\providecommand \Eprint [0]{\href }%
\providecommand \doibase [0]{https://doi.org/}%
\providecommand \selectlanguage [0]{\@gobble}%
\providecommand \bibinfo  [0]{\@secondoftwo}%
\providecommand \bibfield  [0]{\@secondoftwo}%
\providecommand \translation [1]{[#1]}%
\providecommand \BibitemOpen [0]{}%
\providecommand \bibitemStop [0]{}%
\providecommand \bibitemNoStop [0]{.\EOS\space}%
\providecommand \EOS [0]{\spacefactor3000\relax}%
\providecommand \BibitemShut  [1]{\csname bibitem#1\endcsname}%
\let\auto@bib@innerbib\@empty
\bibitem [{\citenamefont {Zhai}\ \emph {et~al.}(2003)\citenamefont {Zhai},
  \citenamefont {Alexandrova}, \citenamefont {Birch}, \citenamefont
  {Boldyrev},\ and\ \citenamefont {Wang}}]{Zhai:2003a}%
  \BibitemOpen
  \bibfield  {author} {\bibinfo {author} {\bibfnamefont {H.~J.}\ \bibnamefont
  {Zhai}}, \bibinfo {author} {\bibfnamefont {A.~N.}\ \bibnamefont
  {Alexandrova}}, \bibinfo {author} {\bibfnamefont {K.~A.}\ \bibnamefont
  {Birch}}, \bibinfo {author} {\bibfnamefont {A.~I.}\ \bibnamefont
  {Boldyrev}},\ and\ \bibinfo {author} {\bibfnamefont {L.~S.}\ \bibnamefont
  {Wang}},\ }\bibfield  {title} {\bibinfo {title} {Hepta- and octacoordinate
  boron in molecular wheels of eight- and nine-atom boron clusters; observation
  and confirmation},\ }\href@noop {} {\bibfield  {journal} {\bibinfo  {journal}
  {Angew. Chem. Int. Ed.}\ }\textbf {\bibinfo {volume} {42}},\ \bibinfo {pages}
  {6004} (\bibinfo {year} {2003})}\BibitemShut {NoStop}%
\bibitem [{\citenamefont {Alexandrova}\ \emph {et~al.}(2006)\citenamefont
  {Alexandrova}, \citenamefont {Boldyrev}, \citenamefont {Zhai},\ and\
  \citenamefont {Wang}}]{Alexandrova:2006a}%
  \BibitemOpen
  \bibfield  {author} {\bibinfo {author} {\bibfnamefont {A.~N.}\ \bibnamefont
  {Alexandrova}}, \bibinfo {author} {\bibfnamefont {A.~I.}\ \bibnamefont
  {Boldyrev}}, \bibinfo {author} {\bibfnamefont {H.~J.}\ \bibnamefont {Zhai}},\
  and\ \bibinfo {author} {\bibfnamefont {L.~S.}\ \bibnamefont {Wang}},\
  }\bibfield  {title} {\bibinfo {title} {All-boron aromatic clusters as
  potential new inorganic ligands and building blocks in chemistry},\
  }\href@noop {} {\bibfield  {journal} {\bibinfo  {journal} {Coord. Chem.
  Rev.}\ }\textbf {\bibinfo {volume} {250}},\ \bibinfo {pages} {2811} (\bibinfo
  {year} {2006})}\BibitemShut {NoStop}%
\bibitem [{\citenamefont {Li}\ \emph {et~al.}(2019)\citenamefont {Li},
  \citenamefont {Gao}, \citenamefont {Cheng}, \citenamefont {He}, \citenamefont
  {Yin}, \citenamefont {Hu}, \citenamefont {Chen}, \citenamefont {Cheng},\ and\
  \citenamefont {Zhao}}]{Li:2019a}%
  \BibitemOpen
  \bibfield  {author} {\bibinfo {author} {\bibfnamefont {D.}~\bibnamefont
  {Li}}, \bibinfo {author} {\bibfnamefont {J.}~\bibnamefont {Gao}}, \bibinfo
  {author} {\bibfnamefont {P.}~\bibnamefont {Cheng}}, \bibinfo {author}
  {\bibfnamefont {J.}~\bibnamefont {He}}, \bibinfo {author} {\bibfnamefont
  {Y.}~\bibnamefont {Yin}}, \bibinfo {author} {\bibfnamefont {Y.}~\bibnamefont
  {Hu}}, \bibinfo {author} {\bibfnamefont {L.}~\bibnamefont {Chen}}, \bibinfo
  {author} {\bibfnamefont {Y.}~\bibnamefont {Cheng}},\ and\ \bibinfo {author}
  {\bibfnamefont {J.}~\bibnamefont {Zhao}},\ }\bibfield  {title} {\bibinfo
  {title} {2d boron sheets: Structure, growth, and electronic and thermal
  transport properties},\ }\href@noop {} {\bibfield  {journal} {\bibinfo
  {journal} {Adv. Funct. Mater.}\ ,\ \bibinfo {pages} {1904349}} (\bibinfo
  {year} {2019})}\BibitemShut {NoStop}%
\bibitem [{\citenamefont {Sun}\ \emph {et~al.}(2017)\citenamefont {Sun},
  \citenamefont {Liu}, \citenamefont {Yin}, \citenamefont {Yu}, \citenamefont
  {Li}, \citenamefont {Hang}, \citenamefont {Zhou}, \citenamefont {Yu},
  \citenamefont {Li}, \citenamefont {Tai},\ and\ \citenamefont
  {Guo}}]{Sun:2017a}%
  \BibitemOpen
  \bibfield  {author} {\bibinfo {author} {\bibfnamefont {X.}~\bibnamefont
  {Sun}}, \bibinfo {author} {\bibfnamefont {X.}~\bibnamefont {Liu}}, \bibinfo
  {author} {\bibfnamefont {J.}~\bibnamefont {Yin}}, \bibinfo {author}
  {\bibfnamefont {J.}~\bibnamefont {Yu}}, \bibinfo {author} {\bibfnamefont
  {Y.}~\bibnamefont {Li}}, \bibinfo {author} {\bibfnamefont {Y.}~\bibnamefont
  {Hang}}, \bibinfo {author} {\bibfnamefont {X.}~\bibnamefont {Zhou}}, \bibinfo
  {author} {\bibfnamefont {M.}~\bibnamefont {Yu}}, \bibinfo {author}
  {\bibfnamefont {J.}~\bibnamefont {Li}}, \bibinfo {author} {\bibfnamefont
  {G.}~\bibnamefont {Tai}},\ and\ \bibinfo {author} {\bibfnamefont
  {W.}~\bibnamefont {Guo}},\ }\bibfield  {title} {\bibinfo {title}
  {Two-dimensional boron crystals: Structural stabilitiy, tunable properties,
  fabrications and applications},\ }\href@noop {} {\bibfield  {journal}
  {\bibinfo  {journal} {Adv. Funct. Mater.}\ ,\ \bibinfo {pages} {1603300}}
  (\bibinfo {year} {2017})}\BibitemShut {NoStop}%
\bibitem [{\citenamefont {Albert}\ and\ \citenamefont
  {Hillebrecht}(2009)}]{Albert:2009}%
  \BibitemOpen
  \bibfield  {author} {\bibinfo {author} {\bibfnamefont {B.}~\bibnamefont
  {Albert}}\ and\ \bibinfo {author} {\bibfnamefont {H.}~\bibnamefont
  {Hillebrecht}},\ }\bibfield  {title} {\bibinfo {title} {Boron: Elementary
  challenge for experimenters and theoreticians},\ }\href@noop {} {\bibfield
  {journal} {\bibinfo  {journal} {Angew. Chem. Int. Ed.}\ }\textbf {\bibinfo
  {volume} {48}},\ \bibinfo {pages} {8640} (\bibinfo {year}
  {2009})}\BibitemShut {NoStop}%
\bibitem [{\citenamefont {Ogitsu}\ \emph {et~al.}(2013)\citenamefont {Ogitsu},
  \citenamefont {Schwegler},\ and\ \citenamefont {Galli}}]{Ogitsu:2013}%
  \BibitemOpen
  \bibfield  {author} {\bibinfo {author} {\bibfnamefont {T.}~\bibnamefont
  {Ogitsu}}, \bibinfo {author} {\bibfnamefont {E.}~\bibnamefont {Schwegler}},\
  and\ \bibinfo {author} {\bibfnamefont {G.}~\bibnamefont {Galli}},\ }\bibfield
   {title} {\bibinfo {title} {$\beta$-rhombohedral boron: At the crossraods of
  the chemistry of boron and the physics of frustration},\ }\href@noop {}
  {\bibfield  {journal} {\bibinfo  {journal} {Chem. Rev.}\ }\textbf {\bibinfo
  {volume} {113}},\ \bibinfo {pages} {3425} (\bibinfo {year}
  {2013})}\BibitemShut {NoStop}%
\bibitem [{\citenamefont {Shirai}(2017)}]{Shirai:2017a}%
  \BibitemOpen
  \bibfield  {author} {\bibinfo {author} {\bibfnamefont {K.}~\bibnamefont
  {Shirai}},\ }\bibfield  {title} {\bibinfo {title} {Phase diagram of boron
  crystals},\ }\href@noop {} {\bibfield  {journal} {\bibinfo  {journal} {Jpn.
  J. Appl. Phys.}\ }\textbf {\bibinfo {volume} {56}},\ \bibinfo {pages}
  {05FA06} (\bibinfo {year} {2017})}\BibitemShut {NoStop}%
\bibitem [{\citenamefont {Veprek}\ \emph {et~al.}(2011)\citenamefont {Veprek},
  \citenamefont {Zhang},\ and\ \citenamefont {Argon}}]{Veprek:2011}%
  \BibitemOpen
  \bibfield  {author} {\bibinfo {author} {\bibfnamefont {S.}~\bibnamefont
  {Veprek}}, \bibinfo {author} {\bibfnamefont {R.}~\bibnamefont {Zhang}},\ and\
  \bibinfo {author} {\bibfnamefont {A.}~\bibnamefont {Argon}},\ }\bibfield
  {title} {\bibinfo {title} {Mechanical properties and hardness of boron and
  boron-rich solids},\ }\href@noop {} {\bibfield  {journal} {\bibinfo
  {journal} {J. Superhard Mater.}\ }\textbf {\bibinfo {volume} {33}},\ \bibinfo
  {pages} {409} (\bibinfo {year} {2011})}\BibitemShut {NoStop}%
\bibitem [{\citenamefont {Zarechnaya}\ \emph
  {et~al.}(2009{\natexlab{a}})\citenamefont {Zarechnaya}, \citenamefont
  {Dubrovinsky}, \citenamefont {Dubrovinskaia}, \citenamefont {Filinchuk},
  \citenamefont {Chernyshov}, \citenamefont {Dmitriev}, \citenamefont
  {Miyajima}, \citenamefont {El~Goresy}, \citenamefont {Braun}, \citenamefont
  {Van~Smaalen}, \citenamefont {Kantor}, \citenamefont {Kantor}, \citenamefont
  {Prakapenka}, \citenamefont {Hanfland}, \citenamefont {Mikhaylushkin},
  \citenamefont {Abrikosov},\ and\ \citenamefont {Simak}}]{Simak:2009}%
  \BibitemOpen
  \bibfield  {author} {\bibinfo {author} {\bibfnamefont {E.~Y.}\ \bibnamefont
  {Zarechnaya}}, \bibinfo {author} {\bibfnamefont {L.}~\bibnamefont
  {Dubrovinsky}}, \bibinfo {author} {\bibfnamefont {N.}~\bibnamefont
  {Dubrovinskaia}}, \bibinfo {author} {\bibfnamefont {Y.}~\bibnamefont
  {Filinchuk}}, \bibinfo {author} {\bibfnamefont {D.}~\bibnamefont
  {Chernyshov}}, \bibinfo {author} {\bibfnamefont {V.}~\bibnamefont
  {Dmitriev}}, \bibinfo {author} {\bibfnamefont {N.}~\bibnamefont {Miyajima}},
  \bibinfo {author} {\bibfnamefont {A.}~\bibnamefont {El~Goresy}}, \bibinfo
  {author} {\bibfnamefont {H.~F.}\ \bibnamefont {Braun}}, \bibinfo {author}
  {\bibfnamefont {S.}~\bibnamefont {Van~Smaalen}}, \bibinfo {author}
  {\bibfnamefont {I.}~\bibnamefont {Kantor}}, \bibinfo {author} {\bibfnamefont
  {A.}~\bibnamefont {Kantor}}, \bibinfo {author} {\bibfnamefont
  {V.}~\bibnamefont {Prakapenka}}, \bibinfo {author} {\bibfnamefont
  {M.}~\bibnamefont {Hanfland}}, \bibinfo {author} {\bibfnamefont {A.~S.}\
  \bibnamefont {Mikhaylushkin}}, \bibinfo {author} {\bibfnamefont {I.~A.}\
  \bibnamefont {Abrikosov}},\ and\ \bibinfo {author} {\bibfnamefont {S.~I.}\
  \bibnamefont {Simak}},\ }\bibfield  {title} {\bibinfo {title} {Superhard
  semiconducting optically transparent high pressure phase of boron},\
  }\href@noop {} {\bibfield  {journal} {\bibinfo  {journal} {Phys. Rev. Lett}\
  }\textbf {\bibinfo {volume} {102}},\ \bibinfo {pages} {185501} (\bibinfo
  {year} {2009}{\natexlab{a}})}\BibitemShut {NoStop}%
\bibitem [{\citenamefont {Zhou}\ \emph {et~al.}(2010)\citenamefont {Zhou},
  \citenamefont {Sun},\ and\ \citenamefont {Chen}}]{Zhou:2010a}%
  \BibitemOpen
  \bibfield  {author} {\bibinfo {author} {\bibfnamefont {W.}~\bibnamefont
  {Zhou}}, \bibinfo {author} {\bibfnamefont {H.}~\bibnamefont {Sun}},\ and\
  \bibinfo {author} {\bibfnamefont {C.}~\bibnamefont {Chen}},\ }\bibfield
  {title} {\bibinfo {title} {Soft bond-deformation paths in superhard
  $\gamma$-boron},\ }\href@noop {} {\bibfield  {journal} {\bibinfo  {journal}
  {Phys. Rev. Lett.}\ }\textbf {\bibinfo {volume} {105}},\ \bibinfo {pages}
  {215503} (\bibinfo {year} {2010})}\BibitemShut {NoStop}%
\bibitem [{\citenamefont {Zhang}\ \emph
  {et~al.}(2020{\natexlab{a}})\citenamefont {Zhang}, \citenamefont {Du},
  \citenamefont {Lin}, \citenamefont {Bergara}, \citenamefont {Chen},
  \citenamefont {Liu}, \citenamefont {Zhang},\ and\ \citenamefont
  {Yang}}]{Zhang:2020}%
  \BibitemOpen
  \bibfield  {author} {\bibinfo {author} {\bibfnamefont {S.}~\bibnamefont
  {Zhang}}, \bibinfo {author} {\bibfnamefont {X.}~\bibnamefont {Du}}, \bibinfo
  {author} {\bibfnamefont {J.}~\bibnamefont {Lin}}, \bibinfo {author}
  {\bibfnamefont {A.}~\bibnamefont {Bergara}}, \bibinfo {author} {\bibfnamefont
  {X.}~\bibnamefont {Chen}}, \bibinfo {author} {\bibfnamefont {X.}~\bibnamefont
  {Liu}}, \bibinfo {author} {\bibfnamefont {X.}~\bibnamefont {Zhang}},\ and\
  \bibinfo {author} {\bibfnamefont {G.}~\bibnamefont {Yang}},\ }\bibfield
  {title} {\bibinfo {title} {Superconducting boron allotropes},\ }\href@noop {}
  {\bibfield  {journal} {\bibinfo  {journal} {Phys. Rev. B}\ }\textbf {\bibinfo
  {volume} {101}},\ \bibinfo {pages} {174507} (\bibinfo {year}
  {2020}{\natexlab{a}})}\BibitemShut {NoStop}%
\bibitem [{\citenamefont {Eremets}\ \emph {et~al.}(2001)\citenamefont
  {Eremets}, \citenamefont {Struzhkin}, \citenamefont {Mao},\ and\
  \citenamefont {Hemley}}]{Eremets:2001}%
  \BibitemOpen
  \bibfield  {author} {\bibinfo {author} {\bibfnamefont {M.}~\bibnamefont
  {Eremets}}, \bibinfo {author} {\bibfnamefont {V.}~\bibnamefont {Struzhkin}},
  \bibinfo {author} {\bibfnamefont {H.-k.}\ \bibnamefont {Mao}},\ and\ \bibinfo
  {author} {\bibfnamefont {R.}~\bibnamefont {Hemley}},\ }\bibfield  {title}
  {\bibinfo {title} {Superconductivity in boron},\ }\href@noop {} {\bibfield
  {journal} {\bibinfo  {journal} {Science}\ }\textbf {\bibinfo {volume}
  {293}},\ \bibinfo {pages} {272} (\bibinfo {year} {2001})}\BibitemShut
  {NoStop}%
\bibitem [{\citenamefont {Ma}\ \emph {et~al.}(2004)\citenamefont {Ma},
  \citenamefont {Tse}, \citenamefont {Klug},\ and\ \citenamefont
  {Ahuja}}]{Ma:2004}%
  \BibitemOpen
  \bibfield  {author} {\bibinfo {author} {\bibfnamefont {Y.}~\bibnamefont
  {Ma}}, \bibinfo {author} {\bibfnamefont {J.}~\bibnamefont {Tse}}, \bibinfo
  {author} {\bibfnamefont {D.}~\bibnamefont {Klug}},\ and\ \bibinfo {author}
  {\bibfnamefont {R.}~\bibnamefont {Ahuja}},\ }\bibfield  {title} {\bibinfo
  {title} {Electron-phonon coupling of $\alpha$-ga boron},\ }\href@noop {}
  {\bibfield  {journal} {\bibinfo  {journal} {Phys. Rev. B}\ }\textbf {\bibinfo
  {volume} {70}},\ \bibinfo {pages} {214107} (\bibinfo {year}
  {2004})}\BibitemShut {NoStop}%
\bibitem [{\citenamefont {Li}\ \emph {et~al.}(2014)\citenamefont {Li},
  \citenamefont {Bao}, \citenamefont {Tian}, \citenamefont {Jin}, \citenamefont
  {Duan}, \citenamefont {He}, \citenamefont {Liu},\ and\ \citenamefont
  {Cui}}]{Li:2014a}%
  \BibitemOpen
  \bibfield  {author} {\bibinfo {author} {\bibfnamefont {D.}~\bibnamefont
  {Li}}, \bibinfo {author} {\bibfnamefont {K.}~\bibnamefont {Bao}}, \bibinfo
  {author} {\bibfnamefont {F.}~\bibnamefont {Tian}}, \bibinfo {author}
  {\bibfnamefont {X.}~\bibnamefont {Jin}}, \bibinfo {author} {\bibfnamefont
  {D.}~\bibnamefont {Duan}}, \bibinfo {author} {\bibfnamefont {Z.}~\bibnamefont
  {He}}, \bibinfo {author} {\bibfnamefont {B.}~\bibnamefont {Liu}},\ and\
  \bibinfo {author} {\bibfnamefont {T.}~\bibnamefont {Cui}},\ }\bibfield
  {title} {\bibinfo {title} {Hih-pressure close-packed structure of boron},\
  }\href@noop {} {\bibfield  {journal} {\bibinfo  {journal} {RSC Adv.}\
  }\textbf {\bibinfo {volume} {4}},\ \bibinfo {pages} {203} (\bibinfo {year}
  {2014})}\BibitemShut {NoStop}%
\bibitem [{\citenamefont {Gao}\ \emph {et~al.}(2018)\citenamefont {Gao},
  \citenamefont {Xie}, \citenamefont {Chen}, \citenamefont {Gu},\ and\
  \citenamefont {Chen}}]{Gao:2018}%
  \BibitemOpen
  \bibfield  {author} {\bibinfo {author} {\bibfnamefont {Y.}~\bibnamefont
  {Gao}}, \bibinfo {author} {\bibfnamefont {Y.}~\bibnamefont {Xie}}, \bibinfo
  {author} {\bibfnamefont {Y.}~\bibnamefont {Chen}}, \bibinfo {author}
  {\bibfnamefont {J.}~\bibnamefont {Gu}},\ and\ \bibinfo {author}
  {\bibfnamefont {Z.}~\bibnamefont {Chen}},\ }\bibfield  {title} {\bibinfo
  {title} {Spindle nodal chain in three-dimensional $\alpha$' boron},\
  }\href@noop {} {\bibfield  {journal} {\bibinfo  {journal} {Phys. Chem. Chem.
  Phys.}\ }\textbf {\bibinfo {volume} {20}},\ \bibinfo {pages} {23500}
  (\bibinfo {year} {2018})}\BibitemShut {NoStop}%
\bibitem [{\citenamefont {Zhang}\ \emph
  {et~al.}(2018{\natexlab{a}})\citenamefont {Zhang}, \citenamefont {Militzer},
  \citenamefont {Gregor}, \citenamefont {Caspersen}, \citenamefont {Yang},
  \citenamefont {Gaffney}, \citenamefont {Ogitsu}, \citenamefont {Swift},
  \citenamefont {Lazicki}, \citenamefont {Erskine}, \citenamefont {London},
  \citenamefont {Celliers}, \citenamefont {Nilsen}, \citenamefont {Sterne},\
  and\ \citenamefont {Whitley}}]{PhysRevE.98.023205}%
  \BibitemOpen
  \bibfield  {author} {\bibinfo {author} {\bibfnamefont {S.}~\bibnamefont
  {Zhang}}, \bibinfo {author} {\bibfnamefont {B.}~\bibnamefont {Militzer}},
  \bibinfo {author} {\bibfnamefont {M.~C.}\ \bibnamefont {Gregor}}, \bibinfo
  {author} {\bibfnamefont {K.}~\bibnamefont {Caspersen}}, \bibinfo {author}
  {\bibfnamefont {L.~H.}\ \bibnamefont {Yang}}, \bibinfo {author}
  {\bibfnamefont {J.}~\bibnamefont {Gaffney}}, \bibinfo {author} {\bibfnamefont
  {T.}~\bibnamefont {Ogitsu}}, \bibinfo {author} {\bibfnamefont
  {D.}~\bibnamefont {Swift}}, \bibinfo {author} {\bibfnamefont
  {A.}~\bibnamefont {Lazicki}}, \bibinfo {author} {\bibfnamefont
  {D.}~\bibnamefont {Erskine}}, \bibinfo {author} {\bibfnamefont {R.~A.}\
  \bibnamefont {London}}, \bibinfo {author} {\bibfnamefont {P.~M.}\
  \bibnamefont {Celliers}}, \bibinfo {author} {\bibfnamefont {J.}~\bibnamefont
  {Nilsen}}, \bibinfo {author} {\bibfnamefont {P.~A.}\ \bibnamefont {Sterne}},\
  and\ \bibinfo {author} {\bibfnamefont {H.~D.}\ \bibnamefont {Whitley}},\
  }\bibfield  {title} {\bibinfo {title} {Theoretical and experimental
  investigation of the equation of state of boron plasmas},\ }\href@noop {}
  {\bibfield  {journal} {\bibinfo  {journal} {Phys. Rev. E}\ }\textbf {\bibinfo
  {volume} {98}},\ \bibinfo {pages} {023205} (\bibinfo {year}
  {2018}{\natexlab{a}})}\BibitemShut {NoStop}%
\bibitem [{\citenamefont {Zhang}\ \emph
  {et~al.}(2020{\natexlab{b}})\citenamefont {Zhang}, \citenamefont {Whitley},\
  and\ \citenamefont {Ogitsu}}]{ZhangShuai:2020}%
  \BibitemOpen
  \bibfield  {author} {\bibinfo {author} {\bibfnamefont {S.}~\bibnamefont
  {Zhang}}, \bibinfo {author} {\bibfnamefont {H.}~\bibnamefont {Whitley}},\
  and\ \bibinfo {author} {\bibfnamefont {T.}~\bibnamefont {Ogitsu}},\
  }\bibfield  {title} {\bibinfo {title} {Phase transformation in boron under
  shock compression},\ }\href@noop {} {\bibfield  {journal} {\bibinfo
  {journal} {Solid State Sci.}\ }\textbf {\bibinfo {volume} {108}},\ \bibinfo
  {pages} {106376} (\bibinfo {year} {2020}{\natexlab{b}})}\BibitemShut
  {NoStop}%
\bibitem [{\citenamefont {Masago}\ \emph {et~al.}(2006)\citenamefont {Masago},
  \citenamefont {Shirai},\ and\ \citenamefont
  {Katayama-Yoshida}}]{Masago:2006a}%
  \BibitemOpen
  \bibfield  {author} {\bibinfo {author} {\bibfnamefont {A.}~\bibnamefont
  {Masago}}, \bibinfo {author} {\bibfnamefont {K.}~\bibnamefont {Shirai}},\
  and\ \bibinfo {author} {\bibfnamefont {H.}~\bibnamefont {Katayama-Yoshida}},\
  }\bibfield  {title} {\bibinfo {title} {Crystal stability of $\alpha$- and
  $\beta$-boron},\ }\href@noop {} {\bibfield  {journal} {\bibinfo  {journal}
  {Phys. Rev. B.}\ }\textbf {\bibinfo {volume} {73}},\ \bibinfo {pages}
  {104102} (\bibinfo {year} {2006})}\BibitemShut {NoStop}%
\bibitem [{\citenamefont {Masago}\ \emph {et~al.}(2004)\citenamefont {Masago},
  \citenamefont {Shirai},\ and\ \citenamefont
  {Katayama-Yoshida}}]{Masago:2004}%
  \BibitemOpen
  \bibfield  {author} {\bibinfo {author} {\bibfnamefont {A.}~\bibnamefont
  {Masago}}, \bibinfo {author} {\bibfnamefont {K.}~\bibnamefont {Shirai}},\
  and\ \bibinfo {author} {\bibfnamefont {H.}~\bibnamefont {Katayama-Yoshida}},\
  }\bibfield  {title} {\bibinfo {title} {The pressure dependence of solid
  boron},\ }\href@noop {} {\bibfield  {journal} {\bibinfo  {journal} {Molecular
  Simulation}\ }\textbf {\bibinfo {volume} {30}},\ \bibinfo {pages} {935}
  (\bibinfo {year} {2004})}\BibitemShut {NoStop}%
\bibitem [{\citenamefont {Masago}\ \emph {et~al.}(2005)\citenamefont {Masago},
  \citenamefont {Shirai},\ and\ \citenamefont
  {Katayama-Yoshida}}]{Masago:2005}%
  \BibitemOpen
  \bibfield  {author} {\bibinfo {author} {\bibfnamefont {A.}~\bibnamefont
  {Masago}}, \bibinfo {author} {\bibfnamefont {K.}~\bibnamefont {Shirai}},\
  and\ \bibinfo {author} {\bibfnamefont {H.}~\bibnamefont {Katayama-Yoshida}},\
  }\bibfield  {title} {\bibinfo {title} {Crystal stability of $\alpha$- and
  $\beta$-boron},\ }\href@noop {} {\bibfield  {journal} {\bibinfo  {journal}
  {Physics of Semiconductors, Pts A and B}\ }\textbf {\bibinfo {volume}
  {772}},\ \bibinfo {pages} {87} (\bibinfo {year} {2005})}\BibitemShut
  {NoStop}%
\bibitem [{\citenamefont {Prasad}\ \emph {et~al.}(2005)\citenamefont {Prasad},
  \citenamefont {Balakrishnarajan},\ and\ \citenamefont
  {Jemmis}}]{Prasad:2005}%
  \BibitemOpen
  \bibfield  {author} {\bibinfo {author} {\bibfnamefont {D.~L. V.~K.}\
  \bibnamefont {Prasad}}, \bibinfo {author} {\bibfnamefont {M.~M.}\
  \bibnamefont {Balakrishnarajan}},\ and\ \bibinfo {author} {\bibfnamefont
  {E.}~\bibnamefont {Jemmis}},\ }\bibfield  {title} {\bibinfo {title}
  {Electronic structure and bonding of $\beta$-rhombohedral boron using cluster
  fragment approach},\ }\href@noop {} {\bibfield  {journal} {\bibinfo
  {journal} {Phys. Rev. B}\ }\textbf {\bibinfo {volume} {72}},\ \bibinfo
  {pages} {195102} (\bibinfo {year} {2005})}\BibitemShut {NoStop}%
\bibitem [{\citenamefont {Shang}\ \emph {et~al.}(2007)\citenamefont {Shang},
  \citenamefont {Wang}, \citenamefont {Arroyave},\ and\ \citenamefont
  {Liu}}]{Shang:2007}%
  \BibitemOpen
  \bibfield  {author} {\bibinfo {author} {\bibfnamefont {S.}~\bibnamefont
  {Shang}}, \bibinfo {author} {\bibfnamefont {Y.}~\bibnamefont {Wang}},
  \bibinfo {author} {\bibfnamefont {R.}~\bibnamefont {Arroyave}},\ and\
  \bibinfo {author} {\bibfnamefont {Z.-K.}\ \bibnamefont {Liu}},\ }\bibfield
  {title} {\bibinfo {title} {Phase stability in alpha- and beta-rhombohedral
  boron},\ }\href@noop {} {\bibfield  {journal} {\bibinfo  {journal} {Phys.
  Rev. B}\ }\textbf {\bibinfo {volume} {75}},\ \bibinfo {pages} {092101}
  (\bibinfo {year} {2007})}\BibitemShut {NoStop}%
\bibitem [{\citenamefont {Siberchicot}(2009)}]{Siberchicot:2007}%
  \BibitemOpen
  \bibfield  {author} {\bibinfo {author} {\bibfnamefont {B.}~\bibnamefont
  {Siberchicot}},\ }\bibfield  {title} {\bibinfo {title} {Ab initio equation of
  state of alpha- and beta-boron: Possible amorphization of beta-boron under
  high pressure},\ }\href@noop {} {\bibfield  {journal} {\bibinfo  {journal}
  {Phys. Rev. B}\ }\textbf {\bibinfo {volume} {79}},\ \bibinfo {pages} {224101}
  (\bibinfo {year} {2009})}\BibitemShut {NoStop}%
\bibitem [{\citenamefont {van Setten}\ \emph {et~al.}(2007)\citenamefont {van
  Setten}, \citenamefont {Uijttewaal}, \citenamefont {de~Wijs},\ and\
  \citenamefont {de~Groot}}]{deGroot:2007}%
  \BibitemOpen
  \bibfield  {author} {\bibinfo {author} {\bibfnamefont {M.~J.}\ \bibnamefont
  {van Setten}}, \bibinfo {author} {\bibfnamefont {M.~A.}\ \bibnamefont
  {Uijttewaal}}, \bibinfo {author} {\bibfnamefont {G.~A.}\ \bibnamefont
  {de~Wijs}},\ and\ \bibinfo {author} {\bibfnamefont {R.~A.}\ \bibnamefont
  {de~Groot}},\ }\bibfield  {title} {\bibinfo {title} {Thermodynamic stability
  of boron: The role of defects and zero point motion},\ }\href@noop {}
  {\bibfield  {journal} {\bibinfo  {journal} {J. Am. Chem. Soc.}\ }\textbf
  {\bibinfo {volume} {129}},\ \bibinfo {pages} {2458} (\bibinfo {year}
  {2007})}\BibitemShut {NoStop}%
\bibitem [{\citenamefont {M.~Widom}(2008{\natexlab{a}})}]{Mikhalkovic:2008a}%
  \BibitemOpen
  \bibfield  {author} {\bibinfo {author} {\bibfnamefont {M.~M.}\ \bibnamefont
  {M.~Widom}},\ }\bibfield  {title} {\bibinfo {title} {Thermodynamic stability
  of boron: The role of defects and zero point motion},\ }\href@noop {}
  {\bibfield  {journal} {\bibinfo  {journal} {16th International Symposium on
  Boron, Borides, and Related Materials}\ }\textbf {\bibinfo {volume} {176}},\
  \bibinfo {pages} {012024} (\bibinfo {year} {2008}{\natexlab{a}})}\BibitemShut
  {NoStop}%
\bibitem [{\citenamefont {M.~Widom}(2008{\natexlab{b}})}]{Mikhalkovic:2008b}%
  \BibitemOpen
  \bibfield  {author} {\bibinfo {author} {\bibfnamefont {M.~M.}\ \bibnamefont
  {M.~Widom}},\ }\bibfield  {title} {\bibinfo {title} {Symmetry-broken crystal
  structure of elemental boron at low temperature},\ }\href@noop {} {\bibfield
  {journal} {\bibinfo  {journal} {Phys. Rev. B}\ }\textbf {\bibinfo {volume}
  {77}},\ \bibinfo {pages} {064113} (\bibinfo {year}
  {2008}{\natexlab{b}})}\BibitemShut {NoStop}%
\bibitem [{\citenamefont {Ogitsu}\ \emph {et~al.}(2009)\citenamefont {Ogitsu},
  \citenamefont {Gygi}, \citenamefont {Reed}, \citenamefont {Motome},
  \citenamefont {Schwegler},\ and\ \citenamefont {Galli}}]{Ogitsu:2009}%
  \BibitemOpen
  \bibfield  {author} {\bibinfo {author} {\bibfnamefont {T.}~\bibnamefont
  {Ogitsu}}, \bibinfo {author} {\bibfnamefont {F.}~\bibnamefont {Gygi}},
  \bibinfo {author} {\bibfnamefont {J.}~\bibnamefont {Reed}}, \bibinfo {author}
  {\bibfnamefont {Y.}~\bibnamefont {Motome}}, \bibinfo {author} {\bibfnamefont
  {E.}~\bibnamefont {Schwegler}},\ and\ \bibinfo {author} {\bibfnamefont
  {G.}~\bibnamefont {Galli}},\ }\bibfield  {title} {\bibinfo {title} {Imperfect
  crystal and unusual semiconductor: Boron, a frustrated element},\ }\href@noop
  {} {\bibfield  {journal} {\bibinfo  {journal} {J. Am. Chem. Soc.}\ }\textbf
  {\bibinfo {volume} {131}},\ \bibinfo {pages} {1903} (\bibinfo {year}
  {2009})}\BibitemShut {NoStop}%
\bibitem [{\citenamefont {Ogitsu}\ \emph {et~al.}(2010)\citenamefont {Ogitsu},
  \citenamefont {Gygi}, \citenamefont {Reed}, \citenamefont {Udagawa},
  \citenamefont {Motome}, \citenamefont {Schwegler},\ and\ \citenamefont
  {Galli}}]{Ogitsu:2010}%
  \BibitemOpen
  \bibfield  {author} {\bibinfo {author} {\bibfnamefont {T.}~\bibnamefont
  {Ogitsu}}, \bibinfo {author} {\bibfnamefont {F.}~\bibnamefont {Gygi}},
  \bibinfo {author} {\bibfnamefont {J.}~\bibnamefont {Reed}}, \bibinfo {author}
  {\bibfnamefont {M.}~\bibnamefont {Udagawa}}, \bibinfo {author} {\bibfnamefont
  {Y.}~\bibnamefont {Motome}}, \bibinfo {author} {\bibfnamefont
  {E.}~\bibnamefont {Schwegler}},\ and\ \bibinfo {author} {\bibfnamefont
  {G.}~\bibnamefont {Galli}},\ }\bibfield  {title} {\bibinfo {title}
  {Geometrical frustration in an elemental solid: An ising model to explain the
  defect structure of beta -rhombohedral boron},\ }\href@noop {} {\bibfield
  {journal} {\bibinfo  {journal} {Phys. Rev. B}\ }\textbf {\bibinfo {volume}
  {81}},\ \bibinfo {pages} {020102} (\bibinfo {year} {2010})}\BibitemShut
  {NoStop}%
\bibitem [{\citenamefont {Ogitsu}\ and\ \citenamefont
  {Schwegler}(2012)}]{Ogitsu:2012}%
  \BibitemOpen
  \bibfield  {author} {\bibinfo {author} {\bibfnamefont {T.}~\bibnamefont
  {Ogitsu}}\ and\ \bibinfo {author} {\bibfnamefont {E.}~\bibnamefont
  {Schwegler}},\ }\bibfield  {title} {\bibinfo {title} {The $\alpha$–$\beta$
  phase boundary of elemental boron},\ }\href@noop {} {\bibfield  {journal}
  {\bibinfo  {journal} {Solid State Sci.}\ }\textbf {\bibinfo {volume} {14}},\
  \bibinfo {pages} {1598} (\bibinfo {year} {2012})}\BibitemShut {NoStop}%
\bibitem [{\citenamefont {White}\ \emph {et~al.}(2015)\citenamefont {White},
  \citenamefont {Cerqueira}, \citenamefont {Whitman}, \citenamefont {Johnson},\
  and\ \citenamefont {Ogitsu}}]{Ogitsu:2015}%
  \BibitemOpen
  \bibfield  {author} {\bibinfo {author} {\bibfnamefont {M.~A.}\ \bibnamefont
  {White}}, \bibinfo {author} {\bibfnamefont {A.~B.}\ \bibnamefont
  {Cerqueira}}, \bibinfo {author} {\bibfnamefont {C.~A.}\ \bibnamefont
  {Whitman}}, \bibinfo {author} {\bibfnamefont {M.~B.}\ \bibnamefont
  {Johnson}},\ and\ \bibinfo {author} {\bibfnamefont {T.}~\bibnamefont
  {Ogitsu}},\ }\bibfield  {title} {\bibinfo {title} {Determination of phase
  stability of elemental boron},\ }\href@noop {} {\bibfield  {journal}
  {\bibinfo  {journal} {Angew. Chem. Int. Ed.}\ }\textbf {\bibinfo {volume}
  {127}},\ \bibinfo {pages} {3697} (\bibinfo {year} {2015})}\BibitemShut
  {NoStop}%
\bibitem [{\citenamefont {Laubengayer}\ \emph {et~al.}(1943)\citenamefont
  {Laubengayer}, \citenamefont {Hurd}, \citenamefont {Newkirk},\ and\
  \citenamefont {Hoard}}]{Laubengeyer:1943}%
  \BibitemOpen
  \bibfield  {author} {\bibinfo {author} {\bibfnamefont {A.~W.}\ \bibnamefont
  {Laubengayer}}, \bibinfo {author} {\bibfnamefont {D.~T.}\ \bibnamefont
  {Hurd}}, \bibinfo {author} {\bibfnamefont {A.~E.}\ \bibnamefont {Newkirk}},\
  and\ \bibinfo {author} {\bibfnamefont {J.~L.}\ \bibnamefont {Hoard}},\
  }\bibfield  {title} {\bibinfo {title} {Boron. i. preparation and properties
  of pure crystalline boron},\ }\href@noop {} {\bibfield  {journal} {\bibinfo
  {journal} {J. Am. Chem. Soc.}\ }\textbf {\bibinfo {volume} {65}},\ \bibinfo
  {pages} {1924} (\bibinfo {year} {1943})}\BibitemShut {NoStop}%
\bibitem [{\citenamefont {Hoard}\ \emph {et~al.}(1951)\citenamefont {Hoard},
  \citenamefont {Geller},\ and\ \citenamefont {Hughes}}]{Hoard:1951}%
  \BibitemOpen
  \bibfield  {author} {\bibinfo {author} {\bibfnamefont {J.~L.}\ \bibnamefont
  {Hoard}}, \bibinfo {author} {\bibfnamefont {S.}~\bibnamefont {Geller}},\ and\
  \bibinfo {author} {\bibfnamefont {R.~E.}\ \bibnamefont {Hughes}},\ }\bibfield
   {title} {\bibinfo {title} {On the structure of elementary boron},\
  }\href@noop {} {\bibfield  {journal} {\bibinfo  {journal} {J. Am. Chem.
  Soc.}\ }\textbf {\bibinfo {volume} {73}},\ \bibinfo {pages} {1892} (\bibinfo
  {year} {1951})}\BibitemShut {NoStop}%
\bibitem [{\citenamefont {Hoard}\ \emph {et~al.}(1958)\citenamefont {Hoard},
  \citenamefont {Hughes},\ and\ \citenamefont {Sands}}]{Hoard:1958}%
  \BibitemOpen
  \bibfield  {author} {\bibinfo {author} {\bibfnamefont {J.~L.}\ \bibnamefont
  {Hoard}}, \bibinfo {author} {\bibfnamefont {R.~E.}\ \bibnamefont {Hughes}},\
  and\ \bibinfo {author} {\bibfnamefont {D.~E.}\ \bibnamefont {Sands}},\
  }\bibfield  {title} {\bibinfo {title} {The structure of tetragonal boron},\
  }\href@noop {} {\bibfield  {journal} {\bibinfo  {journal} {J. Am. Chem.
  Soc.}\ }\textbf {\bibinfo {volume} {80}},\ \bibinfo {pages} {4507} (\bibinfo
  {year} {1958})}\BibitemShut {NoStop}%
\bibitem [{\citenamefont {Hoard}\ and\ \citenamefont
  {Newkirk}(1960)}]{Hoard:1960}%
  \BibitemOpen
  \bibfield  {author} {\bibinfo {author} {\bibfnamefont {J.~L.}\ \bibnamefont
  {Hoard}}\ and\ \bibinfo {author} {\bibfnamefont {A.~E.}\ \bibnamefont
  {Newkirk}},\ }\bibfield  {title} {\bibinfo {title} {An analysis of
  polymorphism in boron based upon x-ray diffraction results},\ }\href@noop {}
  {\bibfield  {journal} {\bibinfo  {journal} {J. Am. Chem. Soc.}\ }\textbf
  {\bibinfo {volume} {82}},\ \bibinfo {pages} {70} (\bibinfo {year}
  {1960})}\BibitemShut {NoStop}%
\bibitem [{\citenamefont {Talley}(1960)}]{Talley:1960}%
  \BibitemOpen
  \bibfield  {author} {\bibinfo {author} {\bibfnamefont {C.~P.}\ \bibnamefont
  {Talley}},\ }\bibfield  {title} {\bibinfo {title} {A new polymorph of
  boron},\ }\href@noop {} {\bibfield  {journal} {\bibinfo  {journal} {Acta
  Crystallogr.}\ }\textbf {\bibinfo {volume} {13}},\ \bibinfo {pages} {271}
  (\bibinfo {year} {1960})}\BibitemShut {NoStop}%
\bibitem [{\citenamefont {Vlasse}\ \emph
  {et~al.}(1979{\natexlab{a}})\citenamefont {Vlasse}, \citenamefont {Naslain},
  \citenamefont {Kasper},\ and\ \citenamefont {Ploog}}]{Vlasse:1979a}%
  \BibitemOpen
  \bibfield  {author} {\bibinfo {author} {\bibfnamefont {M.}~\bibnamefont
  {Vlasse}}, \bibinfo {author} {\bibfnamefont {R.}~\bibnamefont {Naslain}},
  \bibinfo {author} {\bibfnamefont {J.~S.}\ \bibnamefont {Kasper}},\ and\
  \bibinfo {author} {\bibfnamefont {K.}~\bibnamefont {Ploog}},\ }\bibfield
  {title} {\bibinfo {title} {Crystal-structure of tetragonal boron},\
  }\href@noop {} {\bibfield  {journal} {\bibinfo  {journal} {J. Less-Common
  Met.}\ }\textbf {\bibinfo {volume} {67}},\ \bibinfo {pages} {1} (\bibinfo
  {year} {1979}{\natexlab{a}})}\BibitemShut {NoStop}%
\bibitem [{\citenamefont {Vlasse}\ \emph
  {et~al.}(1979{\natexlab{b}})\citenamefont {Vlasse}, \citenamefont {Naslain},
  \citenamefont {Kasper},\ and\ \citenamefont {Ploog}}]{Vlasse:1979b}%
  \BibitemOpen
  \bibfield  {author} {\bibinfo {author} {\bibfnamefont {M.}~\bibnamefont
  {Vlasse}}, \bibinfo {author} {\bibfnamefont {R.}~\bibnamefont {Naslain}},
  \bibinfo {author} {\bibfnamefont {J.~S.}\ \bibnamefont {Kasper}},\ and\
  \bibinfo {author} {\bibfnamefont {K.}~\bibnamefont {Ploog}},\ }\bibfield
  {title} {\bibinfo {title} {Crystal-structure of tetragonal boron related to
  alpha-b12},\ }\href@noop {} {\bibfield  {journal} {\bibinfo  {journal} {J.
  Solid State Chem.}\ }\textbf {\bibinfo {volume} {28}},\ \bibinfo {pages}
  {289} (\bibinfo {year} {1979}{\natexlab{b}})}\BibitemShut {NoStop}%
\bibitem [{\citenamefont {Parakhonskiy}\ \emph {et~al.}(2013)\citenamefont
  {Parakhonskiy}, \citenamefont {Dubrovinskaia}, \citenamefont {Bykova},
  \citenamefont {Wirth},\ and\ \citenamefont
  {Dubrovinsky}}]{Parakhonskiy:2013}%
  \BibitemOpen
  \bibfield  {author} {\bibinfo {author} {\bibfnamefont {G.}~\bibnamefont
  {Parakhonskiy}}, \bibinfo {author} {\bibfnamefont {N.}~\bibnamefont
  {Dubrovinskaia}}, \bibinfo {author} {\bibfnamefont {E.}~\bibnamefont
  {Bykova}}, \bibinfo {author} {\bibfnamefont {R.}~\bibnamefont {Wirth}},\ and\
  \bibinfo {author} {\bibfnamefont {L.}~\bibnamefont {Dubrovinsky}},\
  }\bibfield  {title} {\bibinfo {title} {High pressure synthesis and
  investigation of single crystals of metastable boron phases},\ }\href@noop {}
  {\bibfield  {journal} {\bibinfo  {journal} {High Pressure Res.}\ }\textbf
  {\bibinfo {volume} {33}},\ \bibinfo {pages} {673} (\bibinfo {year}
  {2013})}\BibitemShut {NoStop}%
\bibitem [{\citenamefont {Karmodak}\ and\ \citenamefont
  {Jemmis}(2017)}]{Karmodak:2017}%
  \BibitemOpen
  \bibfield  {author} {\bibinfo {author} {\bibfnamefont {N.}~\bibnamefont
  {Karmodak}}\ and\ \bibinfo {author} {\bibfnamefont {E.~D.}\ \bibnamefont
  {Jemmis}},\ }\bibfield  {title} {\bibinfo {title} {Fragment approach to the
  electronic structure of $\tau$-boron allotrope},\ }\href@noop {} {\bibfield
  {journal} {\bibinfo  {journal} {Phys. Rev. B}\ }\textbf {\bibinfo {volume}
  {95}},\ \bibinfo {pages} {165128} (\bibinfo {year} {2017})}\BibitemShut
  {NoStop}%
\bibitem [{\citenamefont {An}\ \emph {et~al.}(2016)\citenamefont {An},
  \citenamefont {Reddy}, \citenamefont {Xie}, \citenamefont {Hemker},\ and\
  \citenamefont {Goddard~III}}]{An:2016a}%
  \BibitemOpen
  \bibfield  {author} {\bibinfo {author} {\bibfnamefont {Q.}~\bibnamefont
  {An}}, \bibinfo {author} {\bibfnamefont {M.}~\bibnamefont {Reddy}}, \bibinfo
  {author} {\bibfnamefont {K.~Y.}\ \bibnamefont {Xie}}, \bibinfo {author}
  {\bibfnamefont {K.~J.}\ \bibnamefont {Hemker}},\ and\ \bibinfo {author}
  {\bibfnamefont {W.~A.}\ \bibnamefont {Goddard~III}},\ }\bibfield  {title}
  {\bibinfo {title} {New ground state crystal structure of elemental boron},\
  }\href@noop {} {\bibfield  {journal} {\bibinfo  {journal} {Phys. Rev. Lett.}\
  }\textbf {\bibinfo {volume} {117}},\ \bibinfo {pages} {085501} (\bibinfo
  {year} {2016})}\BibitemShut {NoStop}%
\bibitem [{\citenamefont {Oganov}\ \emph {et~al.}(2009)\citenamefont {Oganov},
  \citenamefont {Chen}, \citenamefont {Gatti}, \citenamefont {Ma},
  \citenamefont {Ma}, \citenamefont {Glass}, \citenamefont {Liu}, \citenamefont
  {Yu}, \citenamefont {Kurakevych},\ and\ \citenamefont
  {Solozhenko}}]{Oganov:2009}%
  \BibitemOpen
  \bibfield  {author} {\bibinfo {author} {\bibfnamefont {A.}~\bibnamefont
  {Oganov}}, \bibinfo {author} {\bibfnamefont {J.}~\bibnamefont {Chen}},
  \bibinfo {author} {\bibfnamefont {C.}~\bibnamefont {Gatti}}, \bibinfo
  {author} {\bibfnamefont {Y.}~\bibnamefont {Ma}}, \bibinfo {author}
  {\bibfnamefont {Y.}~\bibnamefont {Ma}}, \bibinfo {author} {\bibfnamefont
  {C.}~\bibnamefont {Glass}}, \bibinfo {author} {\bibfnamefont
  {Z.}~\bibnamefont {Liu}}, \bibinfo {author} {\bibfnamefont {T.}~\bibnamefont
  {Yu}}, \bibinfo {author} {\bibfnamefont {O.}~\bibnamefont {Kurakevych}},\
  and\ \bibinfo {author} {\bibfnamefont {V.}~\bibnamefont {Solozhenko}},\
  }\bibfield  {title} {\bibinfo {title} {Ionic high-pressure form of elemental
  boron},\ }\href@noop {} {\bibfield  {journal} {\bibinfo  {journal} {Nature}\
  }\textbf {\bibinfo {volume} {457}},\ \bibinfo {pages} {863} (\bibinfo {year}
  {2009})}\BibitemShut {NoStop}%
\bibitem [{\citenamefont {Wentorf}(1965)}]{Wentorf:1965}%
  \BibitemOpen
  \bibfield  {author} {\bibinfo {author} {\bibfnamefont {R.}~\bibnamefont
  {Wentorf}},\ }\bibfield  {title} {\bibinfo {title} {Boron: Another form},\
  }\href@noop {} {\bibfield  {journal} {\bibinfo  {journal} {Science}\ }\textbf
  {\bibinfo {volume} {147}},\ \bibinfo {pages} {49} (\bibinfo {year}
  {1965})}\BibitemShut {NoStop}%
\bibitem [{\citenamefont {Zarechnaya}\ \emph
  {et~al.}(2009{\natexlab{b}})\citenamefont {Zarechnaya}, \citenamefont
  {Dubrovinsky}, \citenamefont {Dubrovinskaia}, \citenamefont {Miyajima},
  \citenamefont {Filinchuk}, \citenamefont {Chernyshov},\ and\ \citenamefont
  {Dmitriev}}]{Zarechnaya:2008}%
  \BibitemOpen
  \bibfield  {author} {\bibinfo {author} {\bibfnamefont {E.}~\bibnamefont
  {Zarechnaya}}, \bibinfo {author} {\bibfnamefont {L.}~\bibnamefont
  {Dubrovinsky}}, \bibinfo {author} {\bibfnamefont {N.}~\bibnamefont
  {Dubrovinskaia}}, \bibinfo {author} {\bibfnamefont {N.}~\bibnamefont
  {Miyajima}}, \bibinfo {author} {\bibfnamefont {Y.}~\bibnamefont {Filinchuk}},
  \bibinfo {author} {\bibfnamefont {D.}~\bibnamefont {Chernyshov}},\ and\
  \bibinfo {author} {\bibfnamefont {V.}~\bibnamefont {Dmitriev}},\ }\bibfield
  {title} {\bibinfo {title} {Synthesis of an orthorhombic high pressure boron
  phase},\ }\href@noop {} {\bibfield  {journal} {\bibinfo  {journal} {Sci.
  Technol. Adv. Mater.}\ }\textbf {\bibinfo {volume} {9}},\ \bibinfo {pages}
  {044209} (\bibinfo {year} {2009}{\natexlab{b}})}\BibitemShut {NoStop}%
\bibitem [{\citenamefont {H{\"a}ussermann}\ and\ \citenamefont
  {Mikhaylushkin}(2010)}]{Haussermann:2010}%
  \BibitemOpen
  \bibfield  {author} {\bibinfo {author} {\bibfnamefont {U.}~\bibnamefont
  {H{\"a}ussermann}}\ and\ \bibinfo {author} {\bibfnamefont {A.}~\bibnamefont
  {Mikhaylushkin}},\ }\bibfield  {title} {\bibinfo {title} {Structure and
  bonding of $\gamma$-b28: Is the high pressure form of elemental boron
  ionic?},\ }\href@noop {} {\bibfield  {journal} {\bibinfo  {journal} {Inorg.
  Chem.}\ }\textbf {\bibinfo {volume} {49}},\ \bibinfo {pages} {11270}
  (\bibinfo {year} {2010})}\BibitemShut {NoStop}%
\bibitem [{\citenamefont {Mondal}\ \emph {et~al.}(2011)\citenamefont {Mondal},
  \citenamefont {van Smaalen}, \citenamefont {Sch\"onleber}, \citenamefont
  {Filinchuk}, \citenamefont {Chernyshov}, \citenamefont {Simak}, \citenamefont
  {Mikhaylushkin}, \citenamefont {Abrikosov}, \citenamefont {Zarechnaya},
  \citenamefont {Dubrovinsky},\ and\ \citenamefont
  {Dubrovinskaia}}]{Mondal:2011}%
  \BibitemOpen
  \bibfield  {author} {\bibinfo {author} {\bibfnamefont {S.}~\bibnamefont
  {Mondal}}, \bibinfo {author} {\bibfnamefont {S.}~\bibnamefont {van Smaalen}},
  \bibinfo {author} {\bibfnamefont {A.}~\bibnamefont {Sch\"onleber}}, \bibinfo
  {author} {\bibfnamefont {Y.}~\bibnamefont {Filinchuk}}, \bibinfo {author}
  {\bibfnamefont {D.}~\bibnamefont {Chernyshov}}, \bibinfo {author}
  {\bibfnamefont {S.~I.}\ \bibnamefont {Simak}}, \bibinfo {author}
  {\bibfnamefont {A.~S.}\ \bibnamefont {Mikhaylushkin}}, \bibinfo {author}
  {\bibfnamefont {I.~A.}\ \bibnamefont {Abrikosov}}, \bibinfo {author}
  {\bibfnamefont {E.}~\bibnamefont {Zarechnaya}}, \bibinfo {author}
  {\bibfnamefont {L.}~\bibnamefont {Dubrovinsky}},\ and\ \bibinfo {author}
  {\bibfnamefont {N.}~\bibnamefont {Dubrovinskaia}},\ }\bibfield  {title}
  {\bibinfo {title} {Electron-deficient and polycenter bonds in the
  high-pressure $\gamma$-b$_{28}$ phase of boron},\ }\href@noop {} {\bibfield
  {journal} {\bibinfo  {journal} {Phys. Rev. Lett.}\ }\textbf {\bibinfo
  {volume} {106}},\ \bibinfo {pages} {215502} (\bibinfo {year}
  {2011})}\BibitemShut {NoStop}%
\bibitem [{\citenamefont {von Schnering}\ and\ \citenamefont
  {Nesper}(1991)}]{vonSchnering:1991}%
  \BibitemOpen
  \bibfield  {author} {\bibinfo {author} {\bibfnamefont {H.}~\bibnamefont {von
  Schnering}}\ and\ \bibinfo {author} {\bibfnamefont {R.}~\bibnamefont
  {Nesper}},\ }\bibfield  {title} {\bibinfo {title} {$\alpha$-gallium: An
  alternative to the boron structure},\ }\href@noop {} {\bibfield  {journal}
  {\bibinfo  {journal} {Acta Chem. Scand.}\ }\textbf {\bibinfo {volume} {45}},\
  \bibinfo {pages} {870} (\bibinfo {year} {1991})}\BibitemShut {NoStop}%
\bibitem [{\citenamefont {H{\"a}ussermann}\ \emph {et~al.}(2003)\citenamefont
  {H{\"a}ussermann}, \citenamefont {Simak}, \citenamefont {Ahuja},\ and\
  \citenamefont {Johansson}}]{Haussermann:2003}%
  \BibitemOpen
  \bibfield  {author} {\bibinfo {author} {\bibfnamefont {U.}~\bibnamefont
  {H{\"a}ussermann}}, \bibinfo {author} {\bibfnamefont {S.}~\bibnamefont
  {Simak}}, \bibinfo {author} {\bibfnamefont {R.}~\bibnamefont {Ahuja}},\ and\
  \bibinfo {author} {\bibfnamefont {B.}~\bibnamefont {Johansson}},\ }\bibfield
  {title} {\bibinfo {title} {Metal-nonmetal transition in the boron group
  elements},\ }\href@noop {} {\bibfield  {journal} {\bibinfo  {journal} {Phys.
  Rev. Lett.}\ }\textbf {\bibinfo {volume} {90}},\ \bibinfo {pages} {065701}
  (\bibinfo {year} {2003})}\BibitemShut {NoStop}%
\bibitem [{\citenamefont {Segall}\ and\ \citenamefont
  {Arias}(2003)}]{Segall:2003}%
  \BibitemOpen
  \bibfield  {author} {\bibinfo {author} {\bibfnamefont {D.}~\bibnamefont
  {Segall}}\ and\ \bibinfo {author} {\bibfnamefont {T.}~\bibnamefont {Arias}},\
  }\bibfield  {title} {\bibinfo {title} {Ab initio approach for high-pressure
  systems with application to high-pressure phases of boron: Perturbative
  momentum-space potentials},\ }\href@noop {} {\bibfield  {journal} {\bibinfo
  {journal} {Phys. Rev. B}\ }\textbf {\bibinfo {volume} {67}},\ \bibinfo
  {pages} {064105} (\bibinfo {year} {2003})}\BibitemShut {NoStop}%
\bibitem [{\citenamefont {Chuvashova}\ \emph {et~al.}(2017)\citenamefont
  {Chuvashova}, \citenamefont {Bykova}, \citenamefont {Bykov}, \citenamefont
  {Prakapenka}, \citenamefont {Glazyrin}, \citenamefont {Mezour}, \citenamefont
  {Dubrovinsky},\ and\ \citenamefont {Dubrovinskaia}}]{Chuvashova:2017}%
  \BibitemOpen
  \bibfield  {author} {\bibinfo {author} {\bibfnamefont {I.}~\bibnamefont
  {Chuvashova}}, \bibinfo {author} {\bibfnamefont {E.}~\bibnamefont {Bykova}},
  \bibinfo {author} {\bibfnamefont {M.}~\bibnamefont {Bykov}}, \bibinfo
  {author} {\bibfnamefont {V.}~\bibnamefont {Prakapenka}}, \bibinfo {author}
  {\bibfnamefont {K.}~\bibnamefont {Glazyrin}}, \bibinfo {author}
  {\bibfnamefont {M.}~\bibnamefont {Mezour}}, \bibinfo {author} {\bibfnamefont
  {L.}~\bibnamefont {Dubrovinsky}},\ and\ \bibinfo {author} {\bibfnamefont
  {N.}~\bibnamefont {Dubrovinskaia}},\ }\bibfield  {title} {\bibinfo {title}
  {Nonicosahedral boron allotrope synthesized at high pressure and high
  temperature},\ }\href@noop {} {\bibfield  {journal} {\bibinfo  {journal}
  {Phys. Rev. B}\ }\textbf {\bibinfo {volume} {95}},\ \bibinfo {pages} {180102}
  (\bibinfo {year} {2017})}\BibitemShut {NoStop}%
\bibitem [{\citenamefont {Sanz}\ \emph {et~al.}(2002)\citenamefont {Sanz},
  \citenamefont {Loubeyre},\ and\ \citenamefont {Mezouar}}]{Sanz:2002}%
  \BibitemOpen
  \bibfield  {author} {\bibinfo {author} {\bibfnamefont {D.~N.}\ \bibnamefont
  {Sanz}}, \bibinfo {author} {\bibfnamefont {P.}~\bibnamefont {Loubeyre}},\
  and\ \bibinfo {author} {\bibfnamefont {M.}~\bibnamefont {Mezouar}},\
  }\bibfield  {title} {\bibinfo {title} {Equation of state and pressure induced
  amorphization of beta -boron from x-ray measurements up to 100 gpa},\
  }\href@noop {} {\bibfield  {journal} {\bibinfo  {journal} {Phys. Rev. Lett}\
  }\textbf {\bibinfo {volume} {89}},\ \bibinfo {pages} {245501} (\bibinfo
  {year} {2002})}\BibitemShut {NoStop}%
\bibitem [{\citenamefont {Shirai}\ \emph {et~al.}(2009)\citenamefont {Shirai},
  \citenamefont {Dekura},\ and\ \citenamefont {Yanase}}]{Shirai:2009}%
  \BibitemOpen
  \bibfield  {author} {\bibinfo {author} {\bibfnamefont {K.}~\bibnamefont
  {Shirai}}, \bibinfo {author} {\bibfnamefont {H.}~\bibnamefont {Dekura}},\
  and\ \bibinfo {author} {\bibfnamefont {A.}~\bibnamefont {Yanase}},\
  }\bibfield  {title} {\bibinfo {title} {Electronic structure and electrical
  resistivity of alpha-boron under high pressure},\ }\href@noop {} {\bibfield
  {journal} {\bibinfo  {journal} {J.Phys. Soc. Jpn.}\ }\textbf {\bibinfo
  {volume} {78}} (\bibinfo {year} {2009})}\BibitemShut {NoStop}%
\bibitem [{\citenamefont {Shirai}\ \emph {et~al.}(2011)\citenamefont {Shirai},
  \citenamefont {Dekura}, \citenamefont {Mori}, \citenamefont {Fujii},
  \citenamefont {Hyodo},\ and\ \citenamefont {Kimura}}]{Shirai:2011}%
  \BibitemOpen
  \bibfield  {author} {\bibinfo {author} {\bibfnamefont {K.}~\bibnamefont
  {Shirai}}, \bibinfo {author} {\bibfnamefont {H.}~\bibnamefont {Dekura}},
  \bibinfo {author} {\bibfnamefont {Y.}~\bibnamefont {Mori}}, \bibinfo {author}
  {\bibfnamefont {Y.}~\bibnamefont {Fujii}}, \bibinfo {author} {\bibfnamefont
  {H.}~\bibnamefont {Hyodo}},\ and\ \bibinfo {author} {\bibfnamefont
  {K.}~\bibnamefont {Kimura}},\ }\bibfield  {title} {\bibinfo {title}
  {Structural study of alpha-rhombohedral boron at high pressures},\
  }\href@noop {} {\bibfield  {journal} {\bibinfo  {journal} {J.Phys. Soc.
  Jpn.}\ }\textbf {\bibinfo {volume} {80}} (\bibinfo {year}
  {2011})}\BibitemShut {NoStop}%
\bibitem [{\citenamefont {Nagatochi}\ \emph {et~al.}(2011)\citenamefont
  {Nagatochi}, \citenamefont {Hyodo}, \citenamefont {Sumiyoshi}, \citenamefont
  {Soga}, \citenamefont {Sato}, \citenamefont {Terauchi}, \citenamefont
  {Esaka},\ and\ \citenamefont {Kimura}}]{Nagatochi:2011}%
  \BibitemOpen
  \bibfield  {author} {\bibinfo {author} {\bibfnamefont {T.}~\bibnamefont
  {Nagatochi}}, \bibinfo {author} {\bibfnamefont {H.}~\bibnamefont {Hyodo}},
  \bibinfo {author} {\bibfnamefont {A.}~\bibnamefont {Sumiyoshi}}, \bibinfo
  {author} {\bibfnamefont {K.}~\bibnamefont {Soga}}, \bibinfo {author}
  {\bibfnamefont {Y.}~\bibnamefont {Sato}}, \bibinfo {author} {\bibfnamefont
  {M.}~\bibnamefont {Terauchi}}, \bibinfo {author} {\bibfnamefont
  {F.}~\bibnamefont {Esaka}},\ and\ \bibinfo {author} {\bibfnamefont
  {K.}~\bibnamefont {Kimura}},\ }\bibfield  {title} {\bibinfo {title}
  {Superconductivity in li-doped $\alpha$-rhombohedral boron},\ }\href@noop {}
  {\bibfield  {journal} {\bibinfo  {journal} {Phys. Rev. B}\ }\textbf {\bibinfo
  {volume} {83}},\ \bibinfo {pages} {184507} (\bibinfo {year}
  {2011})}\BibitemShut {NoStop}%
\bibitem [{\citenamefont {Miao}\ \emph {et~al.}(2020)\citenamefont {Miao},
  \citenamefont {Sun}, \citenamefont {Zurek},\ and\ \citenamefont
  {Lin}}]{Zurek:2019k}%
  \BibitemOpen
  \bibfield  {author} {\bibinfo {author} {\bibfnamefont {M.}~\bibnamefont
  {Miao}}, \bibinfo {author} {\bibfnamefont {Y.}~\bibnamefont {Sun}}, \bibinfo
  {author} {\bibfnamefont {E.}~\bibnamefont {Zurek}},\ and\ \bibinfo {author}
  {\bibfnamefont {H.}~\bibnamefont {Lin}},\ }\bibfield  {title} {\bibinfo
  {title} {Chemistry under high pressure},\ }\href@noop {} {\bibfield
  {journal} {\bibinfo  {journal} {Nat. Rev. Chem.}\ }\textbf {\bibinfo {volume}
  {508-527}},\ \bibinfo {pages} {4} (\bibinfo {year} {2020})}\BibitemShut
  {NoStop}%
\bibitem [{\citenamefont {Marques}\ \emph {et~al.}(2011)\citenamefont
  {Marques}, \citenamefont {McMahon}, \citenamefont {Gregoryanz}, \citenamefont
  {Hanfland}, \citenamefont {Guillaume}, \citenamefont {Pickard}, \citenamefont
  {Ackland},\ and\ \citenamefont {Nelmes}}]{Nelmes:2011a}%
  \BibitemOpen
  \bibfield  {author} {\bibinfo {author} {\bibfnamefont {M.}~\bibnamefont
  {Marques}}, \bibinfo {author} {\bibfnamefont {M.~I.}\ \bibnamefont
  {McMahon}}, \bibinfo {author} {\bibfnamefont {E.}~\bibnamefont {Gregoryanz}},
  \bibinfo {author} {\bibfnamefont {M.}~\bibnamefont {Hanfland}}, \bibinfo
  {author} {\bibfnamefont {C.~L.}\ \bibnamefont {Guillaume}}, \bibinfo {author}
  {\bibfnamefont {C.~J.}\ \bibnamefont {Pickard}}, \bibinfo {author}
  {\bibfnamefont {G.~J.}\ \bibnamefont {Ackland}},\ and\ \bibinfo {author}
  {\bibfnamefont {R.~J.}\ \bibnamefont {Nelmes}},\ }\bibfield  {title}
  {\bibinfo {title} {Crystal structures of dense lithium: A
  metal--semiconductor-metal transition},\ }\href@noop {} {\bibfield  {journal}
  {\bibinfo  {journal} {Phys. Rev. Lett.}\ }\textbf {\bibinfo {volume} {106}},\
  \bibinfo {pages} {095502 (1} (\bibinfo {year} {2011})}\BibitemShut {NoStop}%
\bibitem [{\citenamefont {Tsuppayakorn-aek}\ \emph {et~al.}(2018)\citenamefont
  {Tsuppayakorn-aek}, \citenamefont {Luo}, \citenamefont {Watcharatharapong},
  \citenamefont {Ahuja},\ and\ \citenamefont
  {Bovornratanaraks}}]{Tsuppayakorn-aek:2018}%
  \BibitemOpen
  \bibfield  {author} {\bibinfo {author} {\bibfnamefont {P.}~\bibnamefont
  {Tsuppayakorn-aek}}, \bibinfo {author} {\bibfnamefont {W.}~\bibnamefont
  {Luo}}, \bibinfo {author} {\bibfnamefont {T.}~\bibnamefont
  {Watcharatharapong}}, \bibinfo {author} {\bibfnamefont {R.}~\bibnamefont
  {Ahuja}},\ and\ \bibinfo {author} {\bibfnamefont {T.}~\bibnamefont
  {Bovornratanaraks}},\ }\bibfield  {title} {\bibinfo {title} {Structural
  prediction of host-guest structure in lithium at high pressure},\ }\href@noop
  {} {\bibfield  {journal} {\bibinfo  {journal} {Sci. Rep.}\ }\textbf {\bibinfo
  {volume} {8}},\ \bibinfo {pages} {5278} (\bibinfo {year} {2018})}\BibitemShut
  {NoStop}%
\bibitem [{\citenamefont {Gregoryanz}\ \emph {et~al.}(2008)\citenamefont
  {Gregoryanz}, \citenamefont {Lundegaard}, \citenamefont {McMahon},
  \citenamefont {Guillaume}, \citenamefont {Nelmes},\ and\ \citenamefont
  {Mezouar}}]{Gregoryanz:2008a}%
  \BibitemOpen
  \bibfield  {author} {\bibinfo {author} {\bibfnamefont {E.}~\bibnamefont
  {Gregoryanz}}, \bibinfo {author} {\bibfnamefont {L.~F.}\ \bibnamefont
  {Lundegaard}}, \bibinfo {author} {\bibfnamefont {M.~I.}\ \bibnamefont
  {McMahon}}, \bibinfo {author} {\bibfnamefont {C.}~\bibnamefont {Guillaume}},
  \bibinfo {author} {\bibfnamefont {R.~J.}\ \bibnamefont {Nelmes}},\ and\
  \bibinfo {author} {\bibfnamefont {M.}~\bibnamefont {Mezouar}},\ }\bibfield
  {title} {\bibinfo {title} {Structural diversity of sodium},\ }\href@noop {}
  {\bibfield  {journal} {\bibinfo  {journal} {Science}\ }\textbf {\bibinfo
  {volume} {320}},\ \bibinfo {pages} {1054} (\bibinfo {year}
  {2008})}\BibitemShut {NoStop}%
\bibitem [{\citenamefont {Ma}\ \emph {et~al.}(2009)\citenamefont {Ma},
  \citenamefont {Eremets}, \citenamefont {Oganov}, \citenamefont {Xie},
  \citenamefont {Trojan}, \citenamefont {Medvedev}, \citenamefont {Lyakhov},
  \citenamefont {Valle},\ and\ \citenamefont {Prakapenka}}]{Ma:2009a}%
  \BibitemOpen
  \bibfield  {author} {\bibinfo {author} {\bibfnamefont {Y.~M.}\ \bibnamefont
  {Ma}}, \bibinfo {author} {\bibfnamefont {M.}~\bibnamefont {Eremets}},
  \bibinfo {author} {\bibfnamefont {A.~R.}\ \bibnamefont {Oganov}}, \bibinfo
  {author} {\bibfnamefont {Y.}~\bibnamefont {Xie}}, \bibinfo {author}
  {\bibfnamefont {I.}~\bibnamefont {Trojan}}, \bibinfo {author} {\bibfnamefont
  {S.}~\bibnamefont {Medvedev}}, \bibinfo {author} {\bibfnamefont {A.~O.}\
  \bibnamefont {Lyakhov}}, \bibinfo {author} {\bibfnamefont {M.}~\bibnamefont
  {Valle}},\ and\ \bibinfo {author} {\bibfnamefont {V.}~\bibnamefont
  {Prakapenka}},\ }\bibfield  {title} {\bibinfo {title} {Transparent dense
  sodium},\ }\href@noop {} {\bibfield  {journal} {\bibinfo  {journal} {Nature}\
  }\textbf {\bibinfo {volume} {458}},\ \bibinfo {pages} {182} (\bibinfo {year}
  {2009})}\BibitemShut {NoStop}%
\bibitem [{\citenamefont {Pickard}\ and\ \citenamefont
  {Needs}(2010)}]{Pickard:2010a}%
  \BibitemOpen
  \bibfield  {author} {\bibinfo {author} {\bibfnamefont {C.~J.}\ \bibnamefont
  {Pickard}}\ and\ \bibinfo {author} {\bibfnamefont {R.~J.}\ \bibnamefont
  {Needs}},\ }\bibfield  {title} {\bibinfo {title} {Aluminum at terapascal
  pressures},\ }\href@noop {} {\bibfield  {journal} {\bibinfo  {journal} {Nat.
  Mater.}\ }\textbf {\bibinfo {volume} {9}},\ \bibinfo {pages} {624} (\bibinfo
  {year} {2010})}\BibitemShut {NoStop}%
\bibitem [{\citenamefont {Solozhenko}\ \emph {et~al.}(2008)\citenamefont
  {Solozhenko}, \citenamefont {Kurakevych},\ and\ \citenamefont
  {Oganov}}]{Solozhenko:2008}%
  \BibitemOpen
  \bibfield  {author} {\bibinfo {author} {\bibfnamefont {V.}~\bibnamefont
  {Solozhenko}}, \bibinfo {author} {\bibfnamefont {O.}~\bibnamefont
  {Kurakevych}},\ and\ \bibinfo {author} {\bibfnamefont {A.}~\bibnamefont
  {Oganov}},\ }\bibfield  {title} {\bibinfo {title} {On the hardness of a new
  boron phase, orthorhombic $\gamma$-b$_{28}$},\ }\href@noop {} {\bibfield
  {journal} {\bibinfo  {journal} {J. Superhard Mater.}\ }\textbf {\bibinfo
  {volume} {30}},\ \bibinfo {pages} {428} (\bibinfo {year} {2008})}\BibitemShut
  {NoStop}%
\bibitem [{\citenamefont {Zhang}\ \emph
  {et~al.}(2018{\natexlab{b}})\citenamefont {Zhang}, \citenamefont {Zheng},
  \citenamefont {Jin}, \citenamefont {Zheng}, \citenamefont {Legut},
  \citenamefont {Yu}, \citenamefont {Gou}, \citenamefont {Fu}, \citenamefont
  {Guo}, \citenamefont {Yan}, \citenamefont {Peng}, \citenamefont {Jin},
  \citenamefont {Germann},\ and\ \citenamefont {Zhang}}]{Zhang:2018}%
  \BibitemOpen
  \bibfield  {author} {\bibinfo {author} {\bibfnamefont {S.~H.}\ \bibnamefont
  {Zhang}}, \bibinfo {author} {\bibfnamefont {X.}~\bibnamefont {Zheng}},
  \bibinfo {author} {\bibfnamefont {Q.~Q.}\ \bibnamefont {Jin}}, \bibinfo
  {author} {\bibfnamefont {S.~J.}\ \bibnamefont {Zheng}}, \bibinfo {author}
  {\bibfnamefont {D.}~\bibnamefont {Legut}}, \bibinfo {author} {\bibfnamefont
  {X.~H.}\ \bibnamefont {Yu}}, \bibinfo {author} {\bibfnamefont {H.~Y.}\
  \bibnamefont {Gou}}, \bibinfo {author} {\bibfnamefont {Z.~H.}\ \bibnamefont
  {Fu}}, \bibinfo {author} {\bibfnamefont {Y.~Q.}\ \bibnamefont {Guo}},
  \bibinfo {author} {\bibfnamefont {B.~M.}\ \bibnamefont {Yan}}, \bibinfo
  {author} {\bibfnamefont {C.}~\bibnamefont {Peng}}, \bibinfo {author}
  {\bibfnamefont {C.~Q.}\ \bibnamefont {Jin}}, \bibinfo {author} {\bibfnamefont
  {T.~C.}\ \bibnamefont {Germann}},\ and\ \bibinfo {author} {\bibfnamefont
  {R.~F.}\ \bibnamefont {Zhang}},\ }\bibfield  {title} {\bibinfo {title}
  {Unprecedented plastic flow channel in $\gamma$-b$_{28}$ through ultrasoft
  bonds: A challenge to superhardness},\ }\href@noop {} {\bibfield  {journal}
  {\bibinfo  {journal} {Phys. Rev. Mater.}\ }\textbf {\bibinfo {volume} {2}},\
  \bibinfo {pages} {123602} (\bibinfo {year} {2018}{\natexlab{b}})}\BibitemShut
  {NoStop}%
\bibitem [{\citenamefont {Lonie}\ and\ \citenamefont
  {Zurek}(2011)}]{Lonie:2011}%
  \BibitemOpen
  \bibfield  {author} {\bibinfo {author} {\bibfnamefont {D.}~\bibnamefont
  {Lonie}}\ and\ \bibinfo {author} {\bibfnamefont {E.}~\bibnamefont {Zurek}},\
  }\bibfield  {title} {\bibinfo {title} {Xtalopt: An open-souce evolutionary
  algorithm for crystal structure prediction},\ }\href@noop {} {\bibfield
  {journal} {\bibinfo  {journal} {Comp. Phys. Commun.}\ }\textbf {\bibinfo
  {volume} {182}},\ \bibinfo {pages} {372} (\bibinfo {year}
  {2011})}\BibitemShut {NoStop}%
\bibitem [{\citenamefont {Avery}\ \emph {et~al.}(2018)\citenamefont {Avery},
  \citenamefont {Falls},\ and\ \citenamefont {Zurek}}]{Avery:2018}%
  \BibitemOpen
  \bibfield  {author} {\bibinfo {author} {\bibfnamefont {P.}~\bibnamefont
  {Avery}}, \bibinfo {author} {\bibfnamefont {Z.}~\bibnamefont {Falls}},\ and\
  \bibinfo {author} {\bibfnamefont {E.}~\bibnamefont {Zurek}},\ }\bibfield
  {title} {\bibinfo {title} {Xtalopt version r11: An open-source evolutionary
  algorithm for crystal structure prediction},\ }\href@noop {} {\bibfield
  {journal} {\bibinfo  {journal} {Comput. Phys. Commun.}\ }\textbf {\bibinfo
  {volume} {222}},\ \bibinfo {pages} {418} (\bibinfo {year}
  {2018})}\BibitemShut {NoStop}%
\bibitem [{\citenamefont {Avery}\ and\ \citenamefont
  {Zurek}(2017)}]{Avery:2017}%
  \BibitemOpen
  \bibfield  {author} {\bibinfo {author} {\bibfnamefont {P.}~\bibnamefont
  {Avery}}\ and\ \bibinfo {author} {\bibfnamefont {E.}~\bibnamefont {Zurek}},\
  }\bibfield  {title} {\bibinfo {title} {Randspg: An open-source program for
  generating atomistic crystal structures with specific spacegroups},\
  }\href@noop {} {\bibfield  {journal} {\bibinfo  {journal} {Comput. Phys.
  Commun.}\ }\textbf {\bibinfo {volume} {213}},\ \bibinfo {pages} {208}
  (\bibinfo {year} {2017})}\BibitemShut {NoStop}%
\bibitem [{\citenamefont {Lonie}\ and\ \citenamefont
  {Zurek}(2012)}]{Lonie:2012}%
  \BibitemOpen
  \bibfield  {author} {\bibinfo {author} {\bibfnamefont {D.~C.}\ \bibnamefont
  {Lonie}}\ and\ \bibinfo {author} {\bibfnamefont {E.}~\bibnamefont {Zurek}},\
  }\bibfield  {title} {\bibinfo {title} {Identifying duplicate crystal
  structures: Xtalcomp, an open-source solution},\ }\href@noop {} {\bibfield
  {journal} {\bibinfo  {journal} {Comput. Phys. Commun.}\ }\textbf {\bibinfo
  {volume} {183}},\ \bibinfo {pages} {690} (\bibinfo {year}
  {2012})}\BibitemShut {NoStop}%
\bibitem [{\citenamefont {Savin}(2005)}]{ELF}%
  \BibitemOpen
  \bibfield  {author} {\bibinfo {author} {\bibfnamefont {A.}~\bibnamefont
  {Savin}},\ }\bibfield  {title} {\bibinfo {title} {The electron localization
  function (elf) and its relatives: Interpretations and difficulties},\
  }\href@noop {} {\bibfield  {journal} {\bibinfo  {journal} {J. Mol.
  Struc.-Theochem.}\ }\textbf {\bibinfo {volume} {727}},\ \bibinfo {pages}
  {127} (\bibinfo {year} {2005})}\BibitemShut {NoStop}%
\bibitem [{\citenamefont {Kresse}\ and\ \citenamefont
  {Furthm{\"u}ller}(1996)}]{kresse1996efficient}%
  \BibitemOpen
  \bibfield  {author} {\bibinfo {author} {\bibfnamefont {G.}~\bibnamefont
  {Kresse}}\ and\ \bibinfo {author} {\bibfnamefont {J.}~\bibnamefont
  {Furthm{\"u}ller}},\ }\bibfield  {title} {\bibinfo {title} {Efficient
  iterative schemes for \textit{Ab Initio} total-energy calculations using a
  plane-wave basis set},\ }\href@noop {} {\bibfield  {journal} {\bibinfo
  {journal} {Phys. Rev. B}\ }\textbf {\bibinfo {volume} {54}},\ \bibinfo
  {pages} {11169} (\bibinfo {year} {1996})}\BibitemShut {NoStop}%
\bibitem [{\citenamefont {Kresse}\ and\ \citenamefont
  {Joubert}(1999)}]{kresse1999ultrasoft}%
  \BibitemOpen
  \bibfield  {author} {\bibinfo {author} {\bibfnamefont {G.}~\bibnamefont
  {Kresse}}\ and\ \bibinfo {author} {\bibfnamefont {D.}~\bibnamefont
  {Joubert}},\ }\bibfield  {title} {\bibinfo {title} {From ultrasoft
  pseudopotentials to the projector augmented-wave method},\ }\href@noop {}
  {\bibfield  {journal} {\bibinfo  {journal} {Phys. Rev. B}\ }\textbf {\bibinfo
  {volume} {59}},\ \bibinfo {pages} {1758} (\bibinfo {year}
  {1999})}\BibitemShut {NoStop}%
\bibitem [{\citenamefont {Perdew}\ \emph {et~al.}(1996)\citenamefont {Perdew},
  \citenamefont {Burke},\ and\ \citenamefont {Ernzerhof}}]{Perdew:1996}%
  \BibitemOpen
  \bibfield  {author} {\bibinfo {author} {\bibfnamefont {J.~P.}\ \bibnamefont
  {Perdew}}, \bibinfo {author} {\bibfnamefont {K.}~\bibnamefont {Burke}},\ and\
  \bibinfo {author} {\bibfnamefont {M.}~\bibnamefont {Ernzerhof}},\ }\bibfield
  {title} {\bibinfo {title} {Generalized gradient approximation made simple},\
  }\href@noop {} {\bibfield  {journal} {\bibinfo  {journal} {Phys. Rev. Lett.}\
  }\textbf {\bibinfo {volume} {77}},\ \bibinfo {pages} {3865} (\bibinfo {year}
  {1996})}\BibitemShut {NoStop}%
\bibitem [{\citenamefont {Bl{\"o}chl}(1994)}]{blochl1994projector}%
  \BibitemOpen
  \bibfield  {author} {\bibinfo {author} {\bibfnamefont {P.~E.}\ \bibnamefont
  {Bl{\"o}chl}},\ }\bibfield  {title} {\bibinfo {title} {Projector
  augmented-wave method},\ }\href@noop {} {\bibfield  {journal} {\bibinfo
  {journal} {Phys. Rev. B}\ }\textbf {\bibinfo {volume} {50}},\ \bibinfo
  {pages} {17953} (\bibinfo {year} {1994})}\BibitemShut {NoStop}%
\bibitem [{\citenamefont {Monkhorst}\ and\ \citenamefont
  {Pack}(1976)}]{monkhorst1976special}%
  \BibitemOpen
  \bibfield  {author} {\bibinfo {author} {\bibfnamefont {H.~J.}\ \bibnamefont
  {Monkhorst}}\ and\ \bibinfo {author} {\bibfnamefont {J.~D.}\ \bibnamefont
  {Pack}},\ }\bibfield  {title} {\bibinfo {title} {Special points for
  brillouin-zone integrations},\ }\href@noop {} {\bibfield  {journal} {\bibinfo
   {journal} {Phys. Rev. B}\ }\textbf {\bibinfo {volume} {13}},\ \bibinfo
  {pages} {5188} (\bibinfo {year} {1976})}\BibitemShut {NoStop}%
\bibitem [{\citenamefont {Togo}\ and\ \citenamefont
  {Tanaka}(2015)}]{Togo:2015}%
  \BibitemOpen
  \bibfield  {author} {\bibinfo {author} {\bibfnamefont {A.}~\bibnamefont
  {Togo}}\ and\ \bibinfo {author} {\bibfnamefont {I.}~\bibnamefont {Tanaka}},\
  }\bibfield  {title} {\bibinfo {title} {First principles phonon calculations
  in materials science},\ }\href@noop {} {\bibfield  {journal} {\bibinfo
  {journal} {Scr. Mater.}\ }\textbf {\bibinfo {volume} {108}},\ \bibinfo
  {pages} {1} (\bibinfo {year} {2015})}\BibitemShut {NoStop}%
\bibitem [{\citenamefont {Birch}(1952)}]{Birch:1952}%
  \BibitemOpen
  \bibfield  {author} {\bibinfo {author} {\bibfnamefont {F.}~\bibnamefont
  {Birch}},\ }\href@noop {} {\bibfield  {journal} {\bibinfo  {journal} {J.
  Geophys Res.}\ }\textbf {\bibinfo {volume} {57}},\ \bibinfo {pages} {227}
  (\bibinfo {year} {1952})}\BibitemShut {NoStop}%
\bibitem [{\citenamefont {Deringer}\ \emph {et~al.}(2011)\citenamefont
  {Deringer}, \citenamefont {Tchougr{\'e}eff},\ and\ \citenamefont
  {Dronskowski}}]{Deringer:2011}%
  \BibitemOpen
  \bibfield  {author} {\bibinfo {author} {\bibfnamefont {V.~L.}\ \bibnamefont
  {Deringer}}, \bibinfo {author} {\bibfnamefont {A.~L.}\ \bibnamefont
  {Tchougr{\'e}eff}},\ and\ \bibinfo {author} {\bibfnamefont {R.}~\bibnamefont
  {Dronskowski}},\ }\bibfield  {title} {\bibinfo {title} {Crystal orbital
  hamilton population (cohp) analysis as projected from plane-wave basis
  sets},\ }\href@noop {} {\bibfield  {journal} {\bibinfo  {journal} {J. Phys.
  Chem. A}\ }\textbf {\bibinfo {volume} {115}},\ \bibinfo {pages} {5461}
  (\bibinfo {year} {2011})}\BibitemShut {NoStop}%
\bibitem [{\citenamefont {Maintz}\ \emph {et~al.}(2016)\citenamefont {Maintz},
  \citenamefont {Deringer}, \citenamefont {Tchougr{\'e}eff},\ and\
  \citenamefont {Dronskowski}}]{Maintz:2016}%
  \BibitemOpen
  \bibfield  {author} {\bibinfo {author} {\bibfnamefont {S.}~\bibnamefont
  {Maintz}}, \bibinfo {author} {\bibfnamefont {V.~L.}\ \bibnamefont
  {Deringer}}, \bibinfo {author} {\bibfnamefont {A.~L.}\ \bibnamefont
  {Tchougr{\'e}eff}},\ and\ \bibinfo {author} {\bibfnamefont {R.}~\bibnamefont
  {Dronskowski}},\ }\bibfield  {title} {\bibinfo {title} {Lobster: A tool to
  extract chemical bonding from plane-wave based dft},\ }\href@noop {}
  {\bibfield  {journal} {\bibinfo  {journal} {J. Comp. Chem.}\ }\textbf
  {\bibinfo {volume} {37}},\ \bibinfo {pages} {1030} (\bibinfo {year}
  {2016})}\BibitemShut {NoStop}%
\bibitem [{\citenamefont {Dronskowski}\ and\ \citenamefont
  {Bl{\"o}chl}(1993)}]{Dronskowski:1993}%
  \BibitemOpen
  \bibfield  {author} {\bibinfo {author} {\bibfnamefont {R.}~\bibnamefont
  {Dronskowski}}\ and\ \bibinfo {author} {\bibfnamefont {P.~E.}\ \bibnamefont
  {Bl{\"o}chl}},\ }\bibfield  {title} {\bibinfo {title} {Crystal orbital
  hamilton populations (cohp): energy-resolved visualization of chemical
  bonding in solids based on density-functional calculations},\ }\href@noop {}
  {\bibfield  {journal} {\bibinfo  {journal} {J. Phys. Chem.}\ }\textbf
  {\bibinfo {volume} {97}},\ \bibinfo {pages} {8617} (\bibinfo {year}
  {1993})}\BibitemShut {NoStop}%
\bibitem [{\citenamefont {Galeev}\ \emph {et~al.}(2013)\citenamefont {Galeev},
  \citenamefont {Dunnington}, \citenamefont {Schmidt},\ and\ \citenamefont
  {Boldyrev}}]{Galeev:2013}%
  \BibitemOpen
  \bibfield  {author} {\bibinfo {author} {\bibfnamefont {T.~R.}\ \bibnamefont
  {Galeev}}, \bibinfo {author} {\bibfnamefont {B.~D.}\ \bibnamefont
  {Dunnington}}, \bibinfo {author} {\bibfnamefont {J.~R.}\ \bibnamefont
  {Schmidt}},\ and\ \bibinfo {author} {\bibfnamefont {A.~I.}\ \bibnamefont
  {Boldyrev}},\ }\bibfield  {title} {\bibinfo {title} {Solid state adaptive
  natural density partitioning: a tool for deciphering multi-center bonding in
  periodic systems},\ }\href@noop {} {\bibfield  {journal} {\bibinfo  {journal}
  {Phys. Chem. Chem. Phys.}\ }\textbf {\bibinfo {volume} {15}},\ \bibinfo
  {pages} {5022} (\bibinfo {year} {2013})}\BibitemShut {NoStop}%
\bibitem [{\citenamefont {Zubarev}\ and\ \citenamefont
  {Boldyrev}(2008)}]{Zubarev:2008}%
  \BibitemOpen
  \bibfield  {author} {\bibinfo {author} {\bibfnamefont {D.~Y.}\ \bibnamefont
  {Zubarev}}\ and\ \bibinfo {author} {\bibfnamefont {A.~I.}\ \bibnamefont
  {Boldyrev}},\ }\bibfield  {title} {\bibinfo {title} {Developing paradigms of
  chemical bonding: adaptive natural density partitioning},\ }\href@noop {}
  {\bibfield  {journal} {\bibinfo  {journal} {Phys. Chem. Chem. Phys.}\
  }\textbf {\bibinfo {volume} {10}},\ \bibinfo {pages} {5207} (\bibinfo {year}
  {2008})}\BibitemShut {NoStop}%
\bibitem [{\citenamefont {Weinhold}\ and\ \citenamefont {Landis}(2005)}]{NBO}%
  \BibitemOpen
  \bibfield  {author} {\bibinfo {author} {\bibfnamefont {F.}~\bibnamefont
  {Weinhold}}\ and\ \bibinfo {author} {\bibfnamefont {C.~R.}\ \bibnamefont
  {Landis}},\ }\href@noop {} {\emph {\bibinfo {title} {Valency and Bonding: A
  Natural Bond Orbital Donor-Acceptor Perspective}}}\ (\bibinfo  {publisher}
  {Cambridge University Press},\ \bibinfo {year} {2005})\BibitemShut {NoStop}%
\bibitem [{\citenamefont {Dunnington}\ and\ \citenamefont
  {Schmidt}(2012)}]{Dunnington:2012}%
  \BibitemOpen
  \bibfield  {author} {\bibinfo {author} {\bibfnamefont {B.~D.}\ \bibnamefont
  {Dunnington}}\ and\ \bibinfo {author} {\bibfnamefont {J.~R.}\ \bibnamefont
  {Schmidt}},\ }\bibfield  {title} {\bibinfo {title} {Generalization of natural
  bond orbital analysis to periodic systems: Applications to solids and
  surfaces via plane-wave density functional theory},\ }\href@noop {}
  {\bibfield  {journal} {\bibinfo  {journal} {J. Chem. Theory. Comput.}\
  }\textbf {\bibinfo {volume} {8}},\ \bibinfo {pages} {1902} (\bibinfo {year}
  {2012})}\BibitemShut {NoStop}%
\bibitem [{\citenamefont {Dunning}(1989)}]{Dunning:1989}%
  \BibitemOpen
  \bibfield  {author} {\bibinfo {author} {\bibfnamefont {J.~T.~H.}\
  \bibnamefont {Dunning}},\ }\bibfield  {title} {\bibinfo {title} {Gaussian
  basis sets for use in correlated molecular calculations. i. the atoms boron
  through neon and hydrogen},\ }\href@noop {} {\bibfield  {journal} {\bibinfo
  {journal} {J. Chem. Phys.}\ }\textbf {\bibinfo {volume} {90}},\ \bibinfo
  {pages} {1007} (\bibinfo {year} {1989})}\BibitemShut {NoStop}%
\bibitem [{\citenamefont {Momma}\ and\ \citenamefont
  {Izumi}(2011)}]{Momma:2011}%
  \BibitemOpen
  \bibfield  {author} {\bibinfo {author} {\bibfnamefont {K.}~\bibnamefont
  {Momma}}\ and\ \bibinfo {author} {\bibfnamefont {F.}~\bibnamefont {Izumi}},\
  }\bibfield  {title} {\bibinfo {title} {Vesta 3 for three-dimensional
  visualization of crystal, volumetric, and morphology data},\ }\href@noop {}
  {\bibfield  {journal} {\bibinfo  {journal} {J. Appl. Crystallogr.}\ }\textbf
  {\bibinfo {volume} {44}},\ \bibinfo {pages} {1272} (\bibinfo {year}
  {2011})}\BibitemShut {NoStop}%
\bibitem [{dia()}]{diamond}%
  \BibitemOpen
  \href {http://www.crystalimpact.com/diamond} {\bibinfo {title} {Diamond -
  crystal and molecular structure visualization}}\BibitemShut {NoStop}%
\bibitem [{\citenamefont {Fan}\ \emph {et~al.}(2015)\citenamefont {Fan},
  \citenamefont {Li},\ and\ \citenamefont {Wang}}]{Fan:2014}%
  \BibitemOpen
  \bibfield  {author} {\bibinfo {author} {\bibfnamefont {C.}~\bibnamefont
  {Fan}}, \bibinfo {author} {\bibfnamefont {J.}~\bibnamefont {Li}},\ and\
  \bibinfo {author} {\bibfnamefont {L.}~\bibnamefont {Wang}},\ }\bibfield
  {title} {\bibinfo {title} {Phase transitions, mechanical properties, and
  electronic structures of novel boron phases under high-pressure: A
  first-principles study},\ }\href@noop {} {\bibfield  {journal} {\bibinfo
  {journal} {Sci. Rep.}\ }\textbf {\bibinfo {volume} {4}},\ \bibinfo {pages}
  {6786} (\bibinfo {year} {2015})}\BibitemShut {NoStop}%
\bibitem [{\citenamefont {Parija}\ \emph {et~al.}(2018)\citenamefont {Parija},
  \citenamefont {Waetzig}, \citenamefont {Andrews},\ and\ \citenamefont
  {Banerjee}}]{Parija:2018a}%
  \BibitemOpen
  \bibfield  {author} {\bibinfo {author} {\bibfnamefont {A.}~\bibnamefont
  {Parija}}, \bibinfo {author} {\bibfnamefont {G.~R.}\ \bibnamefont {Waetzig}},
  \bibinfo {author} {\bibfnamefont {J.~L.}\ \bibnamefont {Andrews}},\ and\
  \bibinfo {author} {\bibfnamefont {S.}~\bibnamefont {Banerjee}},\ }\bibfield
  {title} {\bibinfo {title} {Traversing energy landscapes away from
  equilibrium: Strategies for accessing and utilizing metastable phase space},\
  }\href@noop {} {\bibfield  {journal} {\bibinfo  {journal} {J. Phys. Chem. C}\
  }\textbf {\bibinfo {volume} {122}},\ \bibinfo {pages} {25709} (\bibinfo
  {year} {2018})}\BibitemShut {NoStop}%
\bibitem [{\citenamefont {Sun}\ \emph {et~al.}(2016)\citenamefont {Sun},
  \citenamefont {Dacek}, \citenamefont {Ong}, \citenamefont {Hautier},
  \citenamefont {Jain}, \citenamefont {Richards}, \citenamefont {Gamst},
  \citenamefont {Persson},\ and\ \citenamefont {Ceder}}]{materialsproject}%
  \BibitemOpen
  \bibfield  {author} {\bibinfo {author} {\bibfnamefont {W.}~\bibnamefont
  {Sun}}, \bibinfo {author} {\bibfnamefont {S.~T.}\ \bibnamefont {Dacek}},
  \bibinfo {author} {\bibfnamefont {S.~P.}\ \bibnamefont {Ong}}, \bibinfo
  {author} {\bibfnamefont {G.}~\bibnamefont {Hautier}}, \bibinfo {author}
  {\bibfnamefont {A.}~\bibnamefont {Jain}}, \bibinfo {author} {\bibfnamefont
  {W.~D.}\ \bibnamefont {Richards}}, \bibinfo {author} {\bibfnamefont {A.~C.}\
  \bibnamefont {Gamst}}, \bibinfo {author} {\bibfnamefont {K.~A.}\ \bibnamefont
  {Persson}},\ and\ \bibinfo {author} {\bibfnamefont {G.}~\bibnamefont
  {Ceder}},\ }\bibfield  {title} {\bibinfo {title} {The thermodynamic scale of
  inorganic crystalline metastability},\ }\href@noop {} {\bibfield  {journal}
  {\bibinfo  {journal} {Sci. Adv.}\ }\textbf {\bibinfo {volume} {2}},\ \bibinfo
  {pages} {e1600225} (\bibinfo {year} {2016})}\BibitemShut {NoStop}%
\bibitem [{\citenamefont {Drozdov}\ \emph {et~al.}(2015)\citenamefont
  {Drozdov}, \citenamefont {Eremets},\ and\ \citenamefont
  {Troyan}}]{drozdov2015superconductivity}%
  \BibitemOpen
  \bibfield  {author} {\bibinfo {author} {\bibfnamefont {A.}~\bibnamefont
  {Drozdov}}, \bibinfo {author} {\bibfnamefont {M.}~\bibnamefont {Eremets}},\
  and\ \bibinfo {author} {\bibfnamefont {I.}~\bibnamefont {Troyan}},\
  }\bibfield  {title} {\bibinfo {title} {Superconductivity above 100 k in
  ph$_3$ at high pressures},\ }\href@noop {} {\bibfield  {journal} {\bibinfo
  {journal} {arXiv preprint arXiv:1508.06224}\ } (\bibinfo {year}
  {2015})}\BibitemShut {NoStop}%
\bibitem [{\citenamefont {Shamp}\ \emph {et~al.}(2016)\citenamefont {Shamp},
  \citenamefont {Terpstra}, \citenamefont {Bi}, \citenamefont {Falls},
  \citenamefont {Avery},\ and\ \citenamefont {Zurek}}]{Zurek:2015j}%
  \BibitemOpen
  \bibfield  {author} {\bibinfo {author} {\bibfnamefont {A.}~\bibnamefont
  {Shamp}}, \bibinfo {author} {\bibfnamefont {T.}~\bibnamefont {Terpstra}},
  \bibinfo {author} {\bibfnamefont {T.}~\bibnamefont {Bi}}, \bibinfo {author}
  {\bibfnamefont {Z.}~\bibnamefont {Falls}}, \bibinfo {author} {\bibfnamefont
  {P.}~\bibnamefont {Avery}},\ and\ \bibinfo {author} {\bibfnamefont
  {E.}~\bibnamefont {Zurek}},\ }\bibfield  {title} {\bibinfo {title}
  {Decomposition products of phosphine under pressure: Ph$_2$ stable and
  superconducting?},\ }\href@noop {} {\bibfield  {journal} {\bibinfo  {journal}
  {J. Am. Chem. Soc.}\ }\textbf {\bibinfo {volume} {138}},\ \bibinfo {pages}
  {1884} (\bibinfo {year} {2016})}\BibitemShut {NoStop}%
\bibitem [{\citenamefont {Flores-Livas}\ \emph {et~al.}(2016)\citenamefont
  {Flores-Livas}, \citenamefont {Amsler}, \citenamefont {Heil}, \citenamefont
  {Sanna}, \citenamefont {Boeri}, \citenamefont {Profeta}, \citenamefont
  {Wolverton}, \citenamefont {Goedecker},\ and\ \citenamefont
  {Gross}}]{Flores-Livas:2016}%
  \BibitemOpen
  \bibfield  {author} {\bibinfo {author} {\bibfnamefont {J.~A.}\ \bibnamefont
  {Flores-Livas}}, \bibinfo {author} {\bibfnamefont {M.}~\bibnamefont
  {Amsler}}, \bibinfo {author} {\bibfnamefont {C.}~\bibnamefont {Heil}},
  \bibinfo {author} {\bibfnamefont {A.}~\bibnamefont {Sanna}}, \bibinfo
  {author} {\bibfnamefont {L.}~\bibnamefont {Boeri}}, \bibinfo {author}
  {\bibfnamefont {G.}~\bibnamefont {Profeta}}, \bibinfo {author} {\bibfnamefont
  {C.}~\bibnamefont {Wolverton}}, \bibinfo {author} {\bibfnamefont
  {S.}~\bibnamefont {Goedecker}},\ and\ \bibinfo {author} {\bibfnamefont
  {E.~K.~U.}\ \bibnamefont {Gross}},\ }\bibfield  {title} {\bibinfo {title}
  {Superconductivity in metastable phases of phosphorus-hydride compounds under
  high pressure},\ }\href@noop {} {\bibfield  {journal} {\bibinfo  {journal}
  {Phys. Rev. B}\ }\textbf {\bibinfo {volume} {93}},\ \bibinfo {pages} {020508}
  (\bibinfo {year} {2016})}\BibitemShut {NoStop}%
\bibitem [{\citenamefont {Bi}\ \emph {et~al.}(2017)\citenamefont {Bi},
  \citenamefont {Miller}, \citenamefont {Shamp},\ and\ \citenamefont
  {Zurek}}]{Zurek:2017c}%
  \BibitemOpen
  \bibfield  {author} {\bibinfo {author} {\bibfnamefont {T.}~\bibnamefont
  {Bi}}, \bibinfo {author} {\bibfnamefont {D.~P.}\ \bibnamefont {Miller}},
  \bibinfo {author} {\bibfnamefont {A.}~\bibnamefont {Shamp}},\ and\ \bibinfo
  {author} {\bibfnamefont {E.}~\bibnamefont {Zurek}},\ }\bibfield  {title}
  {\bibinfo {title} {Superconducting phases of phosphorus hydride under
  pressure: Stabilization via mobile molecular hydrogen},\ }\href@noop {}
  {\bibfield  {journal} {\bibinfo  {journal} {Angew. Chem. Int. Ed.}\ }\textbf
  {\bibinfo {volume} {56}},\ \bibinfo {pages} {10192} (\bibinfo {year}
  {2017})}\BibitemShut {NoStop}%
\bibitem [{\citenamefont {Snider}\ \emph {et~al.}(2020)\citenamefont {Snider},
  \citenamefont {Dasenbrock-Gammon}, \citenamefont {McBride}, \citenamefont
  {Debessai}, \citenamefont {Vindana}, \citenamefont {Vencatasamy},
  \citenamefont {Lawler}, \citenamefont {Salamat},\ and\ \citenamefont
  {Dias}}]{Snider:2020a}%
  \BibitemOpen
  \bibfield  {author} {\bibinfo {author} {\bibfnamefont {E.}~\bibnamefont
  {Snider}}, \bibinfo {author} {\bibfnamefont {N.}~\bibnamefont
  {Dasenbrock-Gammon}}, \bibinfo {author} {\bibfnamefont {R.}~\bibnamefont
  {McBride}}, \bibinfo {author} {\bibfnamefont {M.}~\bibnamefont {Debessai}},
  \bibinfo {author} {\bibfnamefont {H.}~\bibnamefont {Vindana}}, \bibinfo
  {author} {\bibfnamefont {K.}~\bibnamefont {Vencatasamy}}, \bibinfo {author}
  {\bibfnamefont {K.~V.}\ \bibnamefont {Lawler}}, \bibinfo {author}
  {\bibfnamefont {A.}~\bibnamefont {Salamat}},\ and\ \bibinfo {author}
  {\bibfnamefont {R.~P.}\ \bibnamefont {Dias}},\ }\bibfield  {title} {\bibinfo
  {title} {Room-temperature superconductivity in a carbonaceous sulfur
  hydride},\ }\href@noop {} {\bibfield  {journal} {\bibinfo  {journal}
  {Nature}\ }\textbf {\bibinfo {volume} {586}},\ \bibinfo {pages} {373}
  (\bibinfo {year} {2020})}\BibitemShut {NoStop}%
\bibitem [{\citenamefont {Cui}\ \emph {et~al.}(2020)\citenamefont {Cui},
  \citenamefont {Bi}, \citenamefont {Shi}, \citenamefont {Li}, \citenamefont
  {Liu}, \citenamefont {Zurek},\ and\ \citenamefont {Hemley}}]{Zurek:2020b}%
  \BibitemOpen
  \bibfield  {author} {\bibinfo {author} {\bibfnamefont {W.}~\bibnamefont
  {Cui}}, \bibinfo {author} {\bibfnamefont {T.}~\bibnamefont {Bi}}, \bibinfo
  {author} {\bibfnamefont {J.}~\bibnamefont {Shi}}, \bibinfo {author}
  {\bibfnamefont {Y.}~\bibnamefont {Li}}, \bibinfo {author} {\bibfnamefont
  {H.}~\bibnamefont {Liu}}, \bibinfo {author} {\bibfnamefont {E.}~\bibnamefont
  {Zurek}},\ and\ \bibinfo {author} {\bibfnamefont {R.~J.}\ \bibnamefont
  {Hemley}},\ }\bibfield  {title} {\bibinfo {title} {Route to high-\emph{Tc}
  superconductivity via ch$_4$ intercalated h$_3$s hydride perovskites},\
  }\href@noop {} {\bibfield  {journal} {\bibinfo  {journal} {Phys. Rev. B}\
  }\textbf {\bibinfo {volume} {101}},\ \bibinfo {pages} {134504} (\bibinfo
  {year} {2020})}\BibitemShut {NoStop}%
\bibitem [{\citenamefont {Sun}\ \emph {et~al.}(2020)\citenamefont {Sun},
  \citenamefont {Tian}, \citenamefont {Jiang}, \citenamefont {Li},
  \citenamefont {Li}, \citenamefont {Iitaka}, \citenamefont {Zhong},\ and\
  \citenamefont {Xie}}]{Sun:2020a}%
  \BibitemOpen
  \bibfield  {author} {\bibinfo {author} {\bibfnamefont {Y.}~\bibnamefont
  {Sun}}, \bibinfo {author} {\bibfnamefont {Y.}~\bibnamefont {Tian}}, \bibinfo
  {author} {\bibfnamefont {B.}~\bibnamefont {Jiang}}, \bibinfo {author}
  {\bibfnamefont {X.}~\bibnamefont {Li}}, \bibinfo {author} {\bibfnamefont
  {H.}~\bibnamefont {Li}}, \bibinfo {author} {\bibfnamefont {T.}~\bibnamefont
  {Iitaka}}, \bibinfo {author} {\bibfnamefont {X.}~\bibnamefont {Zhong}},\ and\
  \bibinfo {author} {\bibfnamefont {Y.}~\bibnamefont {Xie}},\ }\bibfield
  {title} {\bibinfo {title} {Computational discovery of a dynamically stable
  cubic sh$_3$-like high-temperature superconductor at 100~gpa via ch$_4$
  intercalation},\ }\href@noop {} {\bibfield  {journal} {\bibinfo  {journal}
  {Phys. Rev. B}\ }\textbf {\bibinfo {volume} {101}},\ \bibinfo {pages}
  {174102} (\bibinfo {year} {2020})}\BibitemShut {NoStop}%
\bibitem [{\citenamefont {Shatruk}(2019)}]{Shatruk:2019a}%
  \BibitemOpen
  \bibfield  {author} {\bibinfo {author} {\bibfnamefont {M.}~\bibnamefont
  {Shatruk}},\ }\bibfield  {title} {\bibinfo {title} {Thcr2si2 structure type:
  The ``perovskite'' of intermetallics},\ }\href@noop {} {\bibfield  {journal}
  {\bibinfo  {journal} {J. Solid State Chem.}\ }\textbf {\bibinfo {volume}
  {272}},\ \bibinfo {pages} {198} (\bibinfo {year} {2019})}\BibitemShut
  {NoStop}%
\bibitem [{\citenamefont {Dong}\ \emph {et~al.}(2018)\citenamefont {Dong},
  \citenamefont {Wu}, \citenamefont {Yazyev}, \citenamefont {He}, \citenamefont
  {Tian}, \citenamefont {Zhou},\ and\ \citenamefont {Wang}}]{Dong:2018}%
  \BibitemOpen
  \bibfield  {author} {\bibinfo {author} {\bibfnamefont {X.}~\bibnamefont
  {Dong}}, \bibinfo {author} {\bibfnamefont {Q.}~\bibnamefont {Wu}}, \bibinfo
  {author} {\bibfnamefont {O.~V.}\ \bibnamefont {Yazyev}}, \bibinfo {author}
  {\bibfnamefont {X.-L.}\ \bibnamefont {He}}, \bibinfo {author} {\bibfnamefont
  {Y.}~\bibnamefont {Tian}}, \bibinfo {author} {\bibfnamefont {X.-F.}\
  \bibnamefont {Zhou}},\ and\ \bibinfo {author} {\bibfnamefont {H.-T.}\
  \bibnamefont {Wang}},\ }\href@noop {} {\bibinfo {title} {Condensed matter
  realization of fermion quasiparticles in minkowski space}} (\bibinfo {year}
  {2018}),\ \Eprint {https://arxiv.org/abs/1804.00570} {arXiv:1804.00570
  [cond-mat.mtrl-sci]} \BibitemShut {NoStop}%
\bibitem [{\citenamefont {Longuet-Higgins}\ and\ \citenamefont
  {d.~V.~Roberts}(1955)}]{Longuet-Higgings:1955}%
  \BibitemOpen
  \bibfield  {author} {\bibinfo {author} {\bibfnamefont {H.~C.}\ \bibnamefont
  {Longuet-Higgins}}\ and\ \bibinfo {author} {\bibfnamefont {M.}~\bibnamefont
  {d.~V.~Roberts}},\ }\bibfield  {title} {\bibinfo {title} {The electronic
  structure of an icosahedron of boron atoms},\ }\href@noop {} {\bibfield
  {journal} {\bibinfo  {journal} {Proc. Royal Soc. London}\ }\textbf {\bibinfo
  {volume} {A230}},\ \bibinfo {pages} {110} (\bibinfo {year}
  {1955})}\BibitemShut {NoStop}%
\bibitem [{\citenamefont {Jemmis}\ and\ \citenamefont
  {M.}(2001)}]{Jemmis:2001}%
  \BibitemOpen
  \bibfield  {author} {\bibinfo {author} {\bibfnamefont {E.~D.}\ \bibnamefont
  {Jemmis}}\ and\ \bibinfo {author} {\bibfnamefont {B.~M.}\ \bibnamefont
  {M.}},\ }\bibfield  {title} {\bibinfo {title} {Polyhedral boranes and
  elemental boron: Direct structural relations and diverse electronic
  requirements},\ }\href@noop {} {\bibfield  {journal} {\bibinfo  {journal} {J.
  Am. Chem. Soc.}\ }\textbf {\bibinfo {volume} {123}},\ \bibinfo {pages} {4324}
  (\bibinfo {year} {2001})}\BibitemShut {NoStop}%
\bibitem [{\citenamefont {Fujimori}\ \emph {et~al.}(1999)\citenamefont
  {Fujimori}, \citenamefont {Nakata}, \citenamefont {Nakayama}, \citenamefont
  {Nishibori}, \citenamefont {Kimura}, \citenamefont {Takata},\ and\
  \citenamefont {Sakata}}]{Fujimori:1999}%
  \BibitemOpen
  \bibfield  {author} {\bibinfo {author} {\bibfnamefont {M.}~\bibnamefont
  {Fujimori}}, \bibinfo {author} {\bibfnamefont {T.}~\bibnamefont {Nakata}},
  \bibinfo {author} {\bibfnamefont {T.}~\bibnamefont {Nakayama}}, \bibinfo
  {author} {\bibfnamefont {E.}~\bibnamefont {Nishibori}}, \bibinfo {author}
  {\bibfnamefont {K.}~\bibnamefont {Kimura}}, \bibinfo {author} {\bibfnamefont
  {M.}~\bibnamefont {Takata}},\ and\ \bibinfo {author} {\bibfnamefont
  {M.}~\bibnamefont {Sakata}},\ }\bibfield  {title} {\bibinfo {title} {Peculiar
  covalent bonds in $\alpha$-rhombohedral boron},\ }\href@noop {} {\bibfield
  {journal} {\bibinfo  {journal} {Phys. Rev. Lett.}\ }\textbf {\bibinfo
  {volume} {82}},\ \bibinfo {pages} {4452} (\bibinfo {year}
  {1999})}\BibitemShut {NoStop}%
\bibitem [{\citenamefont {Wade}(1976)}]{Wade:1976}%
  \BibitemOpen
  \bibfield  {author} {\bibinfo {author} {\bibfnamefont {K.}~\bibnamefont
  {Wade}},\ }\bibfield  {title} {\bibinfo {title} {Structural and bonding
  patterns in cluster chemistry},\ }\href@noop {} {\bibfield  {journal}
  {\bibinfo  {journal} {Adv. Inorg. Chem. Radiochem.}\ }\textbf {\bibinfo
  {volume} {18}},\ \bibinfo {pages} {1} (\bibinfo {year} {1976})}\BibitemShut
  {NoStop}%
\bibitem [{\citenamefont {Mingos}(1984)}]{Mingos:1984}%
  \BibitemOpen
  \bibfield  {author} {\bibinfo {author} {\bibfnamefont {D.~M.~P.}\
  \bibnamefont {Mingos}},\ }\bibfield  {title} {\bibinfo {title} {Polyhedral
  skeletal electron pair approach},\ }\href@noop {} {\bibfield  {journal}
  {\bibinfo  {journal} {Acc. Chem. Res.}\ }\textbf {\bibinfo {volume} {7}},\
  \bibinfo {pages} {311} (\bibinfo {year} {1984})}\BibitemShut {NoStop}%
\bibitem [{\citenamefont {Teter}(1998)}]{Teter:1998a}%
  \BibitemOpen
  \bibfield  {author} {\bibinfo {author} {\bibfnamefont {D.~M.}\ \bibnamefont
  {Teter}},\ }\bibfield  {title} {\bibinfo {title} {Computational alchemy: The
  search for new superhard materials},\ }\href
  {https://doi.org/10.1557/s0883769400031420} {\bibfield  {journal} {\bibinfo
  {journal} {{MRS} Bull.}\ }\textbf {\bibinfo {volume} {23}},\ \bibinfo {pages}
  {22} (\bibinfo {year} {1998})}\BibitemShut {NoStop}%
\bibitem [{\citenamefont {Chen}\ \emph {et~al.}(2011)\citenamefont {Chen},
  \citenamefont {Niu}, \citenamefont {Li},\ and\ \citenamefont
  {Li}}]{Chen:2011a}%
  \BibitemOpen
  \bibfield  {author} {\bibinfo {author} {\bibfnamefont {X.-Q.}\ \bibnamefont
  {Chen}}, \bibinfo {author} {\bibfnamefont {H.}~\bibnamefont {Niu}}, \bibinfo
  {author} {\bibfnamefont {D.}~\bibnamefont {Li}},\ and\ \bibinfo {author}
  {\bibfnamefont {Y.}~\bibnamefont {Li}},\ }\bibfield  {title} {\bibinfo
  {title} {Modeling hardness of polycrystalline materials and bulk metallic
  glasses},\ }\href {https://doi.org/10.1016/j.intermet.2011.03.026} {\bibfield
   {journal} {\bibinfo  {journal} {Intermetallics}\ }\textbf {\bibinfo {volume}
  {19}},\ \bibinfo {pages} {1275} (\bibinfo {year} {2011})}\BibitemShut
  {NoStop}%
\bibitem [{\citenamefont {Isayev}\ \emph {et~al.}(2017)\citenamefont {Isayev},
  \citenamefont {Oses}, \citenamefont {Toher}, \citenamefont {Gossett},
  \citenamefont {Curtarolo},\ and\ \citenamefont {Tropsha}}]{Isayev:2017a}%
  \BibitemOpen
  \bibfield  {author} {\bibinfo {author} {\bibfnamefont {O.}~\bibnamefont
  {Isayev}}, \bibinfo {author} {\bibfnamefont {C.}~\bibnamefont {Oses}},
  \bibinfo {author} {\bibfnamefont {C.}~\bibnamefont {Toher}}, \bibinfo
  {author} {\bibfnamefont {E.}~\bibnamefont {Gossett}}, \bibinfo {author}
  {\bibfnamefont {S.}~\bibnamefont {Curtarolo}},\ and\ \bibinfo {author}
  {\bibfnamefont {A.}~\bibnamefont {Tropsha}},\ }\bibfield  {title} {\bibinfo
  {title} {Universal fragment descriptors for predicting properties of
  inorganic crystals},\ }\href {https://doi.org/10.1038/ncomms15679} {\bibfield
   {journal} {\bibinfo  {journal} {Nat. Commun.}\ }\textbf {\bibinfo {volume}
  {8}},\ \bibinfo {pages} {15679} (\bibinfo {year} {2017})}\BibitemShut
  {NoStop}%
\bibitem [{\citenamefont {Curtarolo}\ \emph {et~al.}(2012)\citenamefont
  {Curtarolo}, \citenamefont {Setyawan}, \citenamefont {Wang}, \citenamefont
  {Xue}, \citenamefont {Yang}, \citenamefont {Taylor}, \citenamefont {Nelson},
  \citenamefont {Hart}, \citenamefont {Sanvito}, \citenamefont
  {Buongiorno-Nardelli}, \citenamefont {Mingo},\ and\ \citenamefont
  {Levy}}]{Curtarolo:2012b}%
  \BibitemOpen
  \bibfield  {author} {\bibinfo {author} {\bibfnamefont {S.}~\bibnamefont
  {Curtarolo}}, \bibinfo {author} {\bibfnamefont {W.}~\bibnamefont {Setyawan}},
  \bibinfo {author} {\bibfnamefont {S.}~\bibnamefont {Wang}}, \bibinfo {author}
  {\bibfnamefont {J.}~\bibnamefont {Xue}}, \bibinfo {author} {\bibfnamefont
  {K.}~\bibnamefont {Yang}}, \bibinfo {author} {\bibfnamefont {R.~H.}\
  \bibnamefont {Taylor}}, \bibinfo {author} {\bibfnamefont {L.~J.}\
  \bibnamefont {Nelson}}, \bibinfo {author} {\bibfnamefont {G.~L.}\
  \bibnamefont {Hart}}, \bibinfo {author} {\bibfnamefont {S.}~\bibnamefont
  {Sanvito}}, \bibinfo {author} {\bibfnamefont {M.}~\bibnamefont
  {Buongiorno-Nardelli}}, \bibinfo {author} {\bibfnamefont {N.}~\bibnamefont
  {Mingo}},\ and\ \bibinfo {author} {\bibfnamefont {O.}~\bibnamefont {Levy}},\
  }\bibfield  {title} {\bibinfo {title} {{AFLOWLIB}.{ORG}: A distributed
  materials properties repository from high-throughput ab initio
  calculations},\ }\href {https://doi.org/10.1016/j.commatsci.2012.02.002}
  {\bibfield  {journal} {\bibinfo  {journal} {Comput. Mater. Sci.}\ }\textbf
  {\bibinfo {volume} {58}},\ \bibinfo {pages} {227} (\bibinfo {year}
  {2012})}\BibitemShut {NoStop}%
\bibitem [{\citenamefont {Toher}\ \emph {et~al.}(2018)\citenamefont {Toher},
  \citenamefont {Oses}, \citenamefont {Hicks}, \citenamefont {Gossett},
  \citenamefont {Rose}, \citenamefont {Nath}, \citenamefont {Usanmaz},
  \citenamefont {Ford}, \citenamefont {Perim}, \citenamefont {Calderon},
  \citenamefont {Plata}, \citenamefont {Lederer}, \citenamefont {Jahnatek},
  \citenamefont {Setyawan}, \citenamefont {Wang}, \citenamefont {Xue},
  \citenamefont {Rasch}, \citenamefont {Chepulskii}, \citenamefont {Taylor},
  \citenamefont {Gomez}, \citenamefont {Shi}, \citenamefont {Supka},
  \citenamefont {Al~Orabi}, \citenamefont {Gopal}, \citenamefont {Cerasoli},
  \citenamefont {Liyanage}, \citenamefont {Wang}, \citenamefont {Siloi},
  \citenamefont {Agapito}, \citenamefont {Syshadham}, \citenamefont {Hart},
  \citenamefont {Carrete}, \citenamefont {Legrain}, \citenamefont {Mingo},
  \citenamefont {Zurek}, \citenamefont {Isayev}, \citenamefont {Tropsha},
  \citenamefont {Sanvito}, \citenamefont {Hanson}, \citenamefont {Takeuchi},
  \citenamefont {Mehl}, \citenamefont {Kolmogorov}, \citenamefont {Yang},
  \citenamefont {D'Amico}, \citenamefont {Calzolari}, \citenamefont {Costa},
  \citenamefont {De~Gennaro}, \citenamefont {Nardelli}, \citenamefont
  {Fornari}, \citenamefont {Levy},\ and\ \citenamefont
  {Curtarolo}}]{Zurek:2017o}%
  \BibitemOpen
  \bibfield  {author} {\bibinfo {author} {\bibfnamefont {C.}~\bibnamefont
  {Toher}}, \bibinfo {author} {\bibfnamefont {C.}~\bibnamefont {Oses}},
  \bibinfo {author} {\bibfnamefont {D.}~\bibnamefont {Hicks}}, \bibinfo
  {author} {\bibfnamefont {E.}~\bibnamefont {Gossett}}, \bibinfo {author}
  {\bibfnamefont {F.}~\bibnamefont {Rose}}, \bibinfo {author} {\bibfnamefont
  {P.}~\bibnamefont {Nath}}, \bibinfo {author} {\bibfnamefont {D.}~\bibnamefont
  {Usanmaz}}, \bibinfo {author} {\bibfnamefont {D.~C.}\ \bibnamefont {Ford}},
  \bibinfo {author} {\bibfnamefont {E.}~\bibnamefont {Perim}}, \bibinfo
  {author} {\bibfnamefont {C.~E.}\ \bibnamefont {Calderon}}, \bibinfo {author}
  {\bibfnamefont {J.~J.}\ \bibnamefont {Plata}}, \bibinfo {author}
  {\bibfnamefont {Y.}~\bibnamefont {Lederer}}, \bibinfo {author} {\bibfnamefont
  {M.}~\bibnamefont {Jahnatek}}, \bibinfo {author} {\bibfnamefont
  {W.}~\bibnamefont {Setyawan}}, \bibinfo {author} {\bibfnamefont
  {S.}~\bibnamefont {Wang}}, \bibinfo {author} {\bibfnamefont {J.}~\bibnamefont
  {Xue}}, \bibinfo {author} {\bibfnamefont {K.}~\bibnamefont {Rasch}}, \bibinfo
  {author} {\bibfnamefont {R.~V.}\ \bibnamefont {Chepulskii}}, \bibinfo
  {author} {\bibfnamefont {R.~H.}\ \bibnamefont {Taylor}}, \bibinfo {author}
  {\bibfnamefont {G.}~\bibnamefont {Gomez}}, \bibinfo {author} {\bibfnamefont
  {H.}~\bibnamefont {Shi}}, \bibinfo {author} {\bibfnamefont {A.~R.}\
  \bibnamefont {Supka}}, \bibinfo {author} {\bibfnamefont {R.~A.~R.}\
  \bibnamefont {Al~Orabi}}, \bibinfo {author} {\bibfnamefont {P.}~\bibnamefont
  {Gopal}}, \bibinfo {author} {\bibfnamefont {F.~T.}\ \bibnamefont {Cerasoli}},
  \bibinfo {author} {\bibfnamefont {L.}~\bibnamefont {Liyanage}}, \bibinfo
  {author} {\bibfnamefont {H.}~\bibnamefont {Wang}}, \bibinfo {author}
  {\bibfnamefont {I.}~\bibnamefont {Siloi}}, \bibinfo {author} {\bibfnamefont
  {L.~A.}\ \bibnamefont {Agapito}}, \bibinfo {author} {\bibfnamefont
  {C.}~\bibnamefont {Syshadham}}, \bibinfo {author} {\bibfnamefont {G.~L.}\
  \bibnamefont {Hart}}, \bibinfo {author} {\bibfnamefont {J.}~\bibnamefont
  {Carrete}}, \bibinfo {author} {\bibfnamefont {F.}~\bibnamefont {Legrain}},
  \bibinfo {author} {\bibfnamefont {N.}~\bibnamefont {Mingo}}, \bibinfo
  {author} {\bibfnamefont {E.}~\bibnamefont {Zurek}}, \bibinfo {author}
  {\bibfnamefont {O.}~\bibnamefont {Isayev}}, \bibinfo {author} {\bibfnamefont
  {A.}~\bibnamefont {Tropsha}}, \bibinfo {author} {\bibfnamefont
  {S.}~\bibnamefont {Sanvito}}, \bibinfo {author} {\bibfnamefont {R.~M.}\
  \bibnamefont {Hanson}}, \bibinfo {author} {\bibfnamefont {I.}~\bibnamefont
  {Takeuchi}}, \bibinfo {author} {\bibfnamefont {M.~J.}\ \bibnamefont {Mehl}},
  \bibinfo {author} {\bibfnamefont {A.~N.}\ \bibnamefont {Kolmogorov}},
  \bibinfo {author} {\bibfnamefont {K.}~\bibnamefont {Yang}}, \bibinfo {author}
  {\bibfnamefont {P.}~\bibnamefont {D'Amico}}, \bibinfo {author} {\bibfnamefont
  {A.}~\bibnamefont {Calzolari}}, \bibinfo {author} {\bibfnamefont
  {M.}~\bibnamefont {Costa}}, \bibinfo {author} {\bibfnamefont
  {R.}~\bibnamefont {De~Gennaro}}, \bibinfo {author} {\bibfnamefont {M.~B.}\
  \bibnamefont {Nardelli}}, \bibinfo {author} {\bibfnamefont {M.}~\bibnamefont
  {Fornari}}, \bibinfo {author} {\bibfnamefont {O.}~\bibnamefont {Levy}},\ and\
  \bibinfo {author} {\bibfnamefont {S.}~\bibnamefont {Curtarolo}},\ }\bibfield
  {title} {\bibinfo {title} {The aflow fleet for materials discovery},\ }in\
  \href@noop {} {\emph {\bibinfo {booktitle} {Handbook of Materials Modeling.
  Volume 1 Methods: Theory and Modeling}}},\ Vol.~\bibinfo {volume} {1},\
  \bibinfo {editor} {edited by\ \bibinfo {editor} {\bibfnamefont
  {W.}~\bibnamefont {Andreoni}}\ and\ \bibinfo {editor} {\bibfnamefont
  {S.}~\bibnamefont {Yip}}}\ (\bibinfo  {publisher} {Springer, Cham},\ \bibinfo
  {address} {Switzerland},\ \bibinfo {year} {2018})\ pp.\ \bibinfo {pages}
  {1--28}\BibitemShut {NoStop}%
\bibitem [{\citenamefont {Avery}\ \emph {et~al.}(2019)\citenamefont {Avery},
  \citenamefont {Wang}, \citenamefont {Oses}, \citenamefont {Gossett},
  \citenamefont {Proserpio}, \citenamefont {Toher}, \citenamefont {Curtarolo},\
  and\ \citenamefont {Zurek}}]{Zurek:2019b}%
  \BibitemOpen
  \bibfield  {author} {\bibinfo {author} {\bibfnamefont {P.}~\bibnamefont
  {Avery}}, \bibinfo {author} {\bibfnamefont {X.}~\bibnamefont {Wang}},
  \bibinfo {author} {\bibfnamefont {C.}~\bibnamefont {Oses}}, \bibinfo {author}
  {\bibfnamefont {E.}~\bibnamefont {Gossett}}, \bibinfo {author} {\bibfnamefont
  {D.}~\bibnamefont {Proserpio}}, \bibinfo {author} {\bibfnamefont
  {C.}~\bibnamefont {Toher}}, \bibinfo {author} {\bibfnamefont
  {S.}~\bibnamefont {Curtarolo}},\ and\ \bibinfo {author} {\bibfnamefont
  {E.}~\bibnamefont {Zurek}},\ }\bibfield  {title} {\bibinfo {title}
  {Predicting superhard materials via a machine learning informed evolutionary
  structure search},\ }\href@noop {} {\bibfield  {journal} {\bibinfo  {journal}
  {Npj Comput. Mater.}\ }\textbf {\bibinfo {volume} {5}},\ \bibinfo {pages} {89
  (1} (\bibinfo {year} {2019})}\BibitemShut {NoStop}%
\bibitem [{\citenamefont {Gao}\ and\ \citenamefont {Gao}(2010)}]{Gao:2010a}%
  \BibitemOpen
  \bibfield  {author} {\bibinfo {author} {\bibfnamefont {F.~M.}\ \bibnamefont
  {Gao}}\ and\ \bibinfo {author} {\bibfnamefont {L.~H.}\ \bibnamefont {Gao}},\
  }\bibfield  {title} {\bibinfo {title} {Microscopic models of hardness},\
  }\href {https://doi.org/10.3103/s1063457610030020} {\bibfield  {journal}
  {\bibinfo  {journal} {J. Superhard Mater.}\ }\textbf {\bibinfo {volume}
  {32}},\ \bibinfo {pages} {148} (\bibinfo {year} {2010})}\BibitemShut
  {NoStop}%
\bibitem [{ccr()}]{ccr}%
  \BibitemOpen
  \href@noop {} {}\bibinfo {note} {Http://hdl.handle.net/10477/79221, July 05,
  2020}\BibitemShut {NoStop}%
\end{thebibliography}%

\end{document}